\documentclass[11pt,a4paper]{article}
\usepackage{jheppub}
\usepackage{indentfirst}
\usepackage{amsmath}
\usepackage{bigstrut}
\usepackage{amssymb}
\usepackage{array}
\usepackage{multirow}
\usepackage{etoolbox}
\usepackage{graphicx}
\usepackage{amsthm}
\usepackage{slashed}
\usepackage{array}
\usepackage{epstopdf}
\usepackage{fancyhdr}
\usepackage[table,xcdraw]{xcolor}
\usepackage{float}
\usepackage{subfigure}
\usepackage{hyperref}
\usepackage{url}

\title{Probing 6D Operators at Future $e^-e^+$ Colliders}
\author[a,b]{Wen Han Chiu,}
\author[a,d]{Sze Ching Leung,}
\author[a]{Tao Liu,}
\author[a]{Kun-Feng Lyu,}
\author[b,c]{Lian-Tao Wang}

\affiliation[a]{Department of Physics, The Hong Kong University of Science and Technology,
Clear Water Bay, Kowloon, Hong Kong S.A.R., P.R.C.}
\affiliation[b]{Enrico Fermi Institute, University of Chicago, Chicago, Illinois 60637, USA}
\affiliation[c]{Kavli Institute for Cosmological Physics, University of Chicago, Chicago, Illinois 60637, USA}
\affiliation[d]{Department of Physics and Astronomy, University of Pittsburgh, PA 15260, USA}

\emailAdd{whchiuaa@connect.ust.hk}
\emailAdd{scleungac@connect.ust.hk}
\emailAdd{taoliu@ust.hk}
\emailAdd{klyuaa@connect.ust.hk}
\emailAdd{liantaow@uchicago.edu}

\abstract{We explore the sensitivities at future $e^-e^+$ colliders to probe a set of six-dimensional operators which can modify the SM predictions on Higgs physics and electroweak precision measurements. We consider the case in which the operators are turned on simultaneously. Such an analysis yields a ``conservative'' interpretation on the collider sensitivities, complementary to the ``optimistic'' scenario where the operators are individually probed. After a detail analysis at CEPC in both ``conservative'' and ``optimistic'' scenarios, we also considered the sensitivities for  FCC-ee and ILC. As an illustration of the potential of constraining new physics models, we applied sensitivity analysis to two benchmarks: holographic composite Higgs model and littlest Higgs model.}

\begin{document}
\maketitle
\unitlength = 1mm
\section{Introduction}

The discovery of the Higgs Boson~\cite{Aad:2012tfa,Chatrchyan:2012xdj} is a great success of the Standard Model (SM). However, there are still many open questions left unanswered, such as hierarchy problem, dark matter, and cosmic baryon asymmetry. They hint at the existence of physics beyond the SM (BSM).  One of the promising venue of probing  BSM physics is through measuring the deviation of Higgs properties. This strongly motivates the construction of an electron-positron collider as a Higgs factory within next decades to study the underlying BSM physics. The proposed Higgs factories include the International Linear Collider (ILC, Japan)~\cite{Weiglein:2004hn,Baer:2013cma,Asner:2013psa}, the Circular Electron Positron Collider (CEPC, China)~\cite{CEPC-SPPCStudyGroup:2015csa} and the Future Electron-Positron Circular Collider (FCC-ee, CERN)~\cite{Gomez-Ceballos:2013zzn}.

In an effective field theory (EFT) of the SM, the leading effects of BSM physics above the electroweak (EW) scale can be parametrized by a set of six-dimensional (6D) operators
\begin{equation}
\mathcal{L}_{\text{eff}} = \mathcal{L}_{\text{SM}} + \sum_{i} \frac{c_i}{\Lambda^2} \mathcal{O}_i  \ .
\end{equation}
Here $\mathcal{L}_{\text{SM}}$ describes physics in the SM. $c_i$ and $\Lambda$ denote dimensionless Wilson coefficients and the cutoff scale defined by the BSM physics, respectively. Among these operators, 59 are CP-even and 17 are CP-odd. The form of the operators depends on the choice of  basis \cite{Hagiwara:1993ck,Grzadkowski:2010es,Giudice:2007fh,Elias-Miro:2013eta,Elias-Miro:2013mua,Henning:2014wua}.

Since the discovery of Higgs boson, the probe of the 6D operators, particularly the ones motivated by Higgs physics, at LHC and future $e^-e^+$ colliders has been extensively studied \cite{Pomarol:2013zra,Craig:2014una,Falkowski:2014tna,Corbett:2015ksa,Ellis:2015sca,Ge:2016zro,Ellis:2016yrj,Durieux:2017rsg,Jiang:2016czg,Amar:2014fpa,Jana:2017hqg}.  There are different strategies in analyzing the sensitivities to new physics. It can be done with only a single operator tuning on at a time, which provides an ``optimistic'' projection of the sensitivities at the future $e^-e^+$ colliders. However, new physics models tend to generate multiple such operators. Without assuming a particular model, one could go to the other extreme by turning on all operators simultaneously without assuming any correlation among them. Such an analysis, a primary effort in this paper, will result in a ``conservative'' interpretation on collider sensitivities due to cancellation effects among the multiple contributions. Despite this, we should keep in mind that while this approach give some information about potential degeneracies and correlations in interpreting the measurements, it is not directly applicable to specific models. New physics models typically generate a smaller set of independent operators, equivalently, predicts correlations between different operators in the complete set. For that case, one can analyze the experimental constraints or the collider sensitivities straightforwardly, utilizing the correlation matrix predicted by the specific models. It is not necessary (and also impossible) to go through all potential new physics models, for the purpose of qualitatively demonstrating the capability of a future collider. As an illustration, we pursued such analyses in two benchmark models: the  holographic composite Higgs model and littlest Higgs model. 

Our study partially overlaps with some recent studies on the sensitivities of probing the SM EFT at future $e^-e^+$ colliders~\cite{Durieux:2017rsg,Barklow:2017suo,Barklow:2017awn,Gu:2017ckc}. The study in ref.~\cite{Durieux:2017rsg} was pursued under a yet-to-be-explicitly-established assumption that the 6D EW operators can be constrained sufficiently well. Different from that, we incorporate the sensitivity analysis for these 6D EW operators, without making any first working assumption about them. This may yield a significant impact for the sensitivity discussions on the triple gauge coupling (TGC) measurement. In addition, a recently proposed operating scenario (see, e.g.,~\cite{FCC-ee}) is assumed for the FCC-ee analysis. Refs~\cite{Barklow:2017suo,Barklow:2017awn} took similar strategies, with the results presented in the ``$\kappa$''-scheme and in the 6D operator-scheme, respectively. Compared to these analysis, we focus more on the comparative studies on the sensitivities in the ``optimistic'' and ``conservative'' scenarios, and the sensitivities at the CEPC, ILC and FCC-ee. More than that, there exist some differences between the operator sets studied and the observables applied. We include the operator $\mathcal{O}^{(3)l}_{LL}$ (as is defined in Table~\ref{tab:ops}) in the analysis which was ignored in~\cite{Barklow:2017awn}. But, unlike~\cite{Barklow:2017awn} (and also~\cite{Durieux:2017rsg}), our analysis does not include the Higgs decay observables, and correspondingly several operators which are sensitive to them. As for the study in~\cite{Gu:2017ckc}, it mainly focused on the interpretation of the collider sensitivities in concrete benchmarks. 

We organize this article in the following way. We will introduce the analysis formalism and the observables applied in Section 2 ad Section 3, respectively. The analysis and its results will be presented in Section 4. In this section, we will pursue a $\chi^2$ fit on the sensitivities of probing the 6D operators at CEPC, in both ``optimistic'' and ``conservative'' interpretations. Then we will make a comparative study on the sensitivities at CEPC, FCC-ee, ILC250 (with data at 250 GeV and below) and ILC (with full data), and look into the operators $\mathcal {O}_{6}$ in details which is difficult to probe. We will apply the analysis to study the theory of SILH in Section 5, analyzing the collider sensitivities to probe its benchmarks: holographic composite Higgs model~\cite{Agashe:2004rs,Contino:2006qr} and littlest Higgs model~\cite{ArkaniHamed:2002qy}. We conclude in Section 6. More technical details and analysis results can be found in Appendix.

\section{Analysis Formalism}

There are 13 6D operators which are relevant to the $e^-e^+ \to ZH$ production: 10 CP-even and 3 CP-odd ones. In this article, we focus only on the CP-even ones.  We also include the triple gauge boson operator since it is often generated together with these ones in new physics scenarios. These 11 operators are summarized in Table~\ref{tab:ops}.
This is a subset of the operators in the so called Warsaw basis~\cite{Grzadkowski:2010es}, omitting operators with quarks. 

\begin{table}[h]
\centering
\begin{tabular}{|l|l|l|}
\hline
$\mathcal{O}_{WW}=g^2|H|^2W^{a}_{\mu \nu}W^{a,\mu \nu}$ & $\mathcal{O}_{T}=\frac{1}{2}(H^{\dagger}\overset{\text{$\leftrightarrow$}}{D}_{\mu}H)^2$  & $\mathcal{O}^{(3)l}_{L}=(iH^{\dagger}{\sigma}^{a}\overset{\text{$\leftrightarrow$}}{D}_{\mu}H)(\bar{L}_L{\gamma}^{\mu}{\sigma}^{a}L_L)$\\
$\mathcal{O}_{WB}=gg'H^{\dagger}{\sigma}^aHW^{a}_{\mu \nu}B^{\mu \nu}$ & $\mathcal{O}_{H}=\frac{1}{2}({\partial}_{\mu}|H|^2)^2$  & $\mathcal{O}^{(3)l}_{LL}=(\bar{L}_L{\gamma}_{\mu}{\sigma}^{a}L_L)(\bar{L}_L{\gamma}^{\mu}{\sigma}^{a}L_L)$  \\
$\mathcal{O}_{BB}={g'}^2|H|^2B_{\mu \nu}B^{\mu \nu}$ & $\mathcal{O}_6 = \lambda |H^\dagger H|^3$ & $\mathcal{O}^l_L=(iH^{\dagger}\overset{\text{$\leftrightarrow$}}{D}_{\mu}H)(\bar{L}_{L}{\gamma}^{\mu}L_L)$   \\
$\mathcal{O}_{3W}=g\dfrac{\varepsilon_{abc}}{3!}W_\mu^{a \nu}W_\nu^{b \rho}W_\rho^{a \mu}$ &     & $\mathcal{O}^e_R=(iH^{\dagger}\overset{\text{$\leftrightarrow$}}{D}_{\mu}H)(\bar{l}_{R}{\gamma}^{\mu}l_R)$   \\ \hline
\end{tabular}
\caption{The 6D operators used in this study, with $\lambda = \frac{3 m_h^3}{v^2}$ in $\mathcal{O}_6$.}  \label{tab:ops}
\end{table}

These 11 operators can influence physics at the EW scale in four ways: (1) renormalizing wave function; (2) shifting the definition of EW parameters; (3) modifying the existing SM couplings (including the charge shifting in the gauge boson currents) and (4) inducing new vertices.  

We begin with wave-function renormalization. $\mathcal{O}_{WW}$, $\mathcal{O}_{WB}$, $\mathcal{O}_{BB}$ and $\mathcal{O}_{H}$ will modify the kinetic terms of the gauge or Higgs fields. First, we note that $\frac{c_{WW} }{2\Lambda^2} g^2 v^2 W^{a\mu\nu} W^a_{\mu\nu}$ and $\frac{c_{BB} }{2\Lambda^2} g'^2 v^2 B^{\mu\nu} B_{\mu\nu}$ can be absorbed into a redefinition of SM electorweak gauge couplings. With this, the canonically normalized SM gauge and Higgs fields are
\begin{equation}
\begin{split}
h & = Z_h h'= \left(1-\frac{v^2}{2\Lambda^2} c_H  \right)h'\\
W^{\mu} & = Z_W W'^\mu =  W'^\mu \\
Z^\mu &= Z_Z Z'^\mu =\Big(1+\dfrac{v^2}{\Lambda^2} c_w s_w g g' c_{WB} \Big) Z'^\mu \\
A^\mu &= Z_A A'^\mu + \delta Z_X Z'^\mu \\
&= \Big( 1- \dfrac{v^2}{\Lambda^2} c_w s_w g g' c_{WB} \Big)A'^\mu-  \dfrac{v^2}{\Lambda^2} (c_w^2 - s_w^2) g g' c_{WB} Z'^\mu
\end{split}
\end{equation}
Here $g$, $g'$ are the $SU(2)$ and $U(1)$ gauge couplings and $c_w$ and $s_w$ are the cosine and sine of the Weinberg angle. $Z_{h,W,Z,A}$ are the rescaling factors.
$\mathcal{O}_{WW}$ and $\mathcal{O}_{BB}$ operators can be probed only via the newly introduced vertices like $h Z_{\mu\nu} Z^{\mu\nu}$.

Similarly, though it does not result in a renormalization of the Higgs field,  the operator $\mathcal{O}_6$ can modify the Higgs potential, yielding a shift in the Higgs VEV and mass.  Such a shift can be absorbed by  the definition of the Fermi constant.  The effect of $\mathcal{O}_6$ can be probed only via its contribution to the cubic and quartic Higgs coupling.

Three input  parameters of the EW sector in the SM, typically chosen to be $\{\alpha, m_Z, G_F\}$, receive shifts induced by the 6D operators
\begin{equation}
\begin{split}
G^{\rm sm}_F &= G^{(r)}_F \left( 1 + \dfrac{2 \left(c^{(3)l}_{LL} - c_L^{(3)l}\right) v_{sm}^2}{\Lambda^2} \right)\\
m^{\rm sm}_Z &=  m^{(r)}_Z \left( 1- \delta Z_Z + \dfrac{c_T v_{sm}^2}{2\Lambda^2} \right)\\
\alpha^{\rm sm} &= \alpha^{(r)}(1-2 \delta Z_A)
\end{split}
\end{equation}
with $\delta Z_Z = Z_Z -1$ and $\delta Z_A = Z_A -1$. Here the superscripts ``$\text{sm}$'' represents the SM definition, and ``$(r)$'' represents the reference or the measured central value used as input for the fit. Then the parameter shifts can be denoted as
\begin{equation}
m^{\rm sm}_Z = m^{(r)}_Z \Big( 1+\frac{\delta m_Z }{m_Z^{(r)}} \Big)\quad G^{\rm sm}_F = G_F^{(r)} \Big( 1+\frac{\delta G_F }{G_F^{(r)}} \Big ) \quad \alpha^{\rm sm} = \alpha^{(r)} \Big( 1+\frac{\delta \alpha }{\alpha^{(r)}} \Big ) \ ,
\end{equation}
with
\begin{eqnarray}
\dfrac{\delta m_Z}{m_Z^{(r)}} = - \delta Z_Z + \dfrac{c_T v^2}{2\Lambda^2}  \quad
\dfrac{\delta G_F}{G_F^{(r)}} = \dfrac{2 (c^{(3)l}_{LL} - c_L^{(3)l}) v_{\rm sm}^2}{\Lambda^2} \quad
\dfrac{\delta \alpha}{\alpha^{(r)}} = -2 \delta Z_A  \ .
\end{eqnarray}
This formalism is independent of the definition of the field renormalization factors $\delta Z_Z$ and $\delta Z_A$. Hence, in addition to affect the observable directly, D6 operators can also contribution to the deviation from SM prediction by shifting the definition of input parameters.

From here on,  we will  suppress the superscript $(r)$ for the measured observables, unless specified. Since $v^2_{sm} / \Lambda^2$ differs with $v^2 / \Lambda^2$ only at $\mathcal O (\frac{v^4}{ \Lambda^{-4}})$ order, we also replace the former with the latter.  The new physics corrections to some observables can be derived directly. One example is
\begin{equation}
s_{2w} = \sin 2 \theta_{w} =  \left( \dfrac{4 \pi \alpha }{\sqrt{2} {G}_{F} m^2_{Z}} \right)^{1/2}
\end{equation}
We find
\begin{equation}
\begin{split}
&\dfrac{\delta s_{2w}}{s_{2w}} = \dfrac{1}{2} \dfrac{\delta \alpha}{\alpha} - \dfrac{1}{2} \dfrac{\delta G_F}{G_F} - \dfrac{\delta m_Z}{m_Z}\\
\Rightarrow \quad& \delta \theta_w = \dfrac{s_w c_w}{2(c_w^2-s_w^2)} \left(\dfrac{\delta \alpha}{\alpha} - \dfrac{\delta G_F}{G_F} - \dfrac{2\delta m_Z}{m_Z} \right) \ .
\end{split}   \label{sintheta}
\end{equation}
Another example is
\begin{equation}
g_Z = \dfrac{g}{c_w} = \dfrac{ 4 \sqrt{\pi \alpha}}{s_{2w}} = 2 (\sqrt{2} G_F m_Z^2)^{1/2}
\end{equation}
We have
\begin{eqnarray}
\dfrac{\delta g_Z}{g_Z} = \dfrac{1}{2}\dfrac{\delta G_F}{G_F} + \dfrac{\delta m_Z}{m_Z} \ .  \label{delta_gZ}
\end{eqnarray}
Both of them receive linear corrections from $\mathcal{O}_{WB}$, $\mathcal{O}_{T}$, $\mathcal{O}^{(3)l}_{LL}$ and $\mathcal{O}_L^{(3)l}$.

\section{Observables for Analysis}

Throughout this paper, we will consider three classes of observables: inclusive signal rates of Higgs events,  angular observables in Higgs events, and electroweak precision observables (EWPOs). 
We will not include the total width of Higgs boson and its decay branching ratios. Correspondingly, we will not consider the operators which do not enter the inclusive production rates at tree level, but modify the Higgs decays, such as $h\to b\bar b, \tau \tau$, only.  The incorporation of the Higgs decays as observables could reveal more information about a larger set of operators. We will leave such an important analysis to a future study. Regarding theoretical predictions, we will use ``$\delta$'' to denote the shift caused by wave function renormalization or by definition shift in the EW input parameters. We will use ``$\Delta$'' to denote the total deviation from the reference value for any given observables. 

\subsection{Higgs Events}

{\bf A. Higgs Strahlung Process}

\begin{figure}[H]
\centering
\includegraphics[height=3cm]{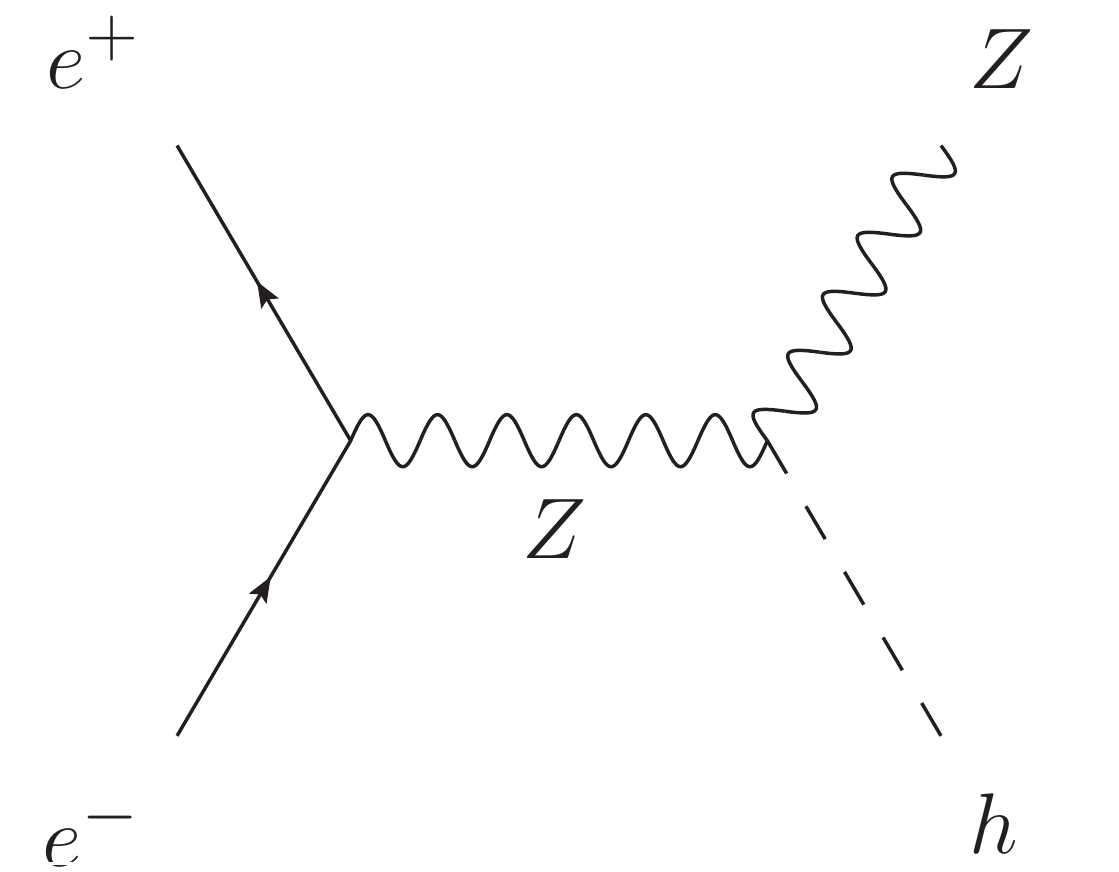}
\includegraphics[height=3cm]{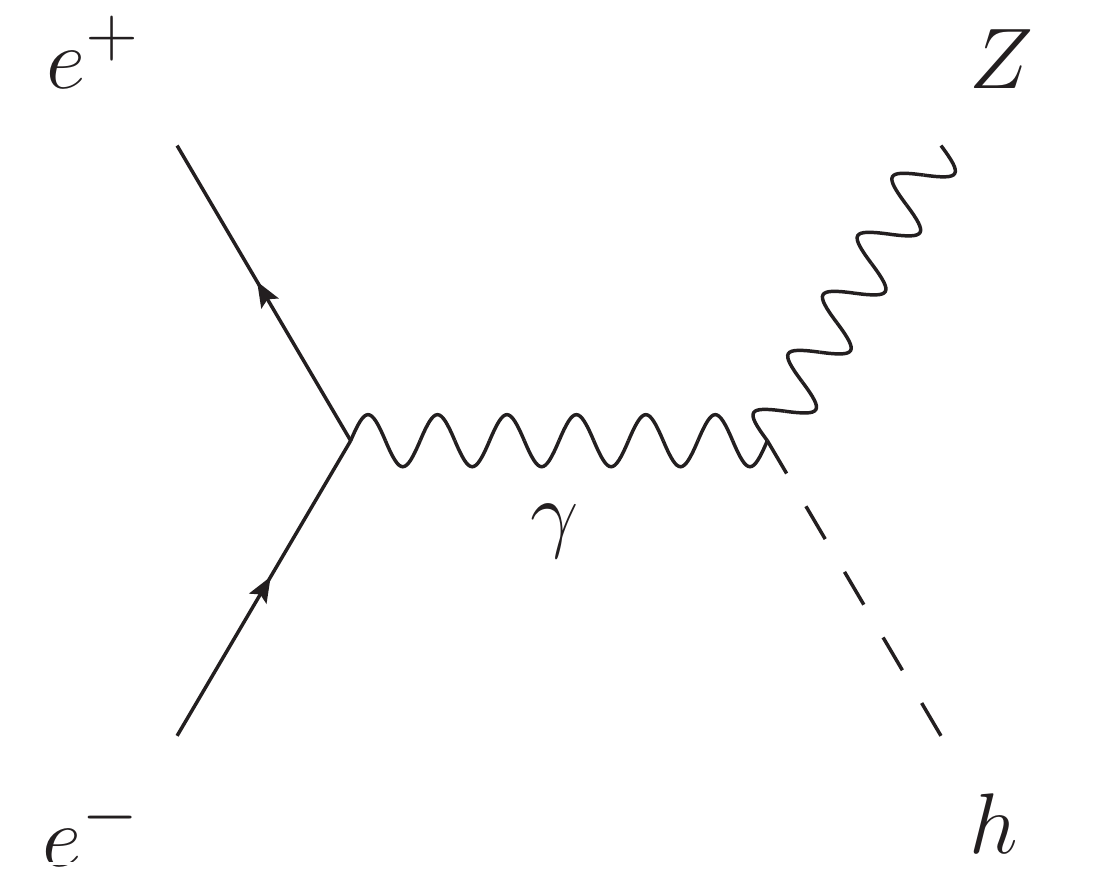}
\includegraphics[height=3cm]{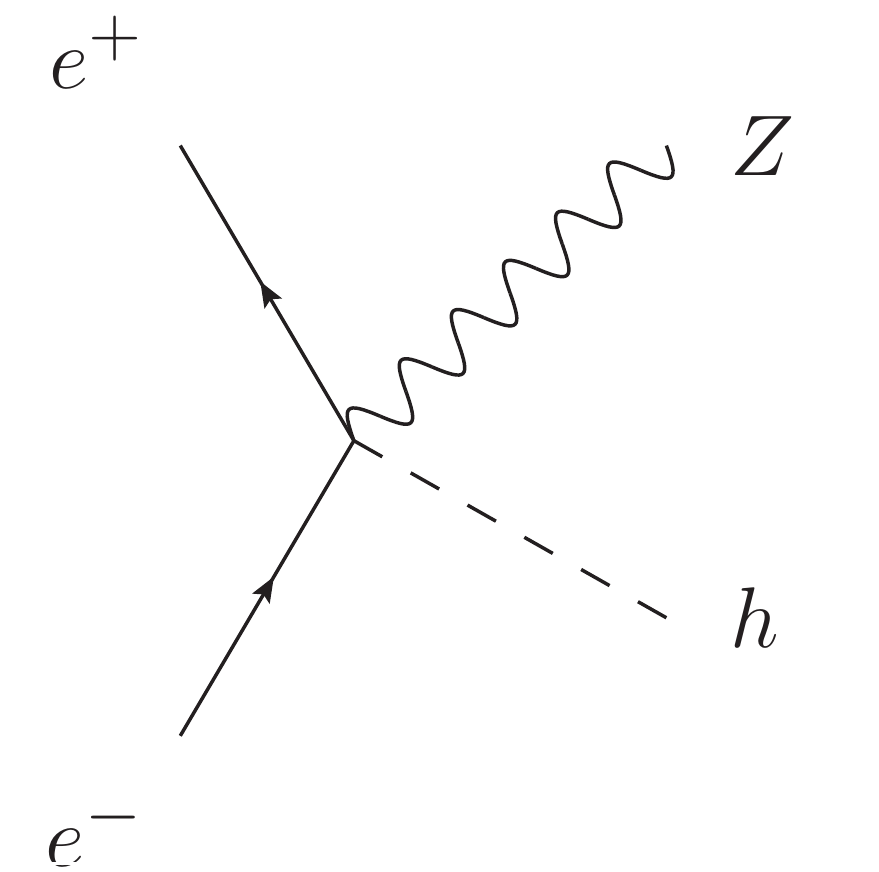}
\caption{Feynman diagrams for Higgsstrahlung process.} \label{ZH_diagram}
\end{figure}

The first important process is $e^+ e^- \rightarrow Z h$, as is shown in Figure~\ref{ZH_diagram}. The signal events can be well-selected using the variable of recoiling mass. At leading order, the relevant Lagrangian is given by
\begin{eqnarray}
\mathcal{L}_{Zh} &\supset&  \dfrac{2 m_Z^2}{v}(1+ c_{ZZ}^{(1)}) hZ_\mu Z^{\mu} + c_{ZZ}^{(2)} hZ_{\mu\nu} Z^{\mu\nu} + c_{AZ} h Z_{\mu\nu} A^{\mu\nu} + g^{(1)}_L Z_\mu \bar{e}_L \gamma^\mu e_L + \\
&&  g^{(1)}_R Z_\mu \bar{e}_R \gamma^\mu e_R + g^{(2)}_L h Z_\mu \bar{e}_L \gamma^\mu e_L +  g^{(2)}_R h Z_\mu \bar{e}_R \gamma^\mu e_R + e A_\mu (\bar{e}_L \gamma^\mu e_L + \bar{e}_R \gamma^\mu e_R)  \nonumber \ ,
\end{eqnarray}
with the coefficients
\begin{equation}
\begin{split}
c_{ZZ}^{(1)} &= \dfrac{1}{2}\dfrac{\delta G_F}{G_F} + \dfrac{2 \delta m_Z}{m_Z} + 2\delta Z_Z + \delta Z_h  \quad\\
c_{ZZ}^{(2)} &= \dfrac{2 v}{\Lambda^2}(c_w^2 g^2 c_{WW} + c_w s_w g g' c_{WB} + s_w^2 g'^2 c_{BB}) \\
c_{AZ} &=\dfrac{2 v}{\Lambda^2} \left(c_w s_w g^2 c_{WW} - \dfrac{1}{2}(c_w^2 - s_w^2) g g' c_{WB} - c_w s_w g'^2 c_{BB}\right) \\
g^{(1)}_L &= g_Z \left(-\dfrac{1}{2}+s_w^2 \right) \left( 1+ \dfrac{\delta g_Z}{g_Z} -\dfrac{2 s_{2w}}{c_{2w}} \delta \theta_w + \delta Z_Z \right) - \dfrac{g_Z v^2}{2\Lambda^2} (c_{L}^{(3)l} + c^l_L) - e \delta Z_X \\
g^{(1)}_R &= g_Z s_w^2 \left( \dfrac{\delta g_Z}{g_Z} + \dfrac{2 c_w}{s_w} \delta \theta_w + \delta Z_Z \right)     -\dfrac{g_Z v^2}{2\Lambda^2} c^e_R - e \delta Z_X \\
g^{(2)}_L &= - g_Z (c^l_L + c^{(3)l}_L)\dfrac{v}{\Lambda^2}  \\
g^{(2)}_R &= - g_Z \dfrac{c^e_R v}{\Lambda^2}
\end{split}
\end{equation}
In this Lagrangian, new vertices appear due to $\mathcal{O}^l_L$, $\mathcal{O}^{(3)l}_L$ and  $\mathcal{O}^e_R$. $\mathcal{O}_{WW}$, $\mathcal{O}_{WB}$ and $\mathcal{O}_{BB}$ also give rise to a term with new Lorentz structure $h Z_{\mu\nu} Z^{\mu\nu}$.  Both yield extra contributions to the production $e^+ e^- \rightarrow Z h$, as is indicated in Figure~\ref{ZH_diagram}.

{\bf B. $WW$ Fusion Process}

\begin{figure}[h]
\centering
\includegraphics[height=2.75cm]{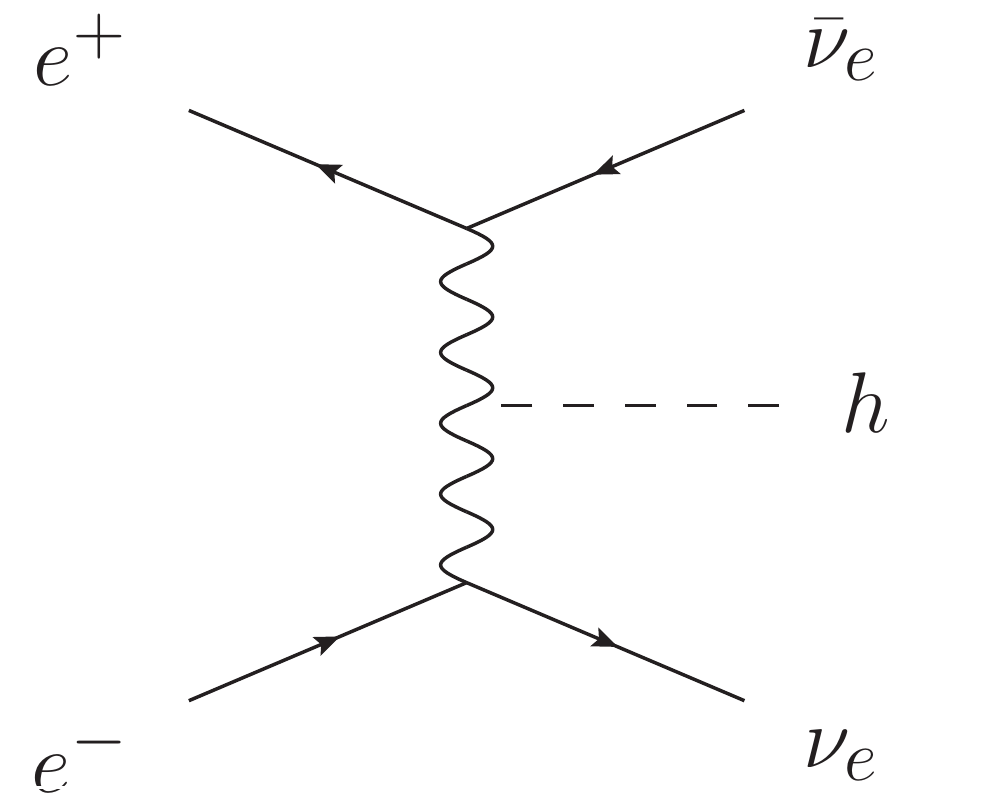}
\includegraphics[height=2.75cm]{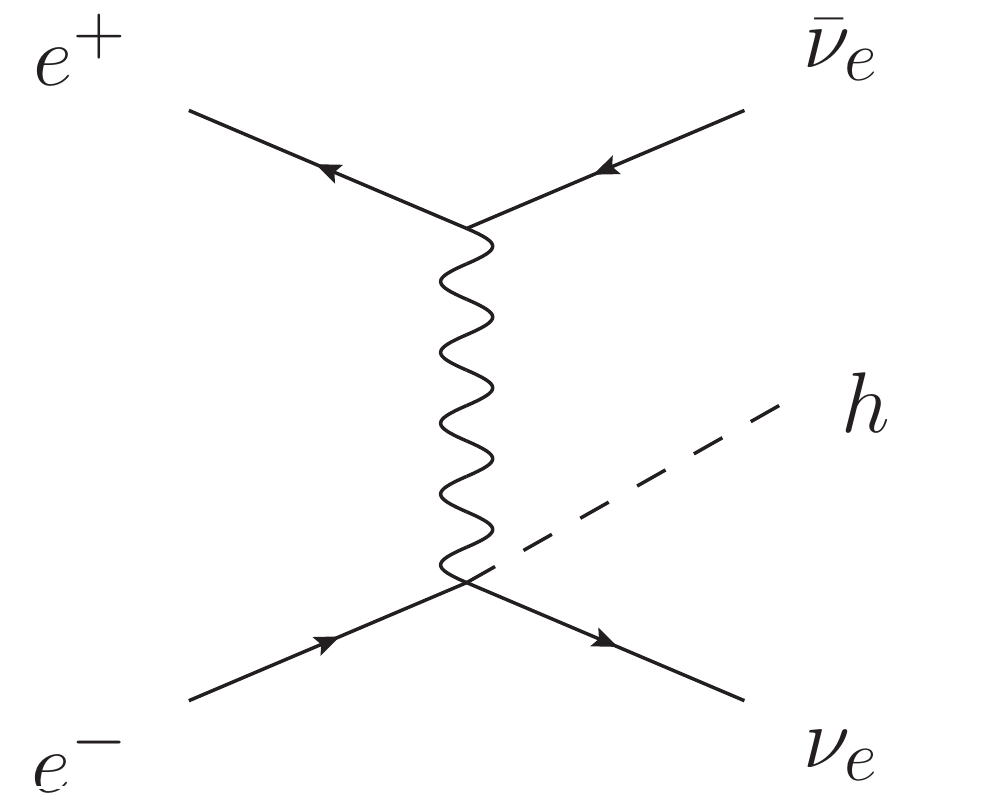}
\includegraphics[height=2.75cm]{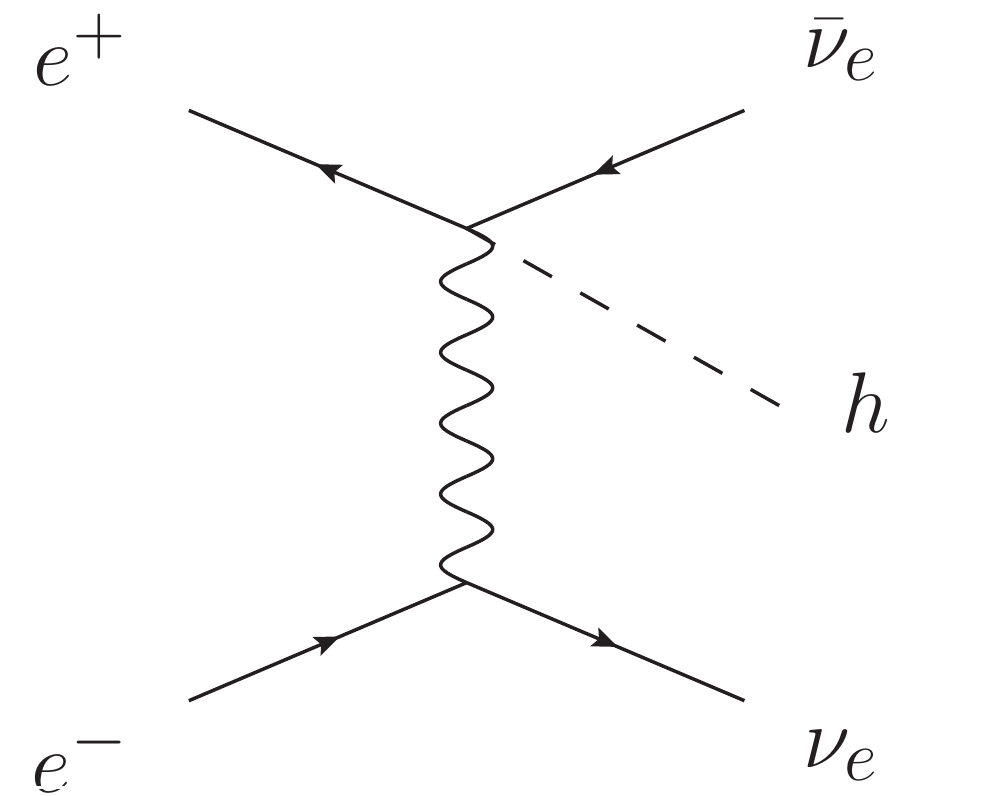}
\includegraphics[height=2.75cm]{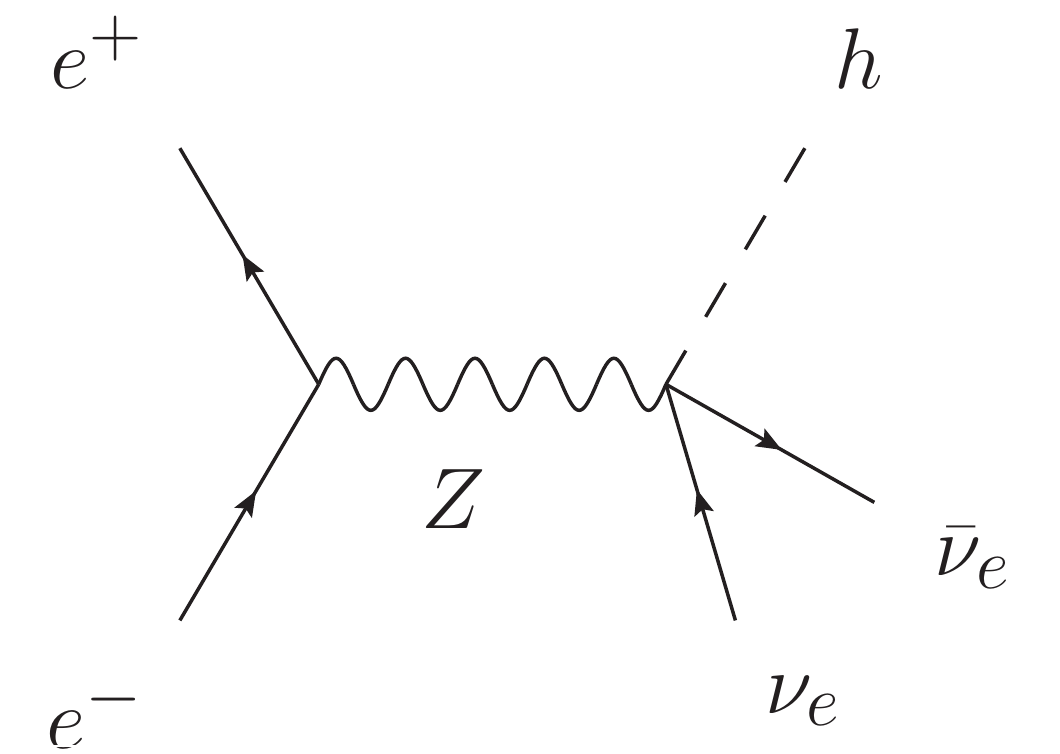}
\caption{Feynman diagrams for $W$-fusion Higgs production.} \label{vvh_diagram}
\end{figure}

Another important process is the $WW$ fusion Higgs production $e^+ e^- \rightarrow \nu_e \bar{\nu}_e h$, as shown in Figure~\ref{vvh_diagram}. Here we didn't take into account the $Z$ associated Higgs production,   with the $Z$ boson decaying into two neutrinos. At leading order, the relevant Lagrangian is given by
\begin{equation}
\begin{split}
\mathcal{L}_{\nu \nu h} \supset \quad & \dfrac{g^2 v}{2}(1+ c_{W}^{(1)}) hW^+_\mu W^{-\mu} + c_{W}^{(2)} hW^+_{\mu\nu} W^{-\mu\nu} + \dfrac{g}{\sqrt{2}} (1+ c^{(3)}_W) (W^+_\mu \bar{\nu}_L \gamma^\mu e_L +\\
& W^-_\mu \bar{e}_L \gamma^\mu \nu_L) + c^{(4)}_W (h W^+_\mu \bar{\nu}_L \gamma^\mu e_L + h W^-_\mu \bar{e}_L \gamma^\mu \nu_L) + c^{(5)} (h Z_\mu \bar{\nu}_L \gamma^\mu \nu_L + h Z_\mu \bar{\nu}_L \gamma^\mu \nu_L)  \ ,
\end{split}
\end{equation}
with the coefficients
\begin{equation}
\begin{split}
c_{W}^{(1)} &= \dfrac{\delta g_Z}{g_Z} - \dfrac{s_w \delta \theta_w}{c_w} -\dfrac{\delta G_F}{2 G_F} + \delta Z_h \quad
c_{W}^{(2)} = \dfrac{2 c_{WW} g^2 v}{\Lambda^2} \\
c_W^{(3)} &= \dfrac{\delta g_Z}{g_Z} - \dfrac{s_w \delta \theta_w}{c_w} + \dfrac{c^{(3)l}_L v^2}{\Lambda^2} \quad
c_W^{(4)} = \dfrac{c^{(3)l}_L g v }{\sqrt{2} \Lambda^2} \quad
c^{(5)} = \dfrac{g_Z}{2}\dfrac{c^l_L -c^{(3)l}_L}{\Lambda^2}
\end{split}
\end{equation}
The Wilson coefficients of ${\mathcal O}_H$, ${\mathcal O}_T$ and ${\mathcal O}^{(3)l}_{LL}$ only appear in $c_{W}^{(1)}$ and $c_{W}^{(3)}$,  resulting in a rescaling of the SM couplings. $\mathcal{O}_{WW}$, $\mathcal{O}^l_L$ and $\mathcal{O}^{(3)l}_L$ yield two new vertices.

{\bf C. $Z$-Associated Di-Higgs Process}

\begin{figure}[h]
\centering
\includegraphics[height=3cm]{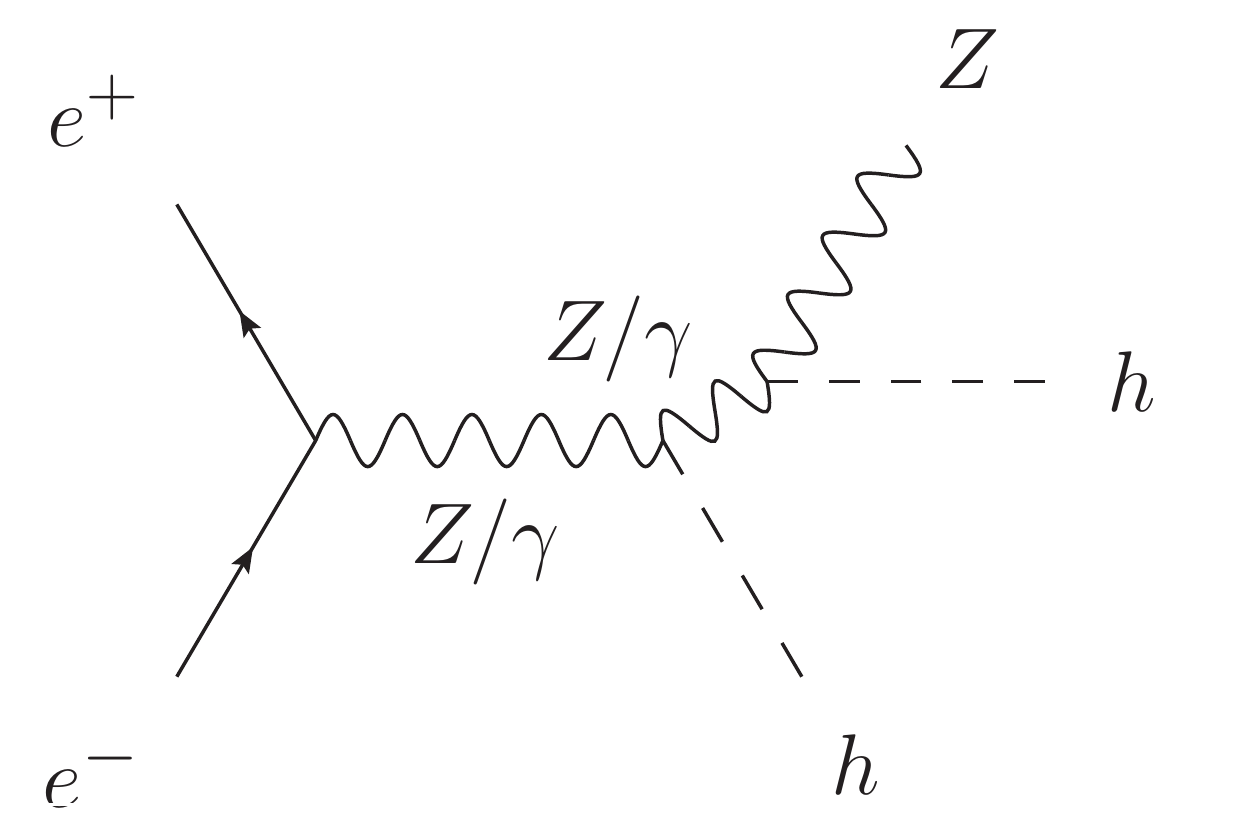}
\includegraphics[height=3cm]{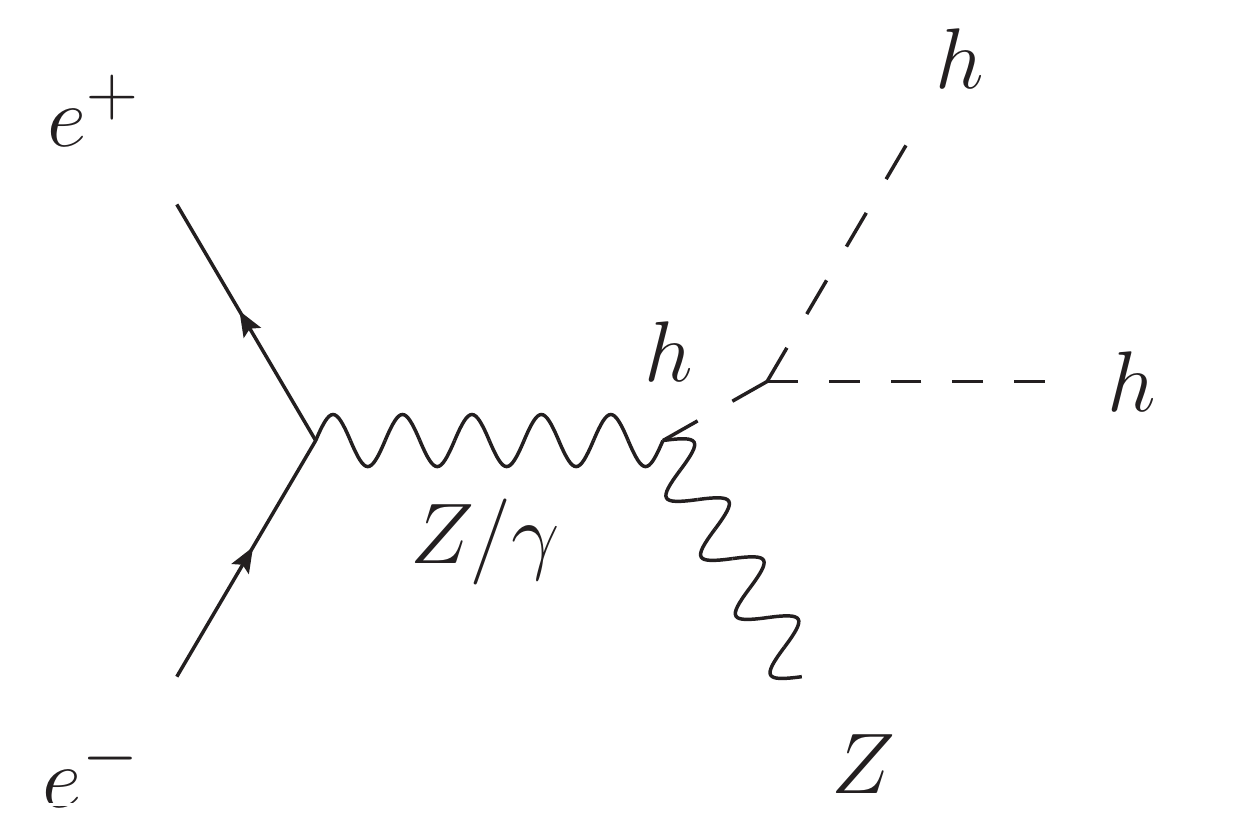}
\includegraphics[height=3cm]{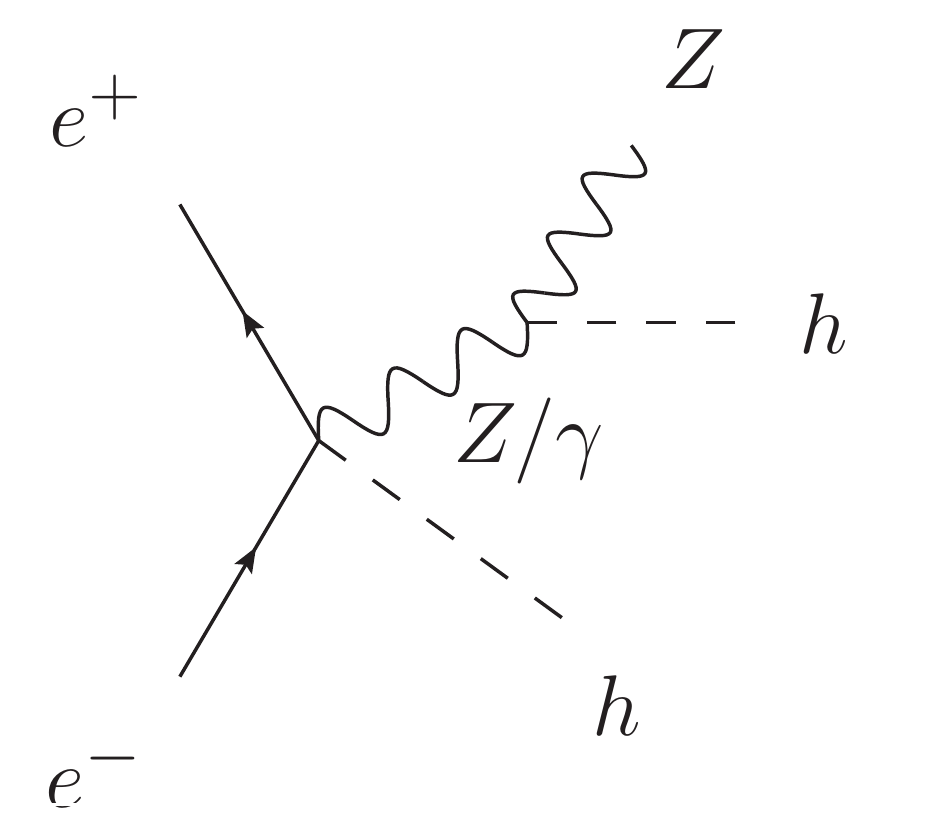}
\includegraphics[height=3cm]{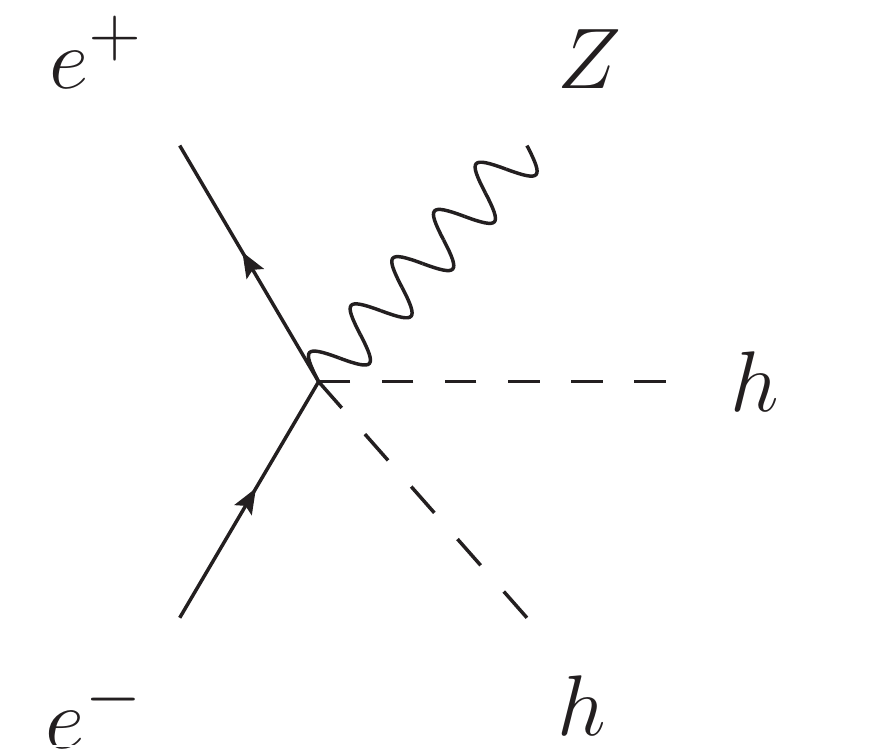}
\includegraphics[height=3cm]{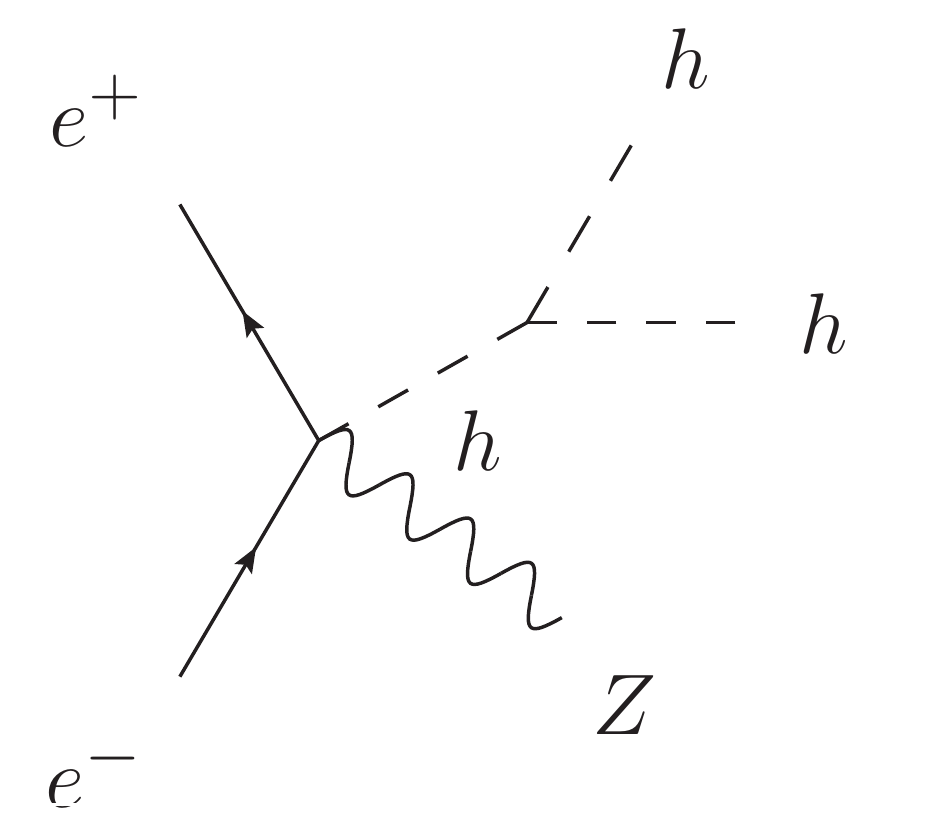}
\includegraphics[height=3cm]{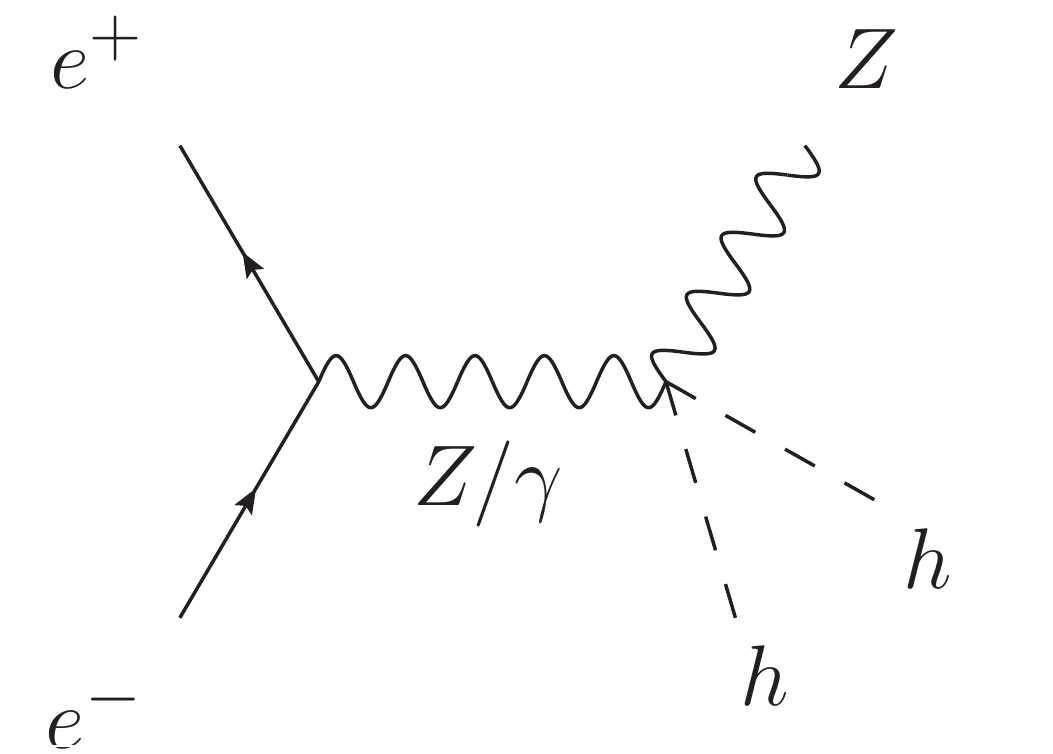}
\caption{Feynman diagrams for di-Higgs production.} \label{zhh_diagram}
\end{figure}

As the beam energy increases, di-Higgs channel switches on. An  important channel is the Z association production process $e^- e^+ \rightarrow Z h h$. 
The relevant Lagrangian for this channel is
\begin{equation}
\begin{split}
\mathcal{L}_{Zhh} \supset \quad & \mathcal{L}_{Zh} + (1+ c_{ZZ}^{(3)} )h h Z^\mu Z_\mu + c_{ZZ}^{(4)} h h Z^{\mu\nu} Z_{\mu\nu}+ c_{AZ}^{(2)} h h Z^{\mu\nu} A_{\mu\nu}+\\
& g^{(3)}_L h h Z_\mu \bar{e}_L \gamma^\mu e_L +  g^{(3)}_R h h Z_\mu \bar{e}_R \gamma^\mu e_R - (1+ \kappa_3) \dfrac{m_h^2}{2v}  h^3 + \dfrac{2 c_H v}{\Lambda^2} h\partial_\mu h \partial^\mu h
\end{split}
\end{equation}
with the coefficients
\begin{equation}\label{k3formula}
\begin{split}
c_{ZZ}^{(3)} &= \dfrac{\delta G_F}{G_F} + \dfrac{2 \delta m_Z}{m_Z} + 2\delta Z_Z + 2\delta Z_h  \quad
c_{ZZ}^{(4)} = \dfrac{\delta Z_Z}{v^2} \quad c_{AZ}^{(2)} = \dfrac{\delta Z_X}{v^2} \\
g^{(3)}_L &= - g_Z \dfrac{c^l_L + c^{(3)l}_L}{\Lambda^2}  \quad
g^{(3)}_R = - g_Z \dfrac{c^e_R}{\Lambda^2} \quad
\kappa_3 = -\dfrac{2\lambda  c_6 v^4}{ m_h^2 \Lambda^2} + \dfrac{\delta G_F}{2 G_F} + 3 \delta Z_h
\end{split}
\end{equation}

\subsection{Higgs Production Angular Observables}

\begin{figure}[h]
\centering
\includegraphics[width=14cm]{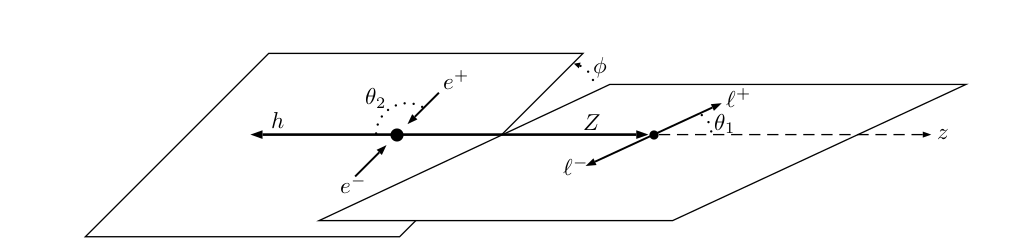}
\caption{The kinematic process $e^- e^+ \rightarrow h Z (\rightarrow l^+ l^-)$\cite{Craig:2015wwr}.} \label{angular}
\end{figure}

A recent discussion on the angular observables for the process $e^- e^+ \rightarrow h Z (\rightarrow l^+ l^-)$ can be found in~\cite{Beneke:2014sba,Craig:2015wwr}. Among the six independent angular observables, four are CP-even, given by
\begin{equation}
\begin{split}
\mathcal{A}_{\theta_1} &= \dfrac{1}{\sigma} \int^1_{-1} d \cos\theta_1 \;   \text{sgn} (\cos (2\theta_1)) \dfrac{d \sigma}{d \cos \theta_1} \\
\mathcal{A}^{(3)}_{\phi} &= \dfrac{1}{\sigma} \int^1_{-1} d \phi \;\text{sgn} (\cos (\phi)) \dfrac{d \sigma}{d \phi}\\
\mathcal{A}^{(4)}_{\phi} &= \dfrac{1}{\sigma} \int^1_{-1} d \phi \; \text{sgn} (\cos (2\phi)) \dfrac{d \sigma}{d \phi}\\
\mathcal{A}_{c\theta_1, c\theta_2} &= \dfrac{1}{\sigma} \int^1_{-1} d \cos\theta_1 \;   \text{sgn} (\cos (\theta_1)) \int^1_{-1} d \cos\theta_2   \; \text{sgn} (\cos (\theta_2)) \dfrac{d^2\sigma}{d \cos \theta_1 d \cos \theta_2}
\end{split}
\end{equation}
Here the angular variables are defined as in Figure~\ref{angular}.

\subsection{Electroweak Precision Tests}

\label{sec:EWPO}

{\bf A. EWPOs at $Z$ Pole}

The EWPOs at $Z$ pole which are relevant to our analysis include
\begin{equation}
\begin{split}
 &R_b = \dfrac{\Gamma_b}{\Gamma_{\text{had}}}, \quad R_l=\dfrac{\Gamma_{\text{had}}}{\Gamma_l} \quad (l = \mu,  \tau), \\
 &A_b,  \quad  A_{FB}^f = \dfrac{3}{4} A_e A_f  \quad (f = b, \mu),  \\
&   N_\nu=\dfrac{\Gamma_{\text{inv}}/\Gamma_l}{(\Gamma_\nu/\Gamma_l)_{\text{SM}}}, \quad  \sin^2 \theta_{\text{eff}}^{\text{lep}}= \dfrac{1}{4} \left( 1- \dfrac{g^l_V}{g^l_A} \right), \quad \Gamma_Z  \ .
\end{split}
\end{equation}
At tree level, the $Z$ partial decay width and the asymmetry are given by
\begin{equation}
\begin{split}
\Gamma_f&  = N_C^f \dfrac{m_Z^{(r)}}{12 \pi} \sqrt{1-\dfrac{4 m_f^2}{m_Z^{(r)2} }} \left[ |g_V^f|^2 +|g_A^f|^2 + \dfrac{2 m_f^2}{m_Z^{(r)2} }  (|g_V^f|^2 - 2|g_A^f|^2) \right]\\
A_f &= \dfrac{2 g_V^f}{g_V^f + g_A^f}
\end{split}
\end{equation}
in terms of vector and axial couplings $g_{V,A}^f$, or by
\begin{equation}
\begin{split}
\Gamma_f &= \Gamma(Z \rightarrow f \bar{f}) \\
&= N_C^f \dfrac{m_Z^{(r)}}{12 \pi} \sqrt{1-\dfrac{4 m_f^2}{m_Z^{(r)2} }} \left[ \dfrac{1}{2}(g_L^2 + g_R^2) + \dfrac{2 m_f^2}{m_Z^{(r)2}} \left(- \dfrac{g_L^2}{4}- \dfrac{g_R^2}{4}- \dfrac{3}{2} g_L g_R\right) \right]\\
A_f &= \dfrac{g_L^2 - g_R^2}{g_L^2 + g_R^2}
\end{split}
\end{equation}
in terms of chiral couplings $g_{L,R}$. $\Gamma_{l, \nu}$ is defined for a single flavor, whereas $\Gamma_{\rm inv}$ includes contribution from all possible flavors. With the 6D operators turned on, the corrections to the chiral couplings of $Z$ boson are given by 
\begin{itemize}
\item  charged lepton \quad
$g_L = g_Z \left(-\dfrac{1}{2}+s_w^2\right) \qquad g_R = g_Z s_w^2$
\begin{equation}\label{lep:eq}
\begin{split}
\Delta g_L &= \delta g_L + \delta \bar g_L =g_L \left( \dfrac{\delta g_Z}{g_Z} -\dfrac{2 s_{2w}}{c_{2w}} \delta \theta_w + \delta Z_Z \right)  - \dfrac{g_Z v^2}{2\Lambda^2} (c_{L}^{(3)l} + c^l_L) - e \:\delta Z_X\\
\Delta g_R &= g_R \left( \dfrac{\delta g_Z}{g_Z} +\dfrac{2 c_w }{s_w}\delta \theta_w + \delta Z_Z \right) -\dfrac{g_Z v^2}{2\Lambda^2} c^e_R  - e \:\delta Z_X
\end{split}
\end{equation}

\item  neutrino \quad
$g_L = \dfrac{g_Z}{2} \qquad g_R = 0$
\begin{equation}\label{nu:eq}
\Delta g_L = g_L \Big( \dfrac{\delta g_Z}{g_Z} + \delta Z_Z \Big)  - \dfrac{g_Z v^2}{2\Lambda^2} (-c_{L}^{(3)l} + c^l_L) \qquad \Delta g_R = 0
\end{equation}

\item up, charm quark \quad
$g_L = g_Z \left(\dfrac{1}{2}-\dfrac{2}{3}s_w^2\right) \qquad g_R =-\dfrac{2}{3} g_Z s_w^2$
\begin{equation}\label{up:eq}
\begin{split}
\Delta g_L &= g_L \Big( \dfrac{\delta g_Z}{g_Z} -\dfrac{8 s_w c_w}{3-4 s_w^2} \delta \theta_w + \delta Z_Z \Big) +\dfrac{2}{3} e \:\delta Z_X\\
\Delta g_R &= g_R \Big( \dfrac{\delta g_Z}{g_Z} +\dfrac{2 c_w }{s_w}\delta \theta_w + \delta Z_Z \Big)+\dfrac{2}{3} e\:\delta Z_X
\end{split}
\end{equation}

\item down, strange, bottom quark \quad
$g_L = g_Z \left(-\dfrac{1}{2}+\dfrac{1}{3}s_w^2\right) \qquad g_R = \dfrac{1}{3} g_Z s_w^2$
\begin{equation}\label{down:eq}
\begin{split}
\Delta g_L &= g_L \Big( \dfrac{\delta g_Z}{g_Z} -\dfrac{4 s_w c_w}{3-2 s_w^2} \delta \theta_w + \delta Z_Z \Big) -\dfrac{1}{3} e\:\delta Z_X\\
\Delta g_R &= g_R \Big( \dfrac{\delta g_Z}{g_Z} +\dfrac{2 c_w }{s_w}\delta \theta_w + \delta Z_Z \Big)-\dfrac{1}{3} e \:\delta Z_X \ .
\end{split}
\end{equation}
\end{itemize}
Here $\delta g_Z$ and $\delta \theta_w$ represent the effect of the EW parameter shift; $\delta Z_Z$ and $\delta Z_X$ represent the effect of field redefinition; and $c_L^{(3)l}$, $c_L^l$ and $c_R^e$ represent the effect of the charge shift in the leptonic $Z$ current. The quark current operators are turned off in this paper, though they may contribute to some of these observables, e.g., $R_b$, in a more general context. For more discussions on this, see, e..g,~\cite{Barklow:2017awn}. 

The formulae for the operator corrections to the EWPOs are presented in Appendix~\ref{App:formulae}, with six Wilson coefficients involved: $c_{WB}$, $c_T$, $c^{(3)l}_{L}$, $c^{(3)l}_{LL}$, $c^l_L$ and $c_R^e$. As is indicated in Eq.(\ref{Nnu}-\ref{leptheta}), the ratio for the coefficients of $c_{WB}$, $c_{T}$ and $c^{(3)l}_{LL}$ in the EWPOs, $N_\nu$, $A_b$, $A_{\text{FB}}^{\mu}$, $A_{\text{FB}}^{b}$, $R_b$, $R_\mu$, $R_\tau$ and $\sin^2 \theta_{\text{eff}}^{\text{lep}}$, is fixed to be $-1.1:2:4$. This is because the three terms in these EWPOs are generated either via
$\Delta g^i_L/g^i_L - \Delta g^i_R / g^i_R$, with $i$ representing charged leptons, up quarks and down quarks, or via $\Delta g^\nu_L/g^\nu_L - \Delta g^l_R/g^l_R$.  Both of them satisfy the relation  
\begin{eqnarray}\label{ratio:com}
 \frac{\Delta g^i_L}{g^i_L} - \frac{\Delta g^i_R }{ g^i_R}, \ \ \frac{\Delta g^\nu_L}{g^\nu_L} - \frac{\Delta g^l_R}{g^l_R}  \  \sim \ s_{2w}  \delta \theta_w - \frac{e}{g_Z} \delta Z_X  + \cdots \cdots
\ \sim \ 2 \delta \theta_w - \delta Z_X + \cdots
\end{eqnarray}
with the combination $2 \delta \theta_w -  \delta Z_X$ fixing this ratio.  This combination also contains a $c^{(3)l}_{L}$ term with its coefficient having a fixed ratio with the other ones, $-1.1:2:-4: 4$. However, this ratio does not hold in  $A_{\text{FB}}^{\mu}$, $A_{\text{FB}}^{b}$, $R_\mu$, $R_\tau$ and $\sin^2 \theta_{\text{eff}}^{\text{lep}}$ due to extra contributions proportional to $c_{L}^{(3)l} + c^l_L$.  Neither does it hold in $N_\nu$ due to to both $c_{L}^{(3)l} \pm c^l_L$  which are caused by the charge shift in the $Z$ boson current. The charge shift can receive contributions from $\mathcal O_R^e$ as well. So the set of EWPOs at $Z$ pole depend on four of the six Wilson coefficients or their linear combinations: $\xi_0=-1.1c_{WB}+2c_T-4c^{(3)l}_{L} + 4c^{(3)l}_{LL}$, $\xi_\pm = c_{L}^{(3)l} \pm c^l_L$ and $c_R^e$, leaving at least two degenerate or approximately degenerate directions. More explicitly, we have 
\begin{itemize}

\item $N_\nu$. It depends on $\xi_0$, $\xi_\pm$ and $c_R^e$.

\item $A_b$ and $R_b$. They only depend on $\xi_0$.

\item $A_{\text{FB}}^{b,\mu}$ and $\sin^2 \theta_{\text{eff}}^{\text{lep}}$. They have the same dependence on $\xi_0$, $\xi_+$ and $c_R^e$.  

\item $R_{\mu, \tau}$. They have the same dependence on $\xi_0$, $\xi_+$ and $c_R^e$.  

\end{itemize}
These degenerate or approximately degenerate directions could be lifted by $\Gamma_Z$, which is approximately proportional to $\Delta g^i_L g^i_L + \Delta g^i_R  g^i_R$, and $m_W$. $\Gamma_Z$ and $m_W$ have different dependences on the variables beyond $\xi_{0,\pm}$ and $c_R^e$. Thus, we have totally six classes of non-degenerate EWPOs to probe the six Wilson coefficients. The entangled dependence of the EWPOs on the six operators also explains the relatively large magnitude for their correlation matrix entries, as are listed in Appendix~\ref{sec:NCM}.

Though $\sin^2 \theta_{\text{eff}}^{\text{lep}}$ and $s_w^2$ are identical in the SM, they represent different measurements. Hence they are influenced by these 6D operators in different ways. $s_w^2$ received corrections via the EW parameter shift only (see Eq.(\ref{sintheta})), whereas $\sin^2 \theta_{\text{eff}}^{\text{lep}}$ receives extra contributions caused by field redefinition (see Eq.(\ref{ratio:com})).

{\bf B. $W$ boson mass}

The $W$ boson mass $m_W= m_Z c_w$ receives contributions via the shift of the EW parameters only, resulting in  
\begin{equation}
\dfrac{\Delta M_W}{M_W} = \dfrac{\delta g_Z}{g_Z} -\dfrac{s_w}{c_w} \delta \theta_w -\dfrac{1}{2}\dfrac{\delta G_F}{G_F} \ .
\end{equation}

{\bf C. Di-boson Process}

The di-boson production $e^- e^+ \rightarrow W^+ W^-$ can be applied to probe the TGC, and hence the operator $\mathcal O_{3W}$.  It is mainly influenced by the coupling shift in $g_Z$ due to $\mathcal{O}_{WB}$, $\mathcal{O}_{T}$, $\mathcal{O}^{(3)l}_{LL}$ and $\mathcal{O}_L^{(3)l}$ (see Eq.(\ref{delta_gZ})), and the charge shift in the electron current of $Z$ boson caused by $\mathcal{O}^l_L$ and $\mathcal{O}^e_R$. Despite this, a full angular analysis might be valuable, given that the total signal rate is dominated by forward transverse $WW$ production and hence less sensitive to anomalous couplings. We leave the latter to a future work.

\section{Analysis of Sensitivity to New Physics }

\begin{table}[ht]
  \centering

  \resizebox{\textwidth}{!}{
    \begin{tabular}{|c|c|c|c|c|}
    \hline
          & Current & CEPC  & FCC-ee & ILC \bigstrut\\
    \hline
    $M_Z$(GeV) & $91.1875 \pm 0.0021$\cite{ALEPH:2005ab} & $\pm 0.0005$\cite{CEPC-SPPCStudyGroup:2015csa} & $\pm 0.0001$  \cite{dEnterria:2016sca} &$\pm 0.0021$ \cite{Baak:2014ora} \bigstrut\\
    \hline
    $G_F(10^{-10} \text{GeV}^{-2})$    & $1166378.7 \pm 0.6$ \cite{Patrignani:2016xqp} & -  & - & - \bigstrut\\
    \hline
    $\alpha (10^{-13})$ & $7297352698 \pm 24$ \cite{Patrignani:2016xqp}&  - & - & - \bigstrut\\
    \hline
    $m_t$[GeV](pole) & $173.34 \pm 0.76_{\text{exp}}\pm 0.5_{\text{th}}$ \cite{Baak:2014ora,ATLAS:2014wva} & $\pm 0.6_{\text{exp}}\pm 0.25_{\text{th}}$ \cite{Baak:2014ora} & $\pm 0.02_{\text{exp}}\pm 0.1_{\text{th}}$\cite{Baak:2014ora}  &$\pm 0.03_{\text{exp}}\pm 0.1_{\text{th}}$ \cite{Baak:2014ora} \bigstrut\\
    \hline
    \end{tabular}%
  }
  
    \caption{Input parameter values for the analysis.}
  \label{tab:input}%
\end{table}%

\begin{table}[ht]
 \small
  \centering
    \begin{tabular}{|c|c|}
    \hline
    Observables & Current  \bigstrut\\
    \hline
    $N_\nu$   & 2.984 $\pm$ 0.008 \cite{dEnterria:2016sca} \bigstrut\\
    \hline
    $A_b$    & 0.923 $\pm$ 0.020 \cite{Patrignani:2016xqp,ALEPH:2005ab} \bigstrut\\
    \hline
    $R_b$    & 0.21629 $\pm$ 0.00066 \cite{dEnterria:2016sca} \bigstrut\\
    \hline
    $R_\mu$   & 20.767 $\pm$ 0.025 \cite{dEnterria:2016sca} \bigstrut\\
    \hline
    $R_\tau$  & 20.767 $\pm$ 0.025 \cite{dEnterria:2016sca} \bigstrut\\
    \hline
    $\Gamma_Z$(MeV) & 2495.2 $\pm$  $2.3 \pm 0.42_{\text{in}}$ \cite{dEnterria:2016sca} \bigstrut\\
    \hline
    $\sin^2 \theta_{\text{eff}}^{\text{lep}} (10^{-5})$  & $(23153 \pm 16 \pm 4_{\text{in}})$ \cite{Fan:2014vta,ALEPH:2005ab} \bigstrut\\
    \hline
    \end{tabular}%
  \caption{Electroweak precision measurements at LEP. The subscript ``in'' denotes an error caused by the input parameter uncertainties which are summarized in Table~\ref{tab:input}. This error is negligibly small for the observables except $\Gamma_Z$ and  $\sin^2 \theta_{\text{eff}}^{\text{lep}}$. }
  \label{tab:data-lep}%
\end{table}%

\begin{table} [ht]
\centering
\resizebox{\textwidth}{!}{
\begin{tabular}{c | c | c | c | c | c | c}
\hline\hline
Observables & \multicolumn{2}{c}{ILC} & \multicolumn{2}{c}{FCC-ee} & \multicolumn{2}{c}{CEPC}\\
\hline\hline
$\sigma(Zh)$ & $2.0\%$ \cite{Barklow:2017suo}& 250GeV,2$\text{ab}^{-1}$ & \textcolor{red}{$0.5\%$} \cite{dEnterria:2016sca}& 240GeV,5$\text{ab}^{-1}$ & $0.5\%$ \cite{CEPC-SPPCStudyGroup:2015csa}& 240GeV,5$\text{ab}^{-1}$ \\
 & $4.2\%$ \cite{Barklow:2017suo}& 500GeV,4$\text{ab}^{-1}$ &- &-  &- & -\\
$\sigma(\nu \bar{\nu} h)$ & \textcolor{red}{3.89$\%$} \cite{Asner:2013psa} & 250GeV,2$\text{ab}^{-1}$ & $ \textcolor{red}{ 0.97\%}$ \cite{Ge:2016zro}& 350GeV,1.5$\text{ab}^{-1}$& 2.86$\%$ \cite{Ge:2016zro}& 240GeV,5$\text{ab}^{-1}$\\
 & \textcolor{red}{1.45$\%$} \cite{Asner:2013psa} & 500GeV,4$\text{ab}^{-1}$ & -& -&-&-\\
$\sigma(Zhh)$ & \textcolor{red}{15.0$\%$} \cite{Asner:2013psa}& 500GeV,4$\text{ab}^{-1}$& - & - & - & -\\
$\sigma(W^+ W^-)$  & \textcolor{red}{0.0200$\%$}\cite{Schael:2013ita} & 250GeV,2$\text{ab}^{-1}$& \textcolor{red}{0.0136$\%$} \cite{Schael:2013ita} & 240GeV,5$\text{ab}^{-1}$ & \textcolor{red}{0.0136$\%$} \cite{Schael:2013ita} & 240GeV,5$\text{ab}^{-1}$\\
 & \textcolor{red}{$0.0191\%$} \cite{Schael:2013ita}& 500GeV,4$\text{ab}^{-1}$ &- & - &- &- \\
\hline
$N_{\nu}$ & 0.0013 \cite{Baer:2013cma}& $Z$ lineshape,100$\text{fb}^{-1}$ & \textcolor{red}{$1.58 \times 10^{-3}$} \cite{dEnterria:2016sca}& $Z$ pole,150$\text{ab}^{-1}$ & 0.0018 \cite{Ge:2016zro}& 240 GeV, 100$\text{fb}^{-1}$\\
$A_{FB}^{b}$ & -& -& -&- & $(\pm 15 \pm 2_{\text{in}})\times 10^{-4}$ \cite{CEPC-SPPCStudyGroup:2015csa}& $Z$ pole, 150$\text{fb}^{-1}$\\
$A_{FB}^{\mu}$ & -& -& \textcolor{red}{$7.1 \times 10^{-4}$} \cite{dEnterria:2016sca,Dam:2016ebi}&$Z$ pole,150$\text{ab}^{-1}$ & -& -\\
$A_{b}$ & 0.001 \cite{Baer:2013cma}&$Z$ pole,100$\text{fb}^{-1}$ & -&- & -& -\\
$R_b$ & $6.5 \times 10^{-4}$ \cite{Baer:2013cma}& $Z$ pole,100$\text{fb}^{-1}$& \textcolor{red}{$3.6 \times 10^{-4}$} \cite{dEnterria:2016sca,Dam:2016ebi}& $Z$ pole,150$\text{ab}^{-1}$& $8 \times 10^{-4}$ \cite{CEPC-SPPCStudyGroup:2015csa}& $Z$ pole, 100$\text{fb}^{-1}$\\
$R_{\mu}$ & $2 \times 10^{-4}$ \cite{Baak:2014ora}& $Z$ pole,100$\text{fb}^{-1}$ & \textcolor{red}{$6.1 \times 10^{-5}$} \cite{dEnterria:2016sca,Dam:2016ebi} &$Z$ pole,150$\text{ab}^{-1}$&$5 \times 10^{-4}$ \cite{CEPC-SPPCStudyGroup:2015csa}& $Z$ pole, 100$\text{fb}^{-1}$\\
$R_{\tau}$ & $2 \times 10^{-4}$ \cite{Baak:2014ora}& $Z$ pole,100$\text{fb}^{-1}$ & \textcolor{red}{$6.1 \times 10^{-5}$} \cite{dEnterria:2016sca,Dam:2016ebi} &$Z$ pole,150$\text{ab}^{-1}$& $5 \times 10^{-4}$ \cite{CEPC-SPPCStudyGroup:2015csa}& $Z$ pole, 100$\text{fb}^{-1}$\\
$\Gamma_Z$(MeV) & $\pm 1 \pm 0.21_{\text{in}}$ \cite{Baer:2013cma,Fan:2014vta}& $Z$ pole,100$\text{fb}^{-1}$& $\pm 0.1 \pm 0.08_{\text{th}} \pm 0.065_{\text{in}}$ \cite{Dam:2016ebi,Fan:2014vta}& $Z$ pole,150$\text{ab}^{-1}$ & $\pm 0.1 \pm 0.08_{\text{th}} \pm 0.13_{\text{in}}$ \cite{CEPC-SPPCStudyGroup:2015csa,Fan:2014vta}& $Z$ pole, 150$\text{fb}^{-1}$\\
$\sin^2\theta_{\text{eff}}^{\text{lep}}(10^{-5})$ & $\pm 1.3 \pm 1.5_{\text{th}} \pm 2.2_{\text{in}}$ \cite{Baer:2013cma,Fan:2014vta}& $Z$ pole,100$\text{fb}^{-1}$& $\pm 0.3 \pm 1.5_{\text{th}} \pm 1.6_{\text{in}}$ \cite{Dam:2016ebi,Fan:2014vta}& $Z$ pole,150$\text{ab}^{-1}$ & $\pm 2.3 \pm 1.5_{\text{th}} \pm 2.5_{\text{in}}$ \cite{CEPC-SPPCStudyGroup:2015csa,Fan:2014vta}& $Z$ pole, 150$\text{fb}^{-1}$\\
\hline
$m_{W}$ (MeV) & $ \textcolor{red}{ \pm 2.5} \pm 1_{\text{th}} \pm 2.8_{\text{in}}$ \cite{Baak:2013fwa,Fan:2014vta}& 250GeV, 2$\text{ab}^{-1}$ & $   \textcolor{red}{ \pm 1.2}  \pm 1_{\text{th}} \pm 0.91_{\text{in}}$ \cite{dEnterria:2016sca,Fan:2014vta}& $WW$ threshold,10$\text{ab}^{-1}$& $ \textcolor{red}{\pm 3} \pm 1_{\text{th}} \pm 3.8_{\text{in}}$ \cite{CEPC-SPPCStudyGroup:2015csa,Fan:2014vta}& 240GeV,5$\text{ab}^{-1}$\\
\hline
$\mathcal{A}_{\theta_1}$ & \textcolor{red}{0.0083}   \cite{Craig:2015wwr} & 250GeV,2$\text{ab}^{-1}$ & 0.0060   \cite{Craig:2015wwr} & 240GeV,5$\text{ab}^{-1}$  & 0.0060   \cite{Craig:2015wwr}& 240GeV,5$\text{ab}^{-1}$\\
$\mathcal{A}_{c\theta_1,c\theta_2}$ & \textcolor{red}{0.0092}   \cite{Craig:2015wwr} & 250GeV,2$\text{ab}^{-1}$ & 0.0067 \cite{Craig:2015wwr} & 240GeV,5$\text{ab}^{-1}$ & 0.0067 \cite{Craig:2015wwr}& 240GeV,5$\text{ab}^{-1}$\\
$\mathcal{A}^{(3)}_{\phi}$ & \textcolor{red}{0.0092}   \cite{Craig:2015wwr} & 250GeV,2$\text{ab}^{-1}$ & 0.0067 \cite{Craig:2015wwr} & 240GeV,5$\text{ab}^{-1}$ & 0.0067 \cite{Craig:2015wwr}& 240GeV,5$\text{ab}^{-1}$\\
$\mathcal{A}^{(4)}_{\phi}$ & \textcolor{red}{0.0092}   \cite{Craig:2015wwr} & 250GeV,2$\text{ab}^{-1}$ & 0.0067 \cite{Craig:2015wwr} & 240GeV,5$\text{ab}^{-1}$ & 0.0067 \cite{Craig:2015wwr}& 240GeV,5$\text{ab}^{-1}$\\
\hline\hline
\end{tabular}
}
\caption{Projected precision of the Higgs and electroweak precision measurements at ILC, FCC-ee and CEPC. A recently proposed operating scenario (see, e.g.,~\cite{FCC-ee}) has been assumed for the FCC-ee analysis. A beam polarization configuration of $(P_{e^-}, P_{e^+}) = (-0.8,0.3)$ is assumed for ILC at 250 and 500 GeV. The errors presented are all relative, except the ones for $M_W,\Gamma_Z$ and $\sin^2\theta^{\text{lep}}_{\text{eff}}$. The subscript ``th'' and ``in'' denotes errors caused by theoretical and input parameter uncertainties, respectively. The numbers in red are obtained by rescaling the experimental errors provided in the referred literatures, which are assumed to be statistical-error-like. As for the precision of measuring $\sigma(\nu\bar \nu h)$ and $\sigma(Zhh)$, we assume that the relevant Higgs decay branching ratios (such as Br$(h\to b\bar b)$) can be precisely measured via $e^-e^+ \to Zh$ at future colliders.} \label{data}
\end{table}

Before performing a full analysis on the sensitivities of probing the 6D operators at future $e^-e^+$ colliders, we will start with a set of analysis using CEPC as an example. We begin with the case in which we turn on one operator at a time.  This simplified approach provides an optimistic estimation on the energy scales that could be probed. It provides a basic idea on how the 6D operators individually contribute to the observables, but the potential cancellations among the contributions from different operators are ignored. The latter could dramatically change the collider sensitivities. To illustrate this point, we will consider several cases with more operators turned on. Finally, we will study the sensitivities at all future $e^-e^+$ colliders.  For each of these future programs, multiple operating scenarios have been suggested. We will focus on a subset of them in the analysis. The input parameter values, and the current and projected measurement precisions used for the analysis are  summarized in Table~\ref{tab:input}, Table~\ref{tab:data-lep} and Table~\ref{data}, respectively.  We will take into account the impact of the input parameter uncertainties for the measurement precisions. This effect was discussed in~\cite{Fan:2014vta} and is denoted as an error with a subscript ``in'' Table~\ref{tab:data-lep} and Table~\ref{data}. Also, a running coupling $\alpha(m_Z)$ in the $\overline{ \rm MS }$ scheme will be used in the analysis. The numerical formulae for the operator corrections to the observables are summarized in Appendix~\ref{App:formulae}. The effective operators are implemented using FeynRules and the cross sections are computed using either CalcHEP or MadGraph5 \cite{Alloul:2013bka, Belyaev:2012qa, Alwall:2011uj}.

\subsection{CEPC Analysis: Turning on Operators Individually }

\begin{table} [h]
  \centering
   \resizebox{\textwidth}{!}{
    \begin{tabular}{|c|c|c|c|c|c|c|c|c|c|c|c|}
    \hline
          & $\mathcal{O}_{WW}$ & $\mathcal{O}_{BB}$ & $\mathcal{O}_{WB}$ & $\mathcal{O}_{T}$ & $\mathcal{O}_{H}$ & $\mathcal{O}_{LL}^{(3)l}$ & $\mathcal{O}_{L}^{(3)l}$ & $\mathcal{O}_{L}^{l}$ & $\mathcal{O}_{R}^{e}$ & $\mathcal{O}_{6}$ & $\mathcal{O}_{3W}$ \bigstrut\\
          \hline
    $\sigma(Zh)$ & \textcolor[rgb]{ 1,  0,  0}{0.0222} & \textcolor[rgb]{ 1,  0,  0}{0.305} & 0.0903 & 0.1   & \textcolor[rgb]{ 1,  0,  0}{0.0825} & 0.0189 & 0.00797 & 0.00561 & 0.0064 & \textcolor[rgb]{ 1,  0,  0}{4.75} & - \bigstrut\\
    \hline
    $\sigma(\nu\bar{\nu}h)$ & 2.17  & -     & 1.01  & 0.0819 & 0.472 & 0.0496 & 0.0392 & 2.1   & -     & 71.7  & - \bigstrut\\
    \hline
    $\sigma(W^+ W^-)$ & -     & -     & 0.00343 & \textcolor[rgb]{ 1,  0,  0}{0.000801} & -     & \textcolor[rgb]{ 1,  0,  0}{0.000401} & 0.0018 & 0.0049 & 0.00744 & -     & \textcolor[rgb]{ 1,  0,  0}{0.197} \bigstrut\\
    \hline
    $N_\nu$ & -     & -     & 0.308 & 0.168 & -     & 0.0845 & 0.129 & 0.0072 & 0.0159 & -     & - \bigstrut\\
    \hline
    $A_{FB}^b$ & -     & -     & \textcolor[rgb]{ 1,  0,  0}{0.00242} & 0.00133 & -     & 0.000664 & 0.00101 & \textcolor[rgb]{ 1,  0,  0}{0.00193} & \textcolor[rgb]{ 1,  0,  0}{0.00169} & -     & - \bigstrut\\
    \hline
    $R_b$ & -     & -     & 0.422 & 0.232 & -     & 0.116 & 0.116 & -     & -     & -     & - \bigstrut\\
    \hline
    $R_\mu$ & -     & -     & 0.0516 & 0.0283 & -     & 0.0141 & 0.00314 & 0.00404 & 0.0046 & -     & - \bigstrut\\
    \hline
    $R_\tau$ & -     & -     & 0.0515 & 0.0283 & -     & 0.0141 & 0.00313 & 0.00403 & 0.0046 & -     & - \bigstrut\\
    \hline
    $\Gamma_Z$ & -     & -     & 0.00653 & 0.000926 & -     & 0.000463 & \textcolor[rgb]{ 1,  0,  0}{0.000604} & 0.00647 & 0.00647 & -     & - \bigstrut\\
    \hline
    $M_W$ & -     & -     & 0.00554 & 0.00142 & -     & 0.00233 & 0.00233 & -     & -     & -     & - \bigstrut\\
    \hline
    $\sin^2\theta_{\text{eff}}^{\text{lep}}$ & -     & -     & 0.00332 & 0.00182 & -     & 0.00091 & 0.00139 & 0.00262 & 0.0023 & -     & - \bigstrut\\
    \hline
    $\mathcal{A}_{\theta_1}$ & 0.894 & 12.7  & 3.33  & -     & -     & -     & 140   & 140   & 163   & -     & - \bigstrut\\
    \hline
    $\mathcal{A}_{c\theta_1,c\theta_2}$ & 0.703 & 2.35  & 0.554 & 0.419 & -     & 0.208 & 0.318 & 0.598 & 0.515 & -     & - \bigstrut\\
    \hline
    $\mathcal{A}^{(3)}_{\phi}$ & 0.444 & 1.45  & 0.554 & 0.23  & -     & 0.115 & 0.183 & 0.312 & 0.302 & -     & - \bigstrut\\
    \hline
    $\mathcal{A}^{(4)}_{\phi}$ & 3.33  & 47.2  & 12.4  & -     & -     & -     & 307   & 307   & 356   & -     & - \bigstrut\\
    \hline
    All   & 0.0222 & 0.296 & 0.00157 & 0.000494 & 0.0813 & 0.000262 & 0.00045 & 0.00124 & 0.00119 & 4.74  & 0.197 \bigstrut\\
    \hline
    \end{tabular}%
    }
     \caption{CEPC sensitivities for measuring the Wilson coefficient, i.e., $\frac{C_i}{\Lambda^2}$(TeV$^{-2}$), of a 6D operator $\mathcal{O}_i$ at $1\sigma$ C.L., with the operators turned on individually. The numbers in red denote the best sensitivity which could be achieved using a single observable, whereas the numbers in the last row represent the sensitivity based on a combination of all observables.}
  \label{tab:addlabel0}%
\end{table}

The sensitivities for probing the 6D operators at CEPC are presented in Table~\ref{tab:addlabel0}, with them turned on individually. 
Each row of the table shows the sensitivity of an observable in probing the operators, with the last row showing the combination. 
$\mathcal{O}_{WB}$, $\mathcal{O}_{T}$, $\mathcal{O}_{LL}^{(3)l}$ and $\mathcal{O}_{L}^{(3)l}$ can be well-probed by the EWPOs, because of the EW parameter shift, the field redefinition and the charge shift in the $Z$ boson current that they caused. $\mathcal{O}_{L}^{l}$ and $\mathcal{O}_{R}^{e}$ contribute to the charge shift in the $Z$ boson current, and hence can be also probed very well. $\mathcal{O}_{3W}$ contributes to TGC directly, and can be probed by the measurement of $e^-e^+ \to W^+W^-$. Probing the other four operators, $\mathcal{O}_{WW}$, $\mathcal{O}_{BB}$, $\mathcal{O}_{H}$ and $\mathcal{O}_{6}$, mainly relies on the measurement of the Higgs observables, such at the signal rate of $e^-e^+ \to Zh$ production.  The angular observables defined in $e^-e^+ \to Zh$  are less sensitive in probing the operators. 
As shown in the last row, the combination of the observables can sizably improve the sensitivities to $\{\mathcal{O}_{WB}, \mathcal{O}_{T}, \mathcal{O}_{LL}^{(3)l}, \mathcal{O}_{L}^{(3)l}, \mathcal{O}_{L}^{l}, \mathcal{O}_{R}^{e}\}$, compared to other operators. This implies that more than one observables are sensitive to each of these operators, as was advertised in Section~\ref{sec:EWPO}.  

\subsection{CEPC Analysis: Turning on Multiple Operators Simultaneously}

\begin{figure}[h]
\centering
\includegraphics[width=9.cm]{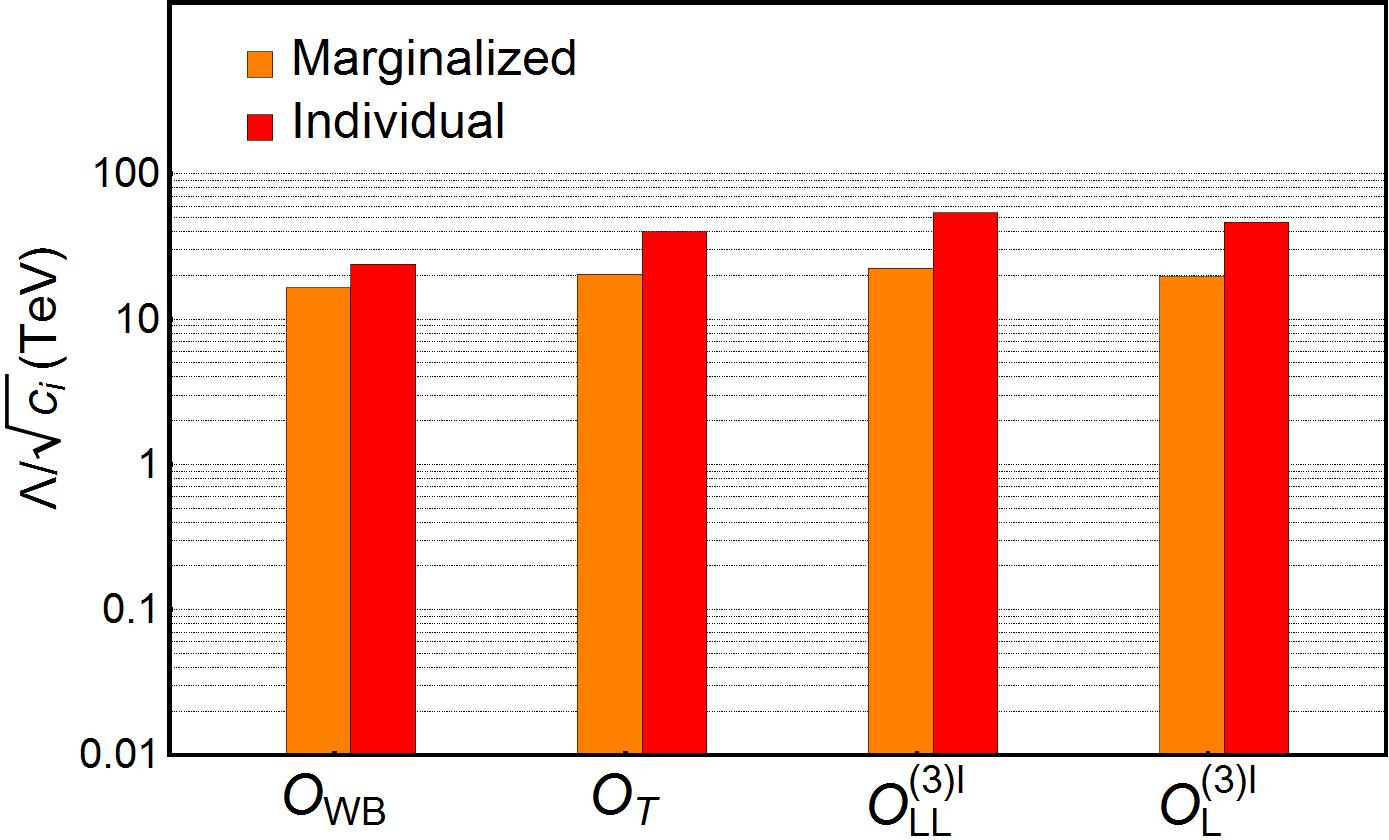}\label{ffig_1D}
\\ \bigskip
\includegraphics[width=9.cm]{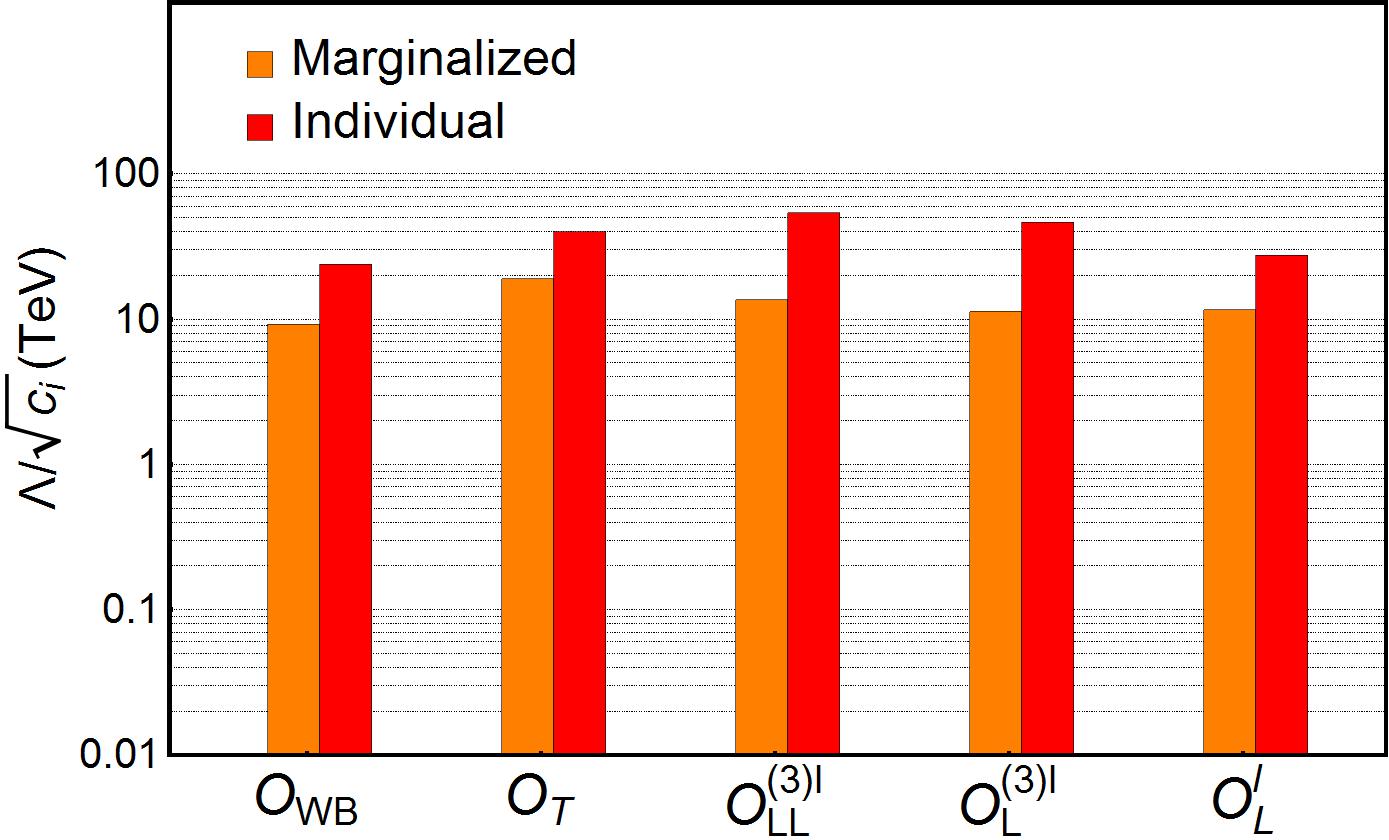}\label{ffig_1D}
\\ \bigskip
\includegraphics[width=9.cm]{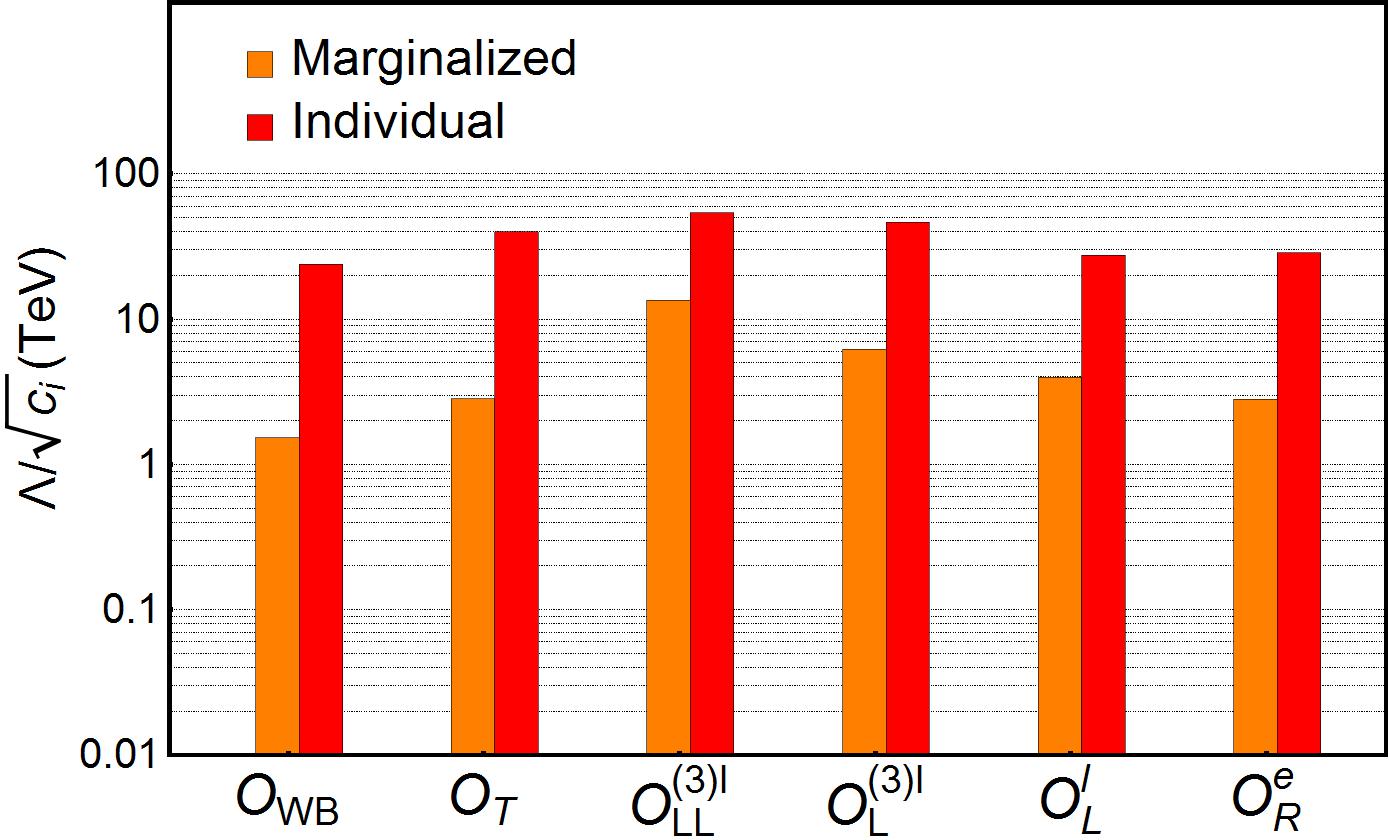}\label{ffig_1D}
\caption{``Optimistic'' (with one operator turned on at a time, denoted by ``Individual'') and ``semi-conservative'' (with multiple operators (instead of all operators) turned on, denoted by ``Marginalized'') sensitivity interpretations for probing each of the set of 6D operators $\{\mathcal{O}_{WB}, \mathcal{O}_{T}, \mathcal{O}_{LL}^{(3)l}, \mathcal{O}_{L}^{(3)l}, \mathcal{O}_{L}^{l}, \mathcal{O}_{R}^{e}\}$, with the EWPOs at CEPC applied. In the top panel, $\{\mathcal{O}_{WB}, \mathcal{O}_{T}, \mathcal{O}_{LL}^{(3)l}, \mathcal{O}_{L}^{(3)l}\}$ are turned on for marginalization. $\mathcal{O}_{L}^{l}$ and $\mathcal{O}_{R}^{e}$ are incorporated subsequently in the middle and bottom panels.} \label{1DEWPO}
\end{figure}

Next let us turn on more 6D operators in Table~\ref{tab:ops}. For a comparison with the results shown in Table~\ref{tab:addlabel}, we need to project the  allowed region in the space of Wilson coefficients to the relevant axis, that is, to ``marginalize'' the irrelevant Wilson coefficients. There is a geometric interpretation regrading this method. The $\chi^2$ defines a 10-dimensional ellipsoid in an 11-dimensional space which is expanded by the set of Wilson coefficients. Marginalizing 10 of the 11 Wilson coefficients is equivalent to imposing the conditions $\frac{\partial \chi^2}{\partial c_i} = 0$, with $i$ running over all of the 10 Wilson Coefficients. It results in a projection of the ellipsoid to the direction defined by the 11th Wilson coefficient. This method can be also generalized to the case with less Wilson coefficients being marginalized. An introduction to this statistical method is given in Appendix~\ref{app:PM}.

We start with the set of six operators $\{\mathcal{O}_{WB}, \mathcal{O}_{T}, \mathcal{O}_{LL}^{(3)l}, \mathcal{O}_{L}^{(3)l}, \mathcal{O}_{L}^{l}, \mathcal{O}_{R}^{e}\}$ which are expected to be constrained by the six classes of EWPOs at tree level (as is discussed in Section~\ref{sec:EWPO}). The CEPC sensitivities for probing each of them are presented in Figure~\ref{1DEWPO}, with the EWPOs applied only, in both the ``optimistic'' and ``semi-conservative'' cases. With the first four operators turned on (top panel), the CEPC sensitivities decrease from dozens of TeV in the  ``optimistic''  case to $\sim \mathcal O(10)$ TeV. The turning on of the fifth operator $\mathcal{O}_{L}^{l}$ doesn't change the results much (middle panel). However, the turning on of the last operator $\mathcal{O}_{R}^{e}$ causes a jump of the CEPC sensitivities for probing these operators except $\mathcal{O}_{LL}^{(3)l}$. This is related to the fact that $R_b$ (one of the six classes of the EWPOs) is a weak observable in probing $\xi_0$. With the sixth operator turned on, the lack of a sixth independent strong EWPOs yields an approximately degenerate direction in the parameter space expanded by the six operators. To break this degeneracy, extra observables (e.g., Bhabha scattering $e^-e^+ \to \mu^-\mu^+$), need to be introduced.

\begin{figure}[h]
\centering
\includegraphics[width=12cm]{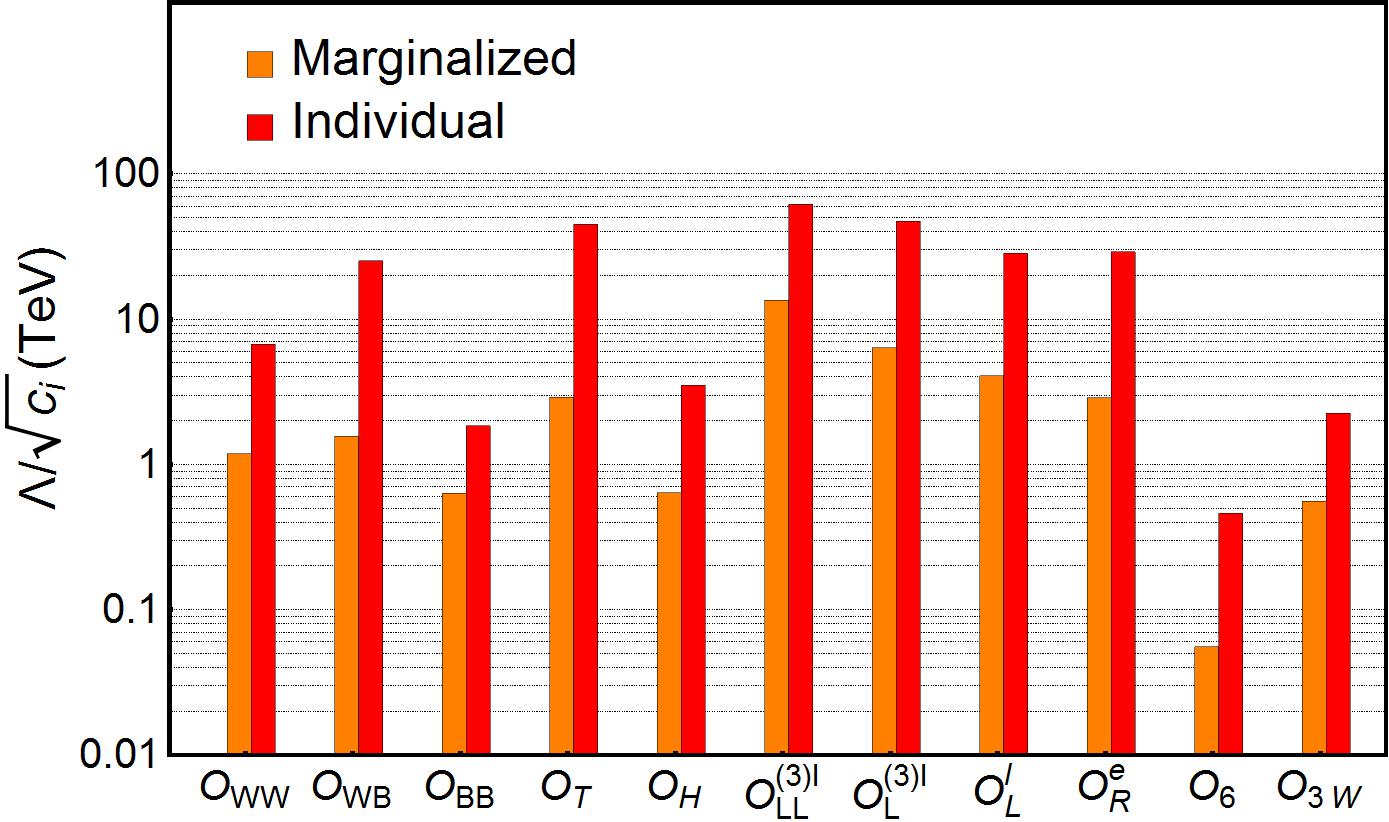}\label{ffig_1D}
\caption{``Optimistic'' (with one operator turned on at a time) and ``conservative'' (with all operators turned on simultaneously) sensitivity projections for probing each of the set of 6D operators at CEPC.} \label{1DCEPC}
\end{figure}

A full analysis for the CEPC sensitivities for probing the whole set of 6D operators is presented in Figure~\ref{1DCEPC}, with all observables in Table~\ref{data} applied.  The normalized correlation matrix for this $\chi^2$ fit is presented in Table~\ref{tab:CEPC_cor} of Appendix~\ref{sec:NCM}.  We have the following observations on the ``marginalization'' results:
\begin{itemize}

\item For the set of operators $\{\mathcal{O}_{WB}, \mathcal{O}_{T}, \mathcal{O}_{LL}^{(3)l}, \mathcal{O}_{L}^{(3)l}, \mathcal{O}_{L}^{l}, \mathcal{O}_{R}^{e}\}$, the CEPC sensitivities are inherited from the ones presented in Figure~\ref{1DEWPO}.  The energy scale that the CEPC is able to probe decreases from dozens of TeV in the ``optimistic''  case to TeV or several TeV, except for $\mathcal{O}_{LL}^{(3)l}$.

\item The operator  $\mathcal{O}_{3W}$ can be weakly probed only via the $e^-e^+ \to W^+ W^-$ production, with the energy scale accessible to the CEPC being decreased from a couple of TeV in the ``optimistic''  case to sub TeV (this feature is also shared by FCC-ee and ILC, as will be shown below). This is a result of the concerted action of (1) the weak dependence of the $e^-e^+ \to W^+ W^-$ production on $\mathcal{O}_{3W}$ due to helicity suppression at linear level~\cite{Azatov:2016sqh}; and (2) the existence of approximate degeneracy for the set of EW operators to which the $e^-e^+ \to W^+ W^-$ production is much more sensitive (see Eq. (\ref{WW1})). This effect yields a sensitivity estimation for probing $\mathcal O_{3W}$ several times weaker than that obtained in~\cite{Durieux:2017rsg}.  

\item The three operators $\{\mathcal{O}_{WW}, \mathcal{O}_{BB}, \mathcal{O}_{H}\}$ contribute to the Higgs events at tree level. The energy scales that the CEPC is able to probe decrease from several TeV/TeV  in the ``optimistic''  case to TeV/sub TeV, with potential cancellation between the operators taken into account. This is related to  the fact that there is only one observable at 240 GeV which is highly sensitive to these operators, say, $\sigma(Zh)$. Though $\sigma(\nu\bar\nu h)$ and the $e^-e^+ \to Zh$ angular observables play a role in constraining the Wilson coefficients, they are too weak to completely break the remaining degeneracies. Despite this, the sensitivities for probing $\mathcal{O}_{WW}, \mathcal{O}_{BB}$ and $\mathcal{O}_{WB}$ could be improved by a couple of times by incorporating the Higgs decay measurements. For example, the decay width of the di-photon mode can be shifted by these operators, yielding 
\begin{eqnarray} 
\frac{\delta \Gamma_{\gamma\gamma}}{\Gamma_{\gamma\gamma}}  \sim 2.95 \frac{c_{BB}}{\Lambda^2} - 2.94 \frac{c_{WB}}{\Lambda^2} + 2.95 \frac{c_{WW}}{\Lambda^2}  - 0.0606 c^{(3)l}_L + 0.0606 c^{(3)l}_{LL}  \ .
\end{eqnarray} 
As is indicated in~\cite{Durieux:2017rsg,Barklow:2017awn}, including the di-photon decay measurement may push the sensitivities of probing the $\mathcal{O}_{WW}$ and $\mathcal{O}_{BB}$ operators up to several TeVs (note, fewer or no relevant EW operators were turned on in~\cite{Durieux:2017rsg,Barklow:2017awn}, which may cause an uncertainty for the estimation).  

\item The operator $\mathcal{O}_{6}$ contributes to the Higgs events at loop level only. The energy scales that the CEPC is able to probe decrease from sub TeV in the ``optimistic''  case to $< \mathcal O(0.1)$TeV.

\end{itemize}
The $\chi^2$ fit sensitivities can be also projected to a 2D plane expanded by two Wilson coefficients, using a marginalization method, as is shown in Figures~\ref{fig_array1} - \ref{fig_array3} in Appendix~\ref{app:2D}.

\subsection{Comparative Study at Future $e^-e^+$ Colliders}

\begin{figure}[H]
\centering
\subfigure{ \includegraphics[width=12cm]{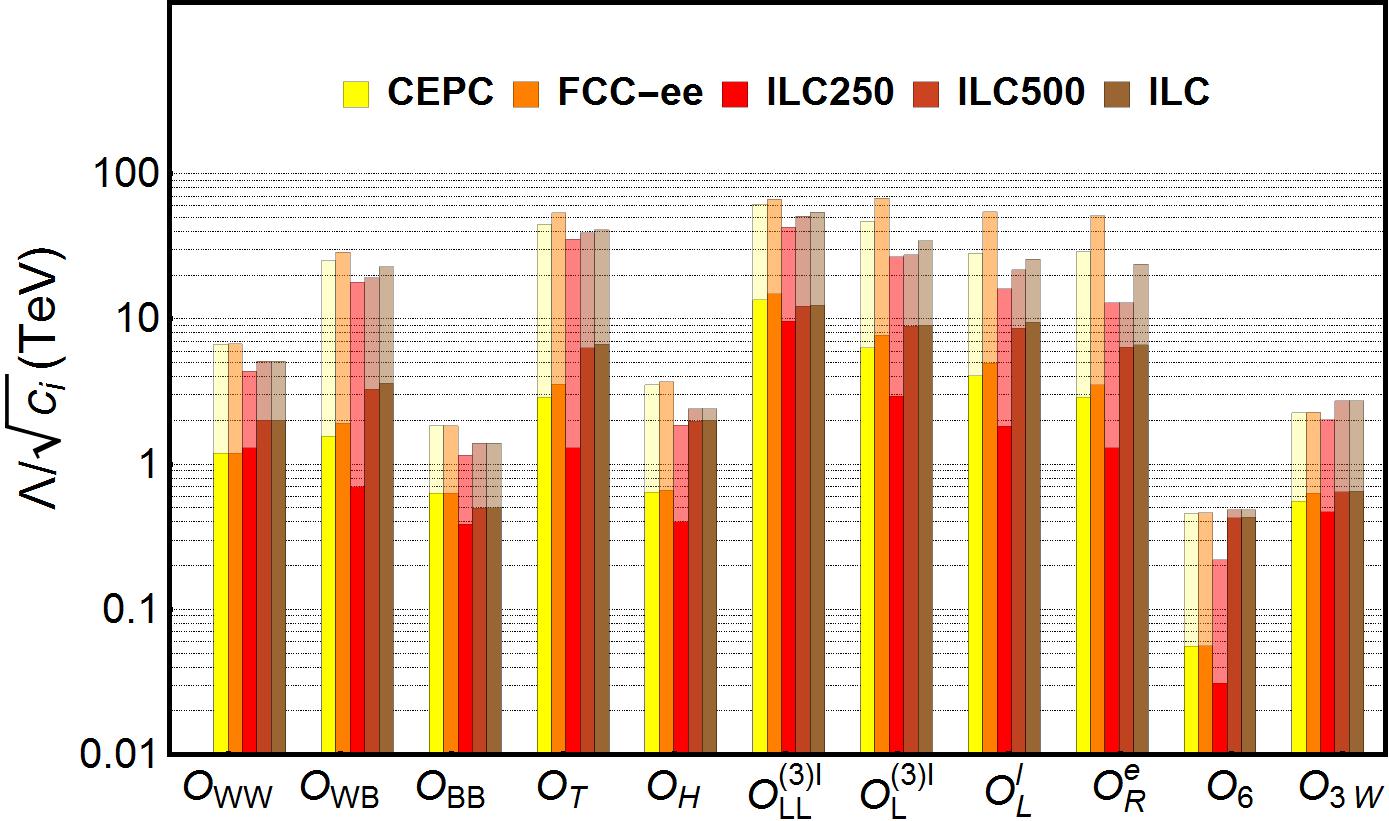}}  
\caption{``Optimistic'' (light) and ``conservative'' (dark) sensitivity projections for  probing each of the set of 6D operators at CEPC, FCC-ee, ILC250, ILC500 and ILC. Here ``ILC250'' refers to a combination of the ILC data at 250 GeV and the EW precision measurements at LEP (see Table~\ref{tab:data-lep}); ``ILC500'' refers to a combination of ``ILC250'' and the ILC data at 500 GeV; and ``ILC'' refers to a more optimistic operating scenario, with the LEP measurements in ``ILC500'' replaced by the Giga-$Z$ data. 
} \label{comparison1D}
\end{figure}

Next let us make a comparison on the sensitivities of probing the 6D operators at the future $e^-e^+$ colliders. For each machine, there exist multiple possibilities for its operating scenario. For concreteness, we consider the measurement precisions at CEPC, FCC-ee and ILC with a subset of possible running scenarios, shown in Table~\ref{data}. The ``optimistic'' and ``conservative'' sensitivity interpretations at each machine are presented Figiure~\ref{comparison1D}. Both CEPC and FCC-ee are circular $e^-e^+$ colliders with non-polarized beams. Benefitting from a larger integrated luminosity at $Z$ pole, the sensitivities at FCC-ee are mildly better than the CEPC ones, in both interpretations. The comparison with  the sensitivities at ILC250, ILC500 and ILC is more involved. The ILC250 is less capable in probing these operators than both CEPC and FCC-ee, because of its relatively small luminosity at 250 GeV and the lack of data at $Z$-pole. However, this can be improved significantly by the data expected to be collected at a higher beam energy\footnote{This feature was also noticed in~\cite{Barklow:2017awn}, but an explicit comparison with the CEPC and the FCC-ee performances was missing.}. With the data at 500 GeV, the ILC500 performance becomes  not much worse than or comparable to the CEPC and FCC-ee ones in the optimistic case. In the conservative case, the ILC500  performance becomes comparable to or even better than the CEPC and FCC-ee ones. This results in a smaller difference between the two kinds of sensitivity interpretations at ILC, compared with the ones at CEPC and FCC-ee, as is indicated in Figure~\ref{comparison1D}. On the other hand, the data at Giga-$Z$ can slightly improve the sensitivities only which could be achieved at ILC500. 

We note that we have oversimplified the beam polarization scenario at ILC, assuming a full-time run for the polarization configuration $(P_{e^-}, P_{e^+}) = (-0.8,0.3)$. Splitting time between different polarization configurations can enhance the power of breaking the operator degeneracies. This effect has been discussed in~\cite{Durieux:2017rsg,Barklow:2017suo}, yielding an improvement of $\sim 20-30 \%$ on the reach of the new physics scale in some of the operators. 

\begin{table}[h]
  \centering
  \resizebox{\textwidth}{!}{
    \begin{tabular}{|c|c|c|c|c|c|c|c|c|c|c|c|}
    \hline
        &  $\mathcal{O}_{WW}$ & $\mathcal{O}_{WB}$ & $\mathcal{O}_{BB}$ & $\mathcal{O}_{T}$ & $\mathcal{O}_{H}$ & $\mathcal{O}_{LL}^{(3)l}$ & $\mathcal{O}_{L}^{(3)l}$ & $\mathcal{O}_{L}^{l}$ & $\mathcal{O}_{R}^{e}$ & $\mathcal{O}_{6}$ & $\mathcal{O}_{3W}$ \bigstrut\\
    \hline 
     ILC250 & 1.30   & 0.697 & 0.384 & 1.29  & 0.401 & 9.62  & 2.92  & 1.83  & 1.29  & 0.0309 & 0.469 \bigstrut\\
    \hline
    $+ \sigma(W^+W^-)$ & 1.30   & 2.17  & 0.386 & 4.08  & 0.468 & 9.63  & 6.78  & 6.11  & 4.08  & 0.0389 & 0.523 \bigstrut\\
    \hline
    $+ \sigma(Zh)$ & 1.75  & 2.21  & 0.493 & 4.16  & 0.897 & 9.78  & 6.89  & 6.21  & 4.16  & 0.0895 & 0.531 \bigstrut\\
    \hline
    $+ \sigma(Zhh)$ & 1.95  & 3.22  & 0.498 & 6.19  & 1.28  & 12.2  & 8.83  & 8.45  & 6.20   & 0.428 & 0.644 \bigstrut\\
    \hline
    $+ \sigma(\nu\nu h) =$ ILC500 & 2.01  & 3.29  & 0.498 & 6.34  & 1.97  & 12.3  & 8.90   & 8.60   & 6.36  & 0.428 & 0.647 \bigstrut\\
        \hline
    \end{tabular}%
 }
  \caption{Projected sensitivities of $\Lambda/\sqrt{c_i}$ (TeV) at ILC250 and ILC500. The four extra observables in the first column are all measured at 500 GeV.}
  \label{tab:addlabel}%
\end{table}%

To get a better picture about the roles played by the observables at 500 GeV, in Table~\ref{tab:addlabel} we present the marginalized fitting results for $\Lambda/\sqrt{c_i}$ (TeV) in the ILC scenarios, varying from ILC250 to ILC500 by adding one more observable at 500 GeV each time. Compared with that at CEPC and FCC-ee, the degeneracy problem for $\{c_{WB}, c_T, c^{(3)l}_{L}, c^{(3)l}_{LL}, c^l_L, c_R^e\}$ at ILC250 is even worse, given the lack of the $Z$-pole data. This problem can be addressed to some extent by the $e^-e^+ \to W^+ W^-$ measurement at ILC500, as is indicated in Table~\ref{tab:addlabel}. $\sigma(W^+W^-)$ does not depend on  $\{\mathcal{O}_{WW}, \mathcal{O}_{BB}, \mathcal{O}_{H}, \mathcal{O}_{6}\}$ at tree level, but it has relatively strong sensitivities to these EW operators (see Eq. (\ref{WW3})). With its help, the constraints for these operators are raised to a level compared to the ones at CEPC and FCC-ee. But this also means that the sensitivity to $\mathcal O_{3W}$ is still weak. A combination of the other three observables at 500 GeV, say,  $\sigma(Zh)$, $\sigma(Zhh)$ and $\sigma(\nu\nu h)$ can help constrain three of $\{\mathcal{O}_{WW}, \mathcal{O}_{BB}, \mathcal{O}_{H}, \mathcal{O}_{6}\}$ which are weakly constrained  at ILC250. Particularly, the ILC500 has a much better performance in probing $\mathcal{O}_{6}$, compare to CEPC and FCC-ee. This is due to the $e^-e^+\to Z hh$ production, an observable which is not available at CEPC and FCC-ee.  Though it is less important in the ``optimistic'' analysis, this observable plays a crucial role in breaking the degeneracy related to $\mathcal{O}_{6}$ in the ``conservative'' scenario. As for CEPC and the FCC-ee , their weakness in probing $\mathcal{O}_{6}$  could be mitigated somewhat by  combining with the LHC data for di-Higgs production, e.g.,  $pp\to h h \rightarrow b \bar{b} \tau \bar{\tau}$~\cite{Goertz:2014qta,Azatov:2015oxa,Barr:2014sga,He:2015spf}. Note, the weak sensitivity to probe $\mathcal{O}_{6}$ below the $Zhh$ thresholds (particularly in the ``conservative'' scenario) may indicate that the non-linear $c_6$ terms, e.g., the one-loop quadratic term induced by the Higgs self-energy correction~\cite{Degrassi:2016wml}, need to be incorporated in the analysis. However, this term, even if being turned on, still fails to yield a bound clearly stronger than the perturbative unitarity one set by the $hh \to hh$ scattering, say, $|\kappa_3| < 5.5$~\cite{DiLuzio:2017tfn}. So, we simply neglect such terms here. 

Such a comparative study can be also extended to a plane expanded by two Wilson coefficients, as is shown in Figure~\ref{comparison2D} in Appendix~\ref{app:2D}.

\section{Application to Two Benchmark Composite Higgs Models}

In this section, we will apply our analysis to a couple of benchmark composite Higgs models. If the composite resonances are heavy, their low energy effects can be captured by a set of correlated EFT operators, named as a ``SILH'' parametrization~\cite{Giudice:2007fh}. The SILH parametrization contains two characteristic parameters: $f$, the decay constant of strong dynamics, and $g_\rho$, the strong coupling. Its Lagrangian is given by \cite{Giudice:2007fh}
\begin{eqnarray}
\label{eq:silh}
\mathcal{L}_{\rm SILH} &=& \dfrac{\tilde c_H}{f^2} \mathcal{O}_H + \dfrac{\tilde c_T}{f^2} \mathcal{O}_T - \dfrac{\tilde c_6}{f^2} \mathcal{O}_6+\dfrac{\tilde c_W}{m_\rho^2} \mathcal{O}_W  + \dfrac{\tilde c_B}{m_\rho^2} \mathcal{O}_B  \\&& + \dfrac{\tilde c_{HW}}{16 \pi^2 f^2} \mathcal{O}_{HW}  + \dfrac{\tilde c_{HB}}{16 \pi^2 f^2} \mathcal{O}_{HB} + \dfrac{\tilde c_{\gamma} g'^2}{16 \pi^2 f^2 g_\rho^2} \mathcal{O}_{BB} + \frac{3! g^2 \tilde c_{3W} }{16\pi^2 m_\rho^2} \mathcal{O}_{3W}  \ .  \label{SILHL}  \nonumber
\end{eqnarray}
Here $m_\rho = g_\rho f$ defines a typical composite resonance mass. To begin with, we neglect the loop-level operators listed in the second line, and rewrite the Lagrangian in the minimal operator basis using the relations \cite{Elias-Miro:2013mua}
\begin{equation}
\label{eq:op_relation}
\begin{split}
\mathcal{O}_W &= g^2 \Big[ -\dfrac{3}{2} \mathcal{O}_H + 2 \mathcal{O}_6 + \dfrac{1}{2} (\mathcal{O}^u_y + \mathcal{O}^d_y + \mathcal{O}^l_y + \text{h.c.} ) + \dfrac{1}{4} \mathcal{O}^{(3)l}_L \Big]\\
\mathcal{O}_B &= g'^2 \Big[-\dfrac{1}{2} \mathcal{O}_T + \dfrac{1}{2} \sum_f(Y_L^f \mathcal{O}^f_L + Y_R^f \mathcal{O}^f_R)\Big]  \ .
\end{split}
\end{equation}
Here $\mathcal{O}^{u,d,l}_y$ denotes the 6D Yukawa operators, say, the product of the Yukawa term and the $H^\dagger H$, and $f$ runs over all fermions in the SM.
These two relations can be further simplified to make connection to our analysis. While substituting $\mathcal{O}_W$ in Eq.~\ref{eq:silh}, we omit the operator $\mathcal{O}_6$, considering its insensitivity to the observables used in the analysis. The 6D Yukawa operators $\mathcal{O}^{u,d,l}_y$ mainly influence the Higgs Yukawa couplings and hence are less relevant for the inclusive observables applied. The case for $\mathcal{O}_B$ is somewhat more complicated. The quark current operators may nontrivially contribute to the $\Gamma_{\rm had}$. So we will exclude all EWPOs involving the Z hadronic width $\Gamma_{\rm had}$ below, in order to safely neglect this subtlety. Then under an assumption of $\Lambda^2 = (4\pi f)^2$, the relevant Lagrangian terms are given by 
\begin{equation}
\mathcal{L_{\text{SILH}}} \supset \frac{c_H}{\Lambda^2} \mathcal{O}_H +  \frac{c_T}{\Lambda^2} \mathcal{O}_T + \frac{c_L^{(3)l} }{\Lambda^2} \mathcal{O}_L^{(3)l} +  \frac{c_L^{l} }{\Lambda^2}  \mathcal{O}_L^{l} +  \frac{c_R^{e} }{\Lambda^2}  \mathcal{O}_R^{e}  \label{SILHL2}
\end{equation}
with
\begin{eqnarray}
c_H  = (4\pi)^2 \left (\tilde c_H - \dfrac{3 g^2 \tilde c_W}{2 g_\rho^2}\right) \ ,  \quad 
c_T   \ = \ (4\pi)^2 \left  (\tilde c_T - \dfrac{g'^2 \tilde c_B}{2 g_\rho^2} \right) \ , \nonumber \\
c_L^{(3)l}  \ = \ (4\pi)^2 \frac{g^2 \tilde c_W}{4 g_\rho^2} \ , \quad 
c_L^{l}   \ = \ -  (4\pi)^2  \dfrac{g'^2 \tilde c_B}{4 g_\rho^2} \ , \quad 
c_R^{e}  \ = \ -  (4\pi)^2 \dfrac{g'^2 \tilde c_B}{2 g_\rho^2} \ .
\end{eqnarray}
The SILH can have different realizations, which are characterized by the values of $\tilde c_i$s. Though the LHC runs are able to constrain the SILH, the experimental bounds are typically model-dependent. One LHC probe is to measure the Higgs couplings such as 
\begin{eqnarray} 
g_{hWW} = g m_W \left( 1 - \frac{\tilde c_H}{2} \frac{v^2}{f^2}  \right)  \ .
\end{eqnarray}
The current LHC runs yield a lower bound $ f > 600-700$ GeV, for $\tilde c_H =1$~\cite{Sanz:2017tco,Banerjee:2017wmg}, under the assumption of no mixing effect with extra scalars. Such a bound could be pushed up to $\sim 1.5$ TeV at HL-LHC. Another LHC probe is to search for the composite resonances. For example, the current searches for the fermionic top partner via its pair production set up an lower limit for the resonance mass $0.9-1.2$ TeV~\cite{ATLAS:2016btu,ATLAS:2017lvm,Sirunyan:2017pks,Sirunyan:2017usq}, and hence yield a constraint $g_\rho f = m_\rho > \mathcal O(1)$ TeV. Below we will consider two benchmark models: holographic composite Higgs model and littlest Higgs model.

{\bf A. Holographic Composite Higgs Model}

\begin{figure}[h]
\centering
\subfigure{
\includegraphics[height=5.5 cm]{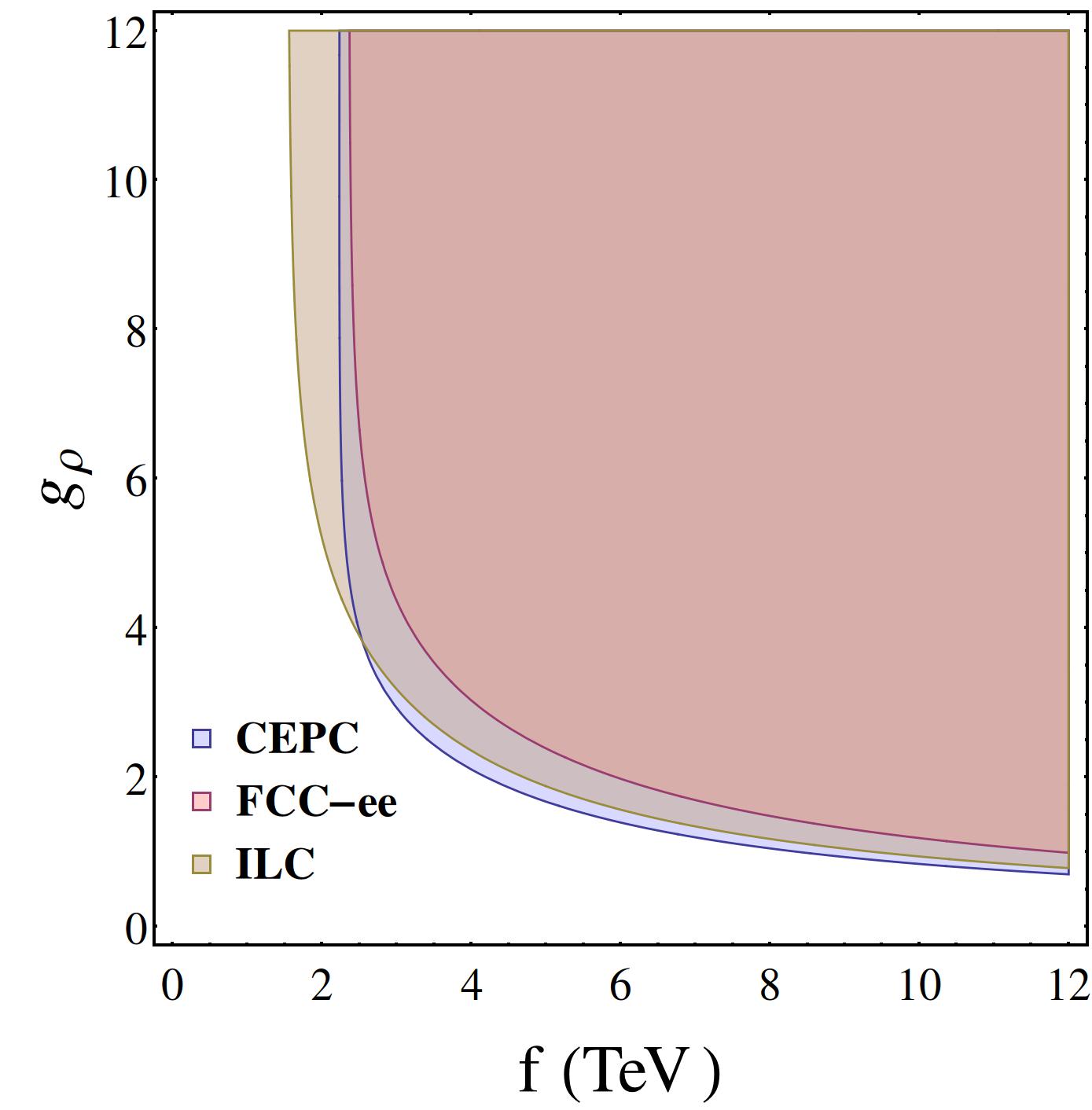}} \hspace{1.2cm}
\subfigure{
\includegraphics[height=5.5 cm]{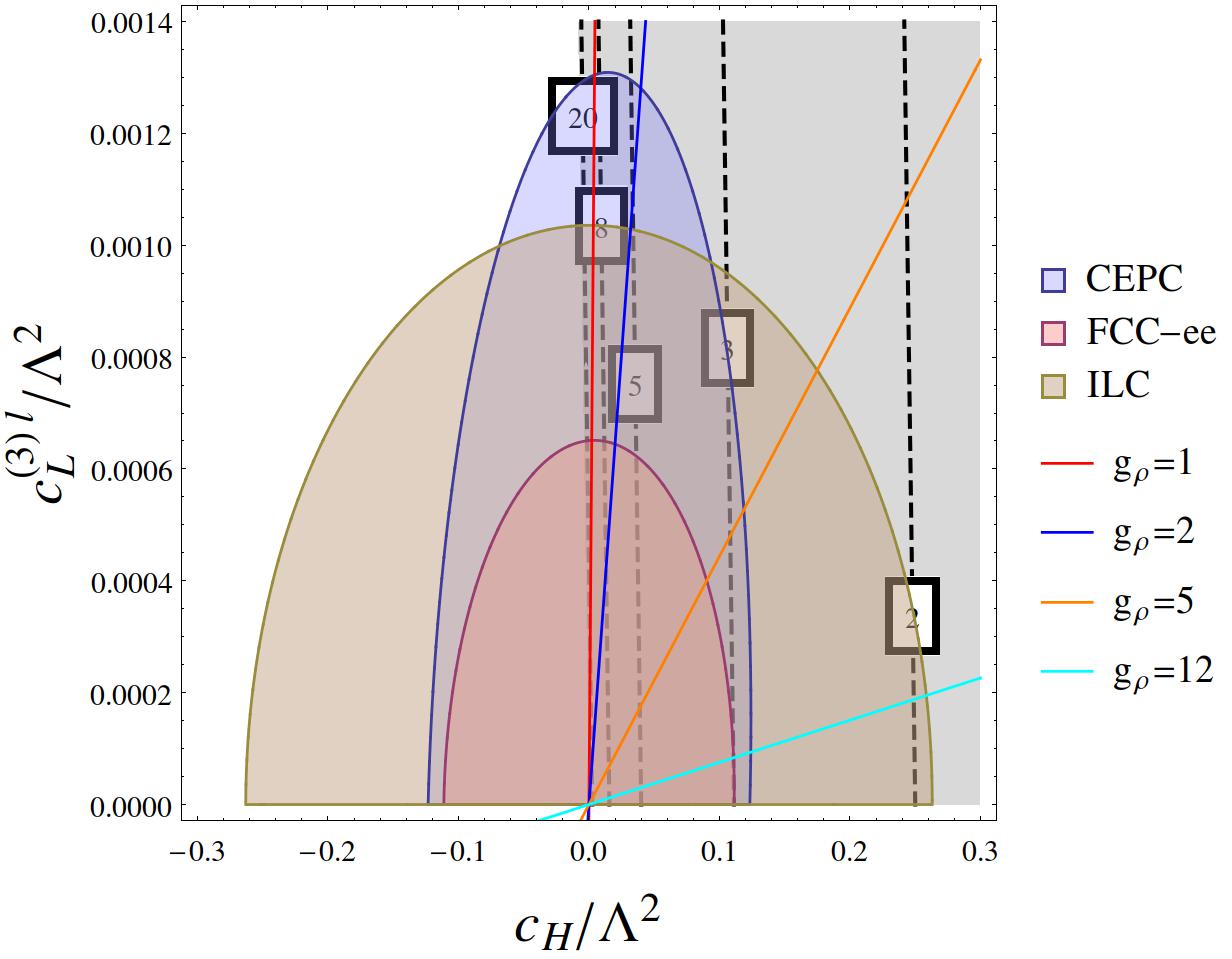}}
\caption{Sensitivities of probing the holographic composite Higgs model at future $e^-e^+$ colliders. In the right panel, the coordinate axes are in the unit of (TeV)$^{-2}$. The solid lines in color and the dashed lines represent the contours of $g_\rho$ and $f$ in strong dynamics, respectively. The gray region indicates the ranges defined by $f>0$ and $0 < g_\rho < 4\pi $.}  \label{holoHmodel}
\end{figure}

The holographic Higgs model~\cite{Agashe:2004rs,Contino:2006qr} is based on a theory over a slice of ADS$_5$ space-time. This space-time, characterized by a constant radius of curvature for its internal space, is compactified with two 4D branes as boundaries. By matching the holographic Higgs model  with the SILH EFT, one obtains the Wilson coefficients in the Lagrangian Eq.(\ref{SILHL}) as~\cite{Giudice:2007fh}
\begin{equation}
\tilde c_T = 0 \quad \tilde c_H = 1 \quad \tilde c_W = \tilde c_B \approx 1  \ .
\end{equation}
This setup yields a coefficient $c_H/\Lambda^2$ in the Lagrangian Eq.(\ref{SILHL2}) which depends on both SILH parameters: $f$ and $g_\rho$. As for the other coefficients, all of them are dependent on $(g_\rho f)^2$ only and hence are identical up to a constant factor.

The sensitivities of probing the holographic composite Higgs model at future $e^-e^+$ colliders are presented in Figure~\ref{holoHmodel}. According to the left panel, the parameter region with a small $f$ or/and a weak $g_\rho$ is relatively easy to probe. This is because it yields relatively light composite resonances and hence  a lower effective interacting scale. This observation is consistent with what one had in~\cite{Gu:2017ckc}, where the ``SILH'' pattern is essentially the holographic composite Higgs model discussed here, except that several more operators were turned on in~\cite{Gu:2017ckc}. Note, as the strong coupling  $g_\rho$ approaches $\sim 4\pi$, the loop-level operators in Eq. (\ref{SILHL}) may not be negligible in the analysis compared to $\mathcal {O}_W$ and $\mathcal {O}_B$. It is straightforward to project the sensitivities to the planes of the Wilson coefficients. For illustration, the projection at the $c_H/\Lambda^2 - c_L^{(3)l}/\Lambda^2$ plane is shown in the right panel in Figure~\ref{holoHmodel}. The projections at the other planes are either a single line (the ones with no axis being defined by $c_H/\Lambda^2$), or a rescaling of this panel along the $c_L^{(3)l}/\Lambda^2$ axis (the ones with the horizontal axis defined by $c_H/\Lambda^2$).

{\bf B. Littlest Higgs Model}

\begin{figure}[h]
\centering
\subfigure{
\includegraphics[height=5. cm]{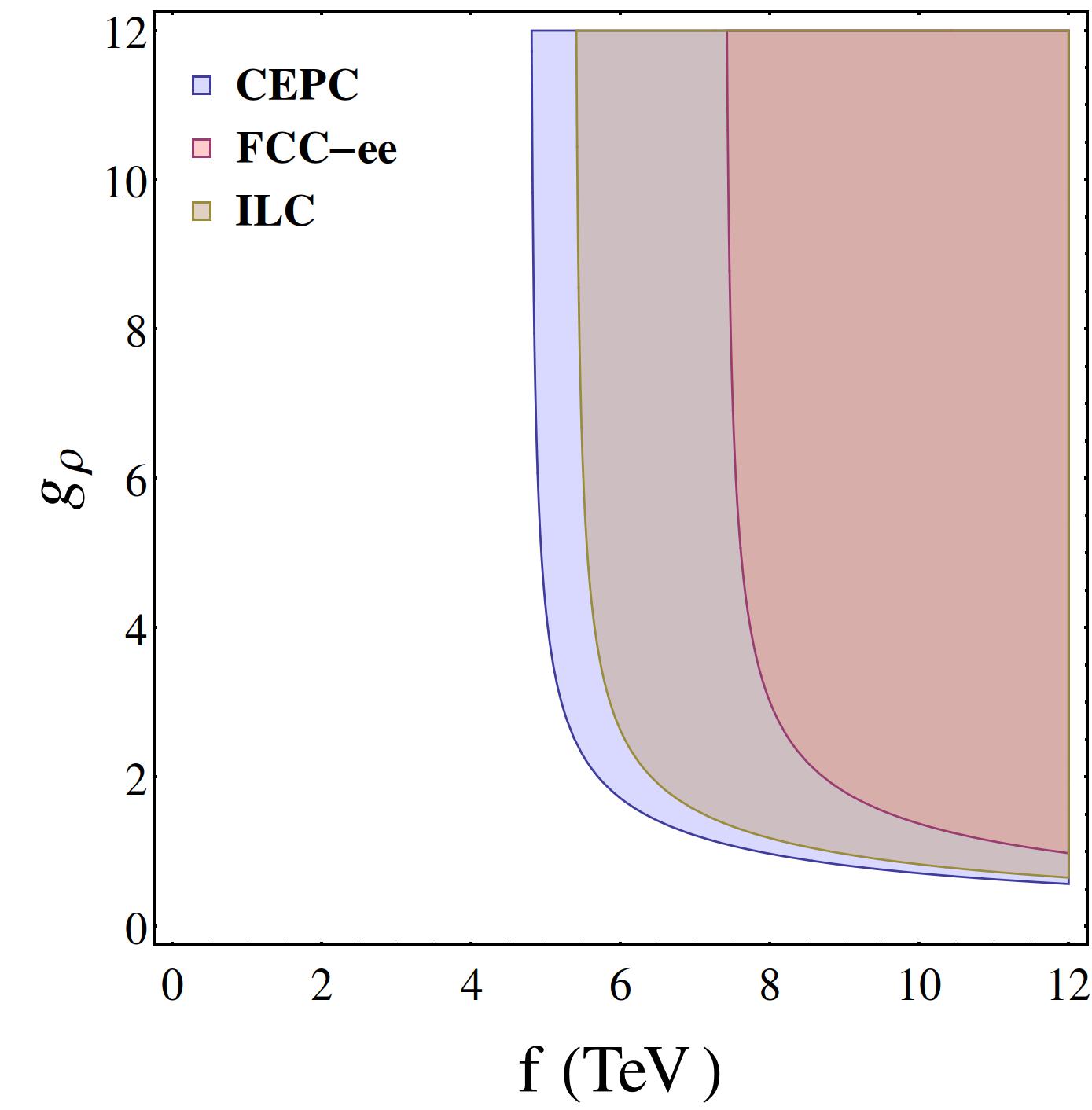}} \hspace{1.2cm}
\subfigure{
\includegraphics[height=5. cm]{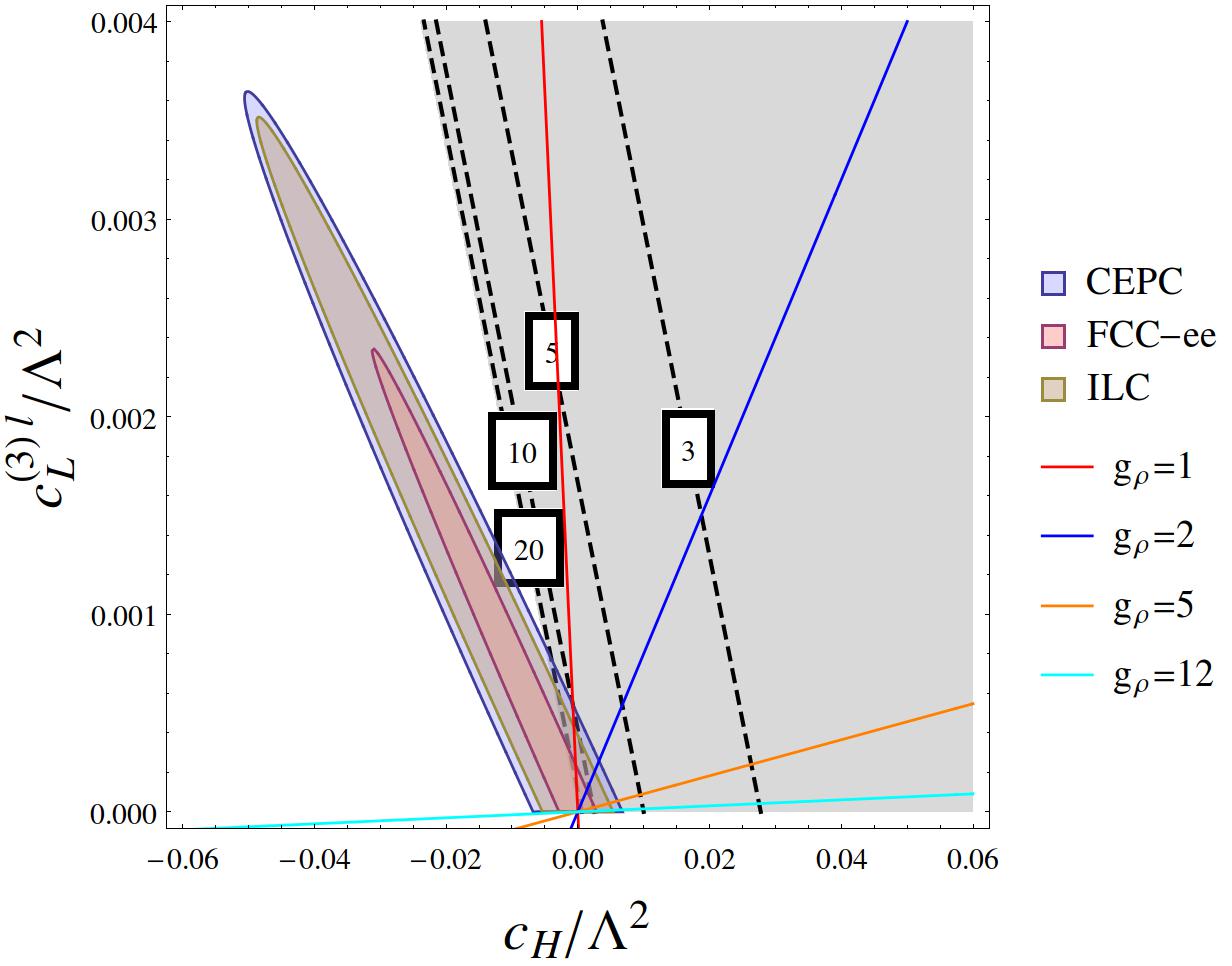}}
\subfigure{
\includegraphics[height=5. cm]{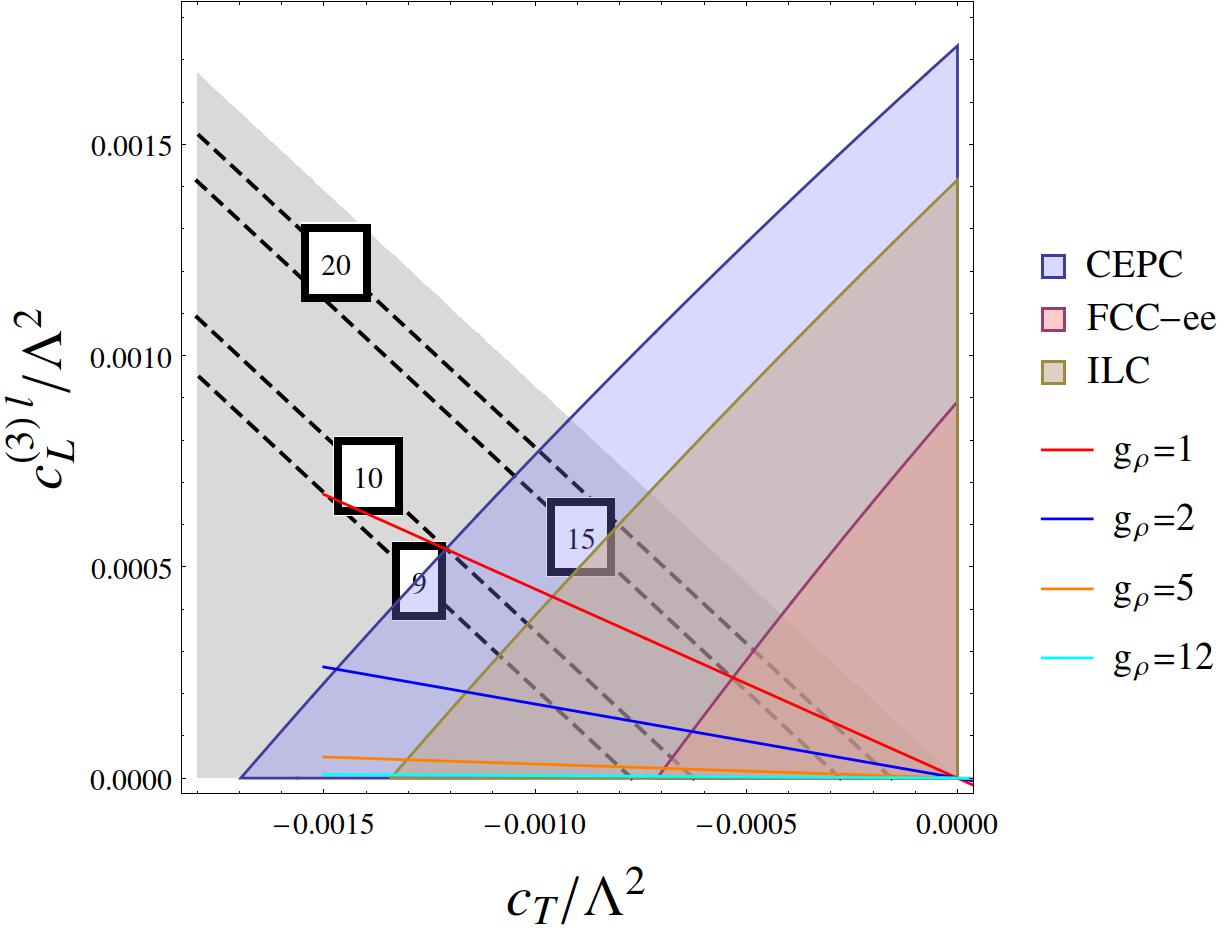}}
\subfigure{
\includegraphics[height=5. cm]{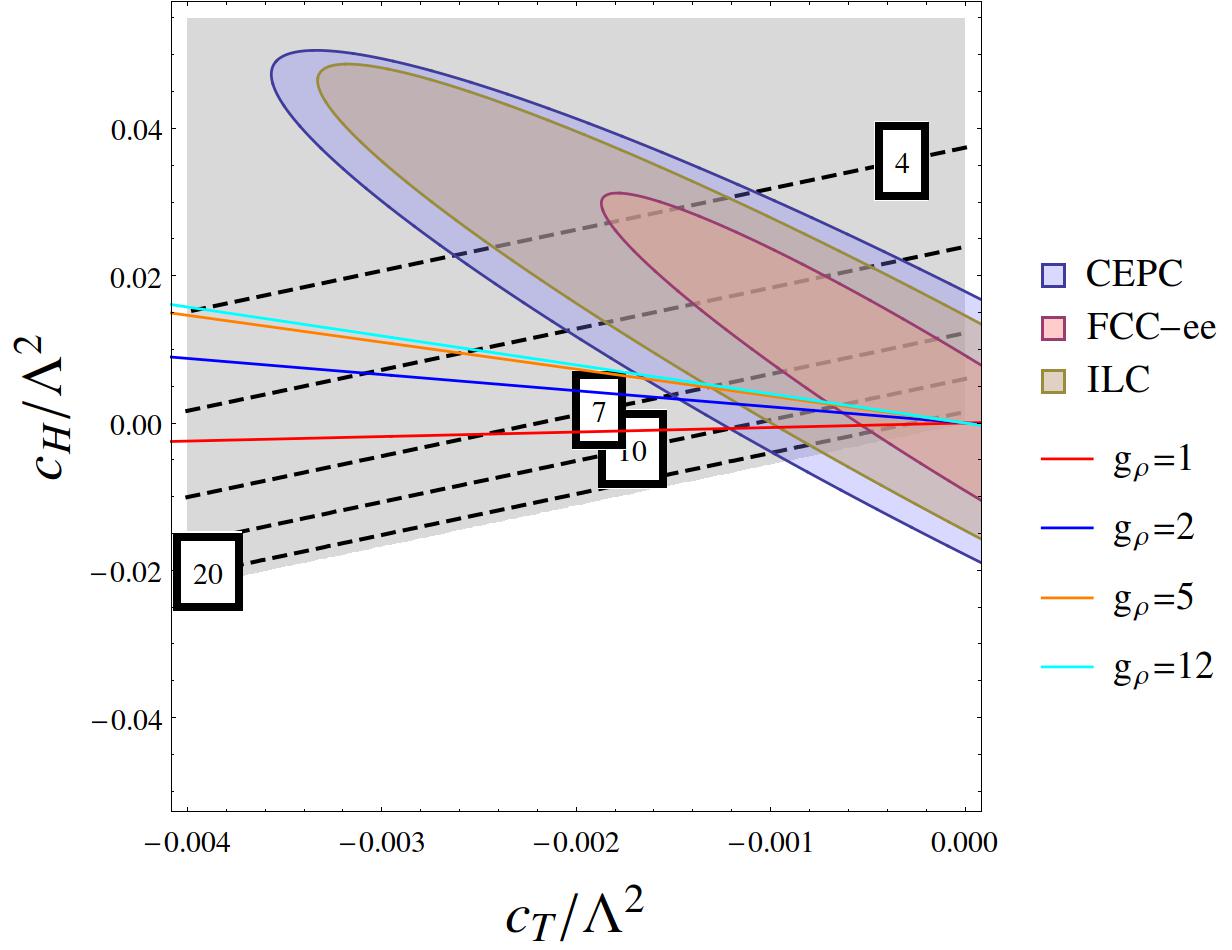}}
\caption{Sensitivities of probing the littlest Higgs model at future $e^-e^+$ colliders. In the panels except the left-upper one, the coordinate axes are in the unit of (TeV)$^{-2}$.  The solid lines in color and the dashed lines represent the contours of $g_\rho$ and $f$ in strong dynamics, respectively. The gray region indicates the ranges defined by $f>0$ and $0 < g_\rho < 4\pi $.}   \label{littHmodel}
\end{figure}

The littlest Higgs model~\cite{ArkaniHamed:2002qy} is a composite Higgs model with collectively symmetry breaking, with a coset group $SU(5)/SO(5)$. By matching the littlest Higgs model  with the SILH EFT, one can figure out the Wilson coefficients in the Lagrangian Eq.(\ref{SILHL}) as~\cite{Giudice:2007fh}
\begin{equation}
\tilde c_T = -\dfrac{1}{16} \quad \tilde c_H = \dfrac{1}{4} \quad \tilde c_W = \dfrac{1}{2} \quad \tilde c_B = 0 \ .
\end{equation}
This setup yields two vanishing coefficients in the Lagrangian Eq.(\ref{SILHL2}): $c_L^l/\Lambda^2$ and $c_R^e/\Lambda^2$. The other three coefficients $c_T/\Lambda^2$, $c_L^{(3)l}/\Lambda^2$ and $c_H$  are dependent on $f$, $g_\rho$, and both of them, respectively.

The sensitivities of probing the littlest Higgs model at future $e^-e^+$ colliders are presented in Figure~\ref{littHmodel}. Similar to the holographic composite Higgs model,  the parameter region with a small $f$ or/and a weak $g_\rho$ will be probed first (left-upper panel). The sensitivity projections to the planes expanded by $c_T/\Lambda^2$, $c_L^{(3)l}/\Lambda^2$ and $c_H$ are also presented.


\section{Conclusions}

In this article we presented a systematic study on the sensitivities of probing the UV physics at the future $e^-e^+$ colliders. The effect of new physics is parametrized by a set of 6D operators at leading order in its EFT. We turned on eleven of these operators simultaneously, which can be probed  by Higgs physics and EW precision measurements. The analysis provides a ``conservative'' projection on the collider sensitivities, complementary to the ``optimistic'' projection presented where these 6D operators are turned on individually. Then we made a comparative study on the sensitivities at CEPC, FCC-ee and ILC. Three running scenarios at ILC were considered: ``ILC250'' (ILC data at 250 GeV + EWPO measurements at LEP), ``ILC500'' (ILC250 + ILC data at 500 GeV) and ``ILC'' (ILC data at 250 and 500 GeV + GigaZ data). As an application, we analyzed two benchmark models in the  composite Higgs scenario. Our results can be briefly summarized as following. 

\begin{itemize}
\item In the ``optimistic'' analysis at CEPC, $\{\mathcal{O}_{WB}, \mathcal{O}_{T}, \mathcal{O}_{LL}^{(3)l}, \mathcal{O}_{L}^{(3)l}, \mathcal{O}_{L}^{l}, \mathcal{O}_{R}^{e}\}$ can be probed up to dozens of TeV by measuring the EWPOs, because of their tree-level contributions to the field redefinition and the coupling and charge shifts in the $Z$ boson current. $\{\mathcal{O}_{WW}, \mathcal{O}_{BB}, \mathcal{O}_{H},\mathcal{O}_{3W} \}$ can be probed up to TeV or several TeVs by measuring the Higgs observables and the $e^-e^+ \to WW$ production, due to their corrections to the Higgs couplings and TGC, respectively.  $\mathcal{O}_{6}$ is difficult to probe because it contributes to $e^-e^+\to Zh$ at loop level only.  These features are shared by FCC-ee and ILC250, ILC500, ILC (though the sensitivities to probe $O_6$ can be improved to some extent by measuring the $e^-e^+\to Zhh$ production at ILC500 and ILC).  

\item In the ``conservative'' analysis where the set of eleven operators are turned on simultaneously,  the energy scales that the CEPC and FCC-ee are able to probe decrease to $\sim \mathcal O(1-10)$TeV for $\{\mathcal{O}_{WB}, \mathcal{O}_{T}, \mathcal{O}_{LL}^{(3)l}, \mathcal{O}_{L}^{(3)l}, \mathcal{O}_{L}^{l}, \mathcal{O}_{R}^{e}\}$. This is mainly due to an approximate degeneracy caused by the weakness of $R_b$. For $\{\mathcal{O}_{WW}, \mathcal{O}_{BB}, \mathcal{O}_{H},  \mathcal{O}_{3W}\}$, the sensitivities decrease  to TeV or sub TeV, and for $\mathcal{O}_{6}$ to $< \mathcal O(0.1)$ TeV. 

\item Benefitting from a larger integrated luminosity at $Z$ pole, the sensitivities at FCC-ee are mildly better than the CEPC ones, in both ``optimistic'' and ``conservative'' projections.

\item An ILC run with $E_{\rm CM} = 500$ GeV (ILC500) is highly beneficial. Limited by its relatively small luminosity at 250 GeV and the lack of data at $Z$-pole,  ILC250 is less capable in probing these operators. However, this can be adequately compensated by the data at 500 GeV. By combining with the 500 GeV data, the ILC performance is comparable to or better than the CEPC and FCC-ee ones. Moreover, compared to CEPC and FCC-ee, ILC500 performs much better in probing the $\mathcal{O}_{6}$ operator or measuring the cubic Higgs coupling in the ``conservative'' analysis. This is mainly because the $e^-e^+ \to Zhh$ production, an observable not available at CEPC and FCC-ee~\cite{Barklow:2017awn}, can break the degeneracy related to $\mathcal O_{6}$. Additionally, the ILC can also benefit from time splitting among different polarization configurations~\cite{Durieux:2017rsg,Barklow:2017suo}.

\item As an application, the ``conservative'' analysis is applied to the simplified model of SILH, with the mutual dependence of the Wilson coefficients taken into account. The analysis indicates that CEPC, FCC-ee and ILC have a potential to probe its decay constant up to $\mathcal O(1-10)$TeV, with the strong coupling varying between $1 - 4\pi$.

\end{itemize}


\begin{acknowledgments}

We would like to thank M. Peskin for valuable comments on the draft, and J. Gu, Y. Jiang and Z. Liu for helpful discussions. We would thank C. Grojean et. al. for coordinated publication of their related work. We would acknowledges the hospitality of the Jockey Club Institute for Advanced Study at Hong Kong University of Science and Technology, where part of this work was performed. T. Liu and K. Lyu are supported by the Collaborative Research Fund (CRF) under Grant No. HUKST4/CRF/13G. T. Liu is also supported by the General Research Fund (GRF) under Grant No 16312716. Both the CRF and GRF grants are issued by the Research Grants Council of Hong Kong S.A.R.. L.-T. Wang is  supported by the U.S. Department of Energy under grant No. DE-SC0013642.

\end{acknowledgments}

\newpage

\appendix

\section{Feynman Rules for the Interaction Vertices}

The modified Feynman rules for the interaction vertices are listed as below \\

\begin{minipage}{0.22\textwidth}
\includegraphics[width=3.3cm]{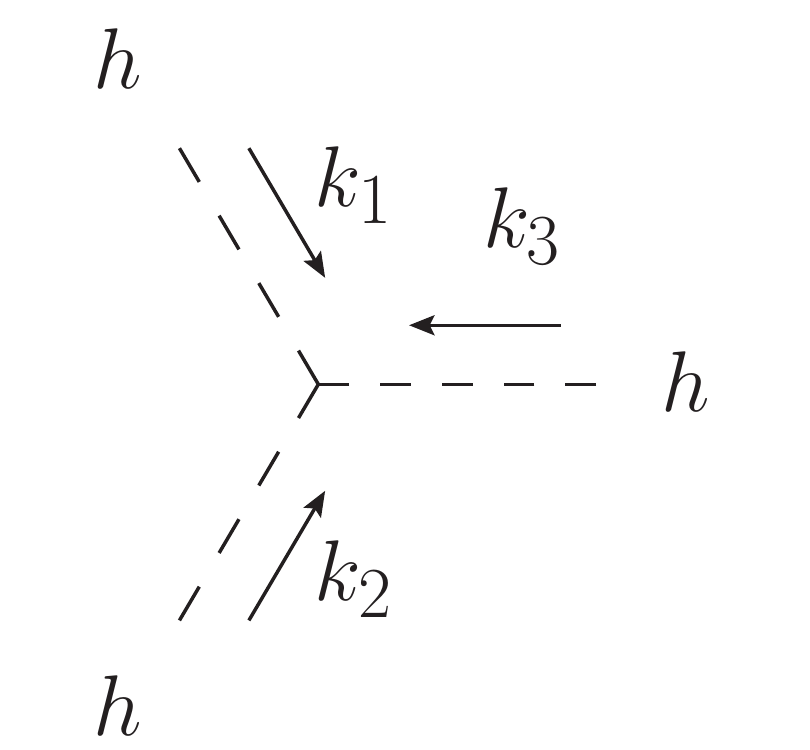}
\end{minipage}
\begin{minipage}{0.75\textwidth}
\begin{equation}
\begin{split}
 =&-3 i \dfrac{ m_h^2}{v}  \left(1 + 3 \delta Z_h + \dfrac{\delta G_F}{2G_F}- \dfrac{2 c_6 v^4 }{\lambda m_h^2 \Lambda^2}\right) \\
  &~-2i \dfrac{c_H}{\Lambda^2} v (k_1 \cdot k_2+ k_1 \cdot k_3 + k_2 \cdot k_3 )
\end{split}
\end{equation}
\end{minipage}
\vspace{0.5cm}

\begin{minipage}{0.22\textwidth}
\includegraphics[width=3.3cm]{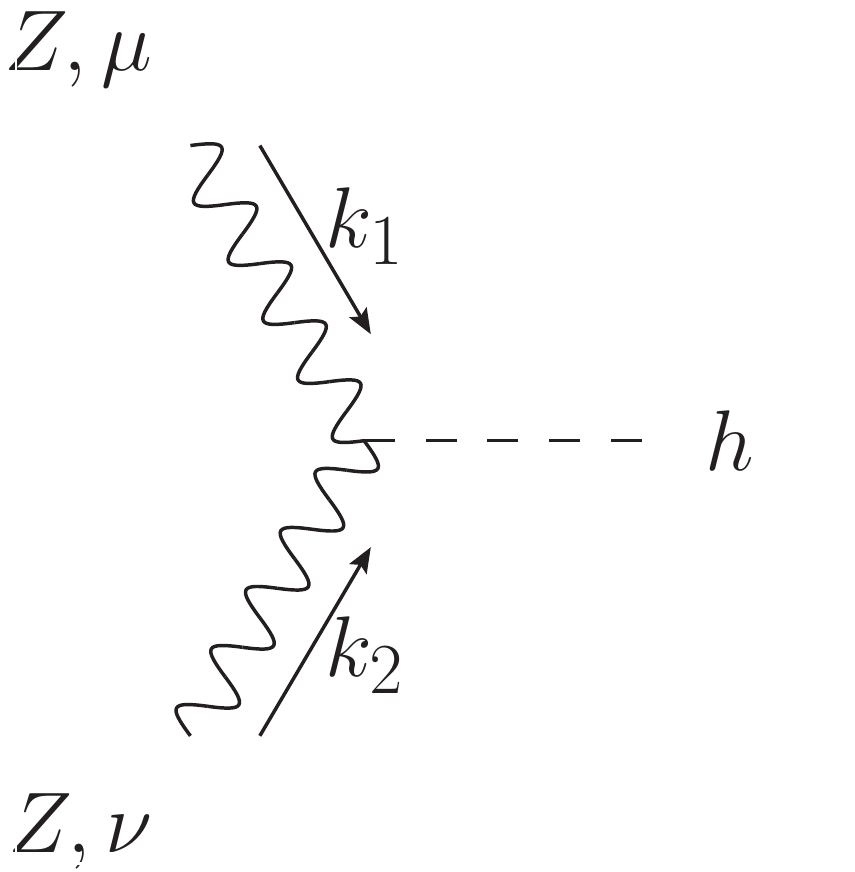}
\end{minipage}
\begin{minipage}{0.75\textwidth}
\begin{equation}
\begin{split}
=&i g^{\mu\nu} \Bigg[ \dfrac{g_z^2 v}{2}  \left(1+2 \delta Z_Z + \delta Z_h + \dfrac{2 \delta g_Z}{g_Z} + \dfrac{\delta G_F}{2G_F}\right) \\
& ~~~~~~~+\dfrac{v}{\Lambda^2}  (v^2 d_1 - (k_1 \cdot k_2) d_2) \Bigg]+
i \dfrac{v d_2}{\Lambda^2} k_1^\nu k_2^\mu
\end{split}
\end{equation}
\end{minipage}
\vspace{0.5cm}

\begin{minipage}{0.22\textwidth}
\includegraphics[width=3.3cm]{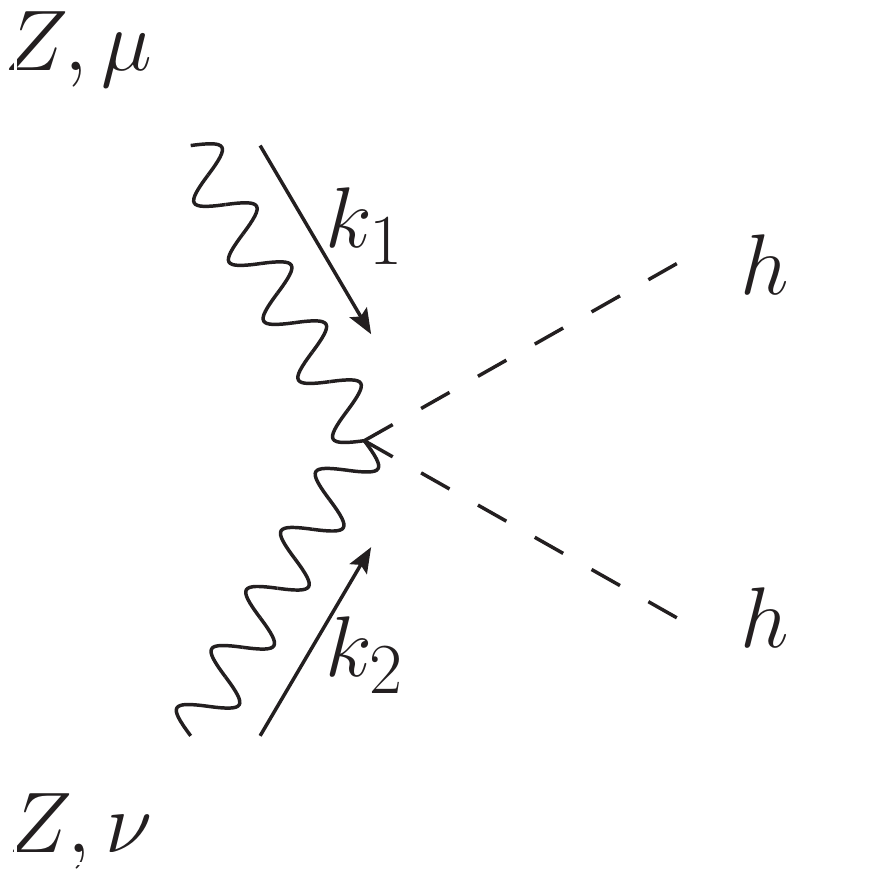}
\end{minipage}
\begin{minipage}{0.75\textwidth}
\begin{equation}
\begin{split}
=&i g^{\mu\nu} \Bigg[ \dfrac{g_z^2}{2}   \left(1+2 \delta Z_Z + \delta Z_h + \dfrac{\delta g_Z}{g_Z}\right) \\
&~~~~~~~+\dfrac{1}{\Lambda^2}  (v^2 d_3 - (k_1 \cdot k_2) d_2) \Bigg]+
i  \dfrac{d_2}{\Lambda^2} k_1^\nu k_2^\mu
\end{split}
\end{equation}
\end{minipage}\\
\vspace{0.5cm}

\begin{minipage}{0.22\textwidth}
\includegraphics[width=3.3cm]{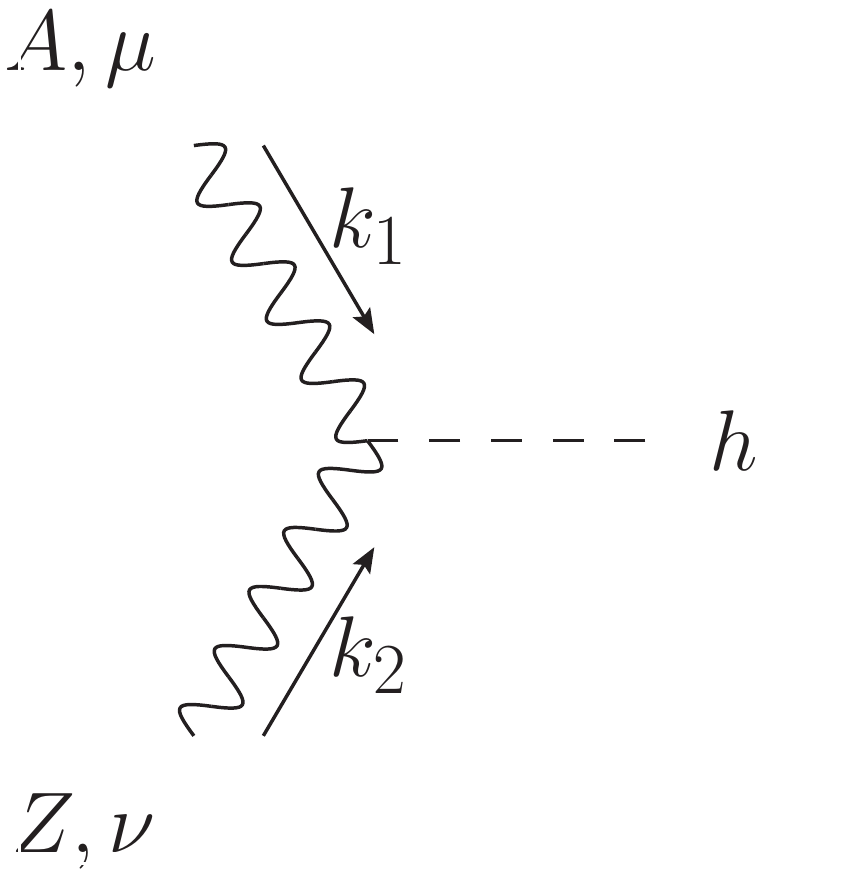}
\end{minipage}
\begin{minipage}{0.75\textwidth}
\begin{equation}
=i \dfrac{v d_4}{\Lambda^2}[-(k_1\cdot k_2) g^{\mu \nu} +k_1^\nu k_2^\mu]
\end{equation}
\end{minipage}\\
\vspace{0.5cm}

\begin{minipage}{0.22\textwidth}
\includegraphics[width=3.3cm]{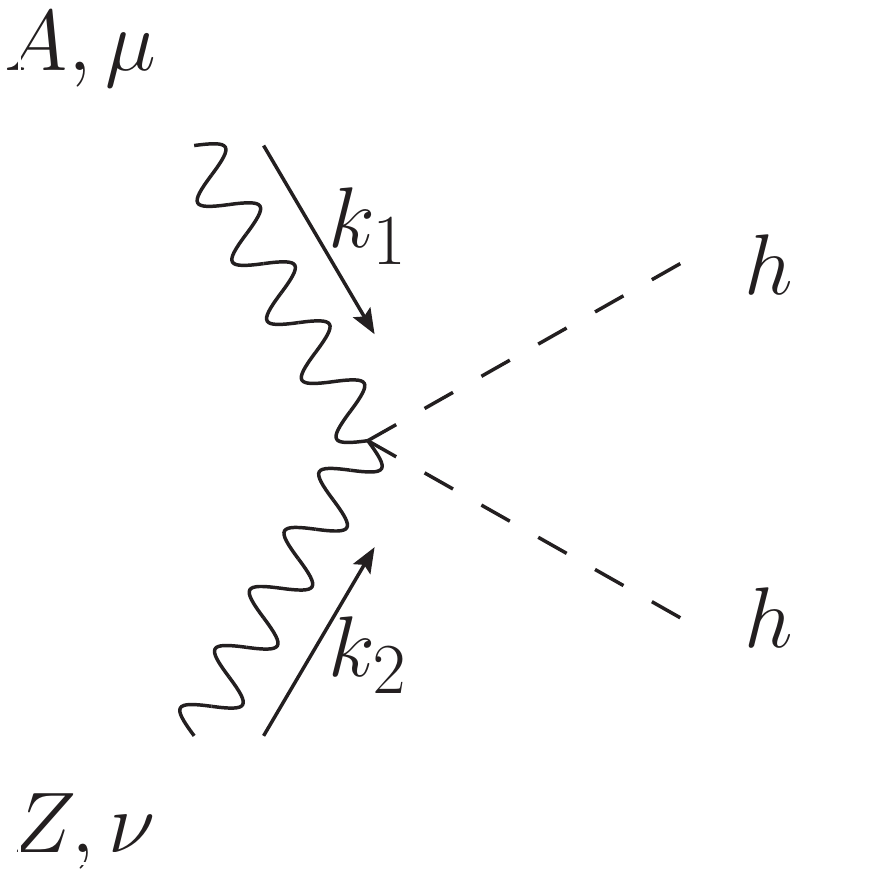}
\end{minipage}
\begin{minipage}{0.75\textwidth}
\begin{equation}
=i \dfrac{ d_4}{\Lambda^2}[-(k_1\cdot k_2) g^{\mu \nu} +k_1^\nu k_2^\mu]
\end{equation}
\end{minipage}\\
\vspace{0.5cm}

\begin{minipage}{0.22\textwidth}
\includegraphics[width=3.3cm]{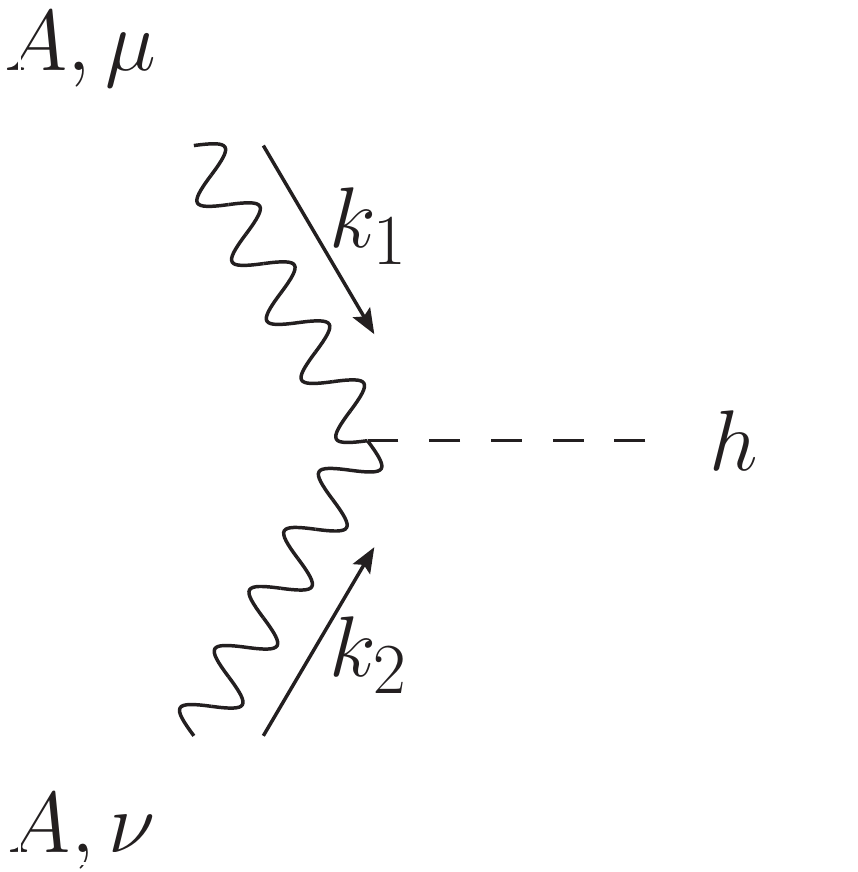}
\end{minipage}
\begin{minipage}{0.75\textwidth}
\begin{equation}
=i \dfrac{v d_7}{\Lambda^2}[-(k_1\cdot k_2) g^{\mu \nu} +k_1^\nu k_2^\mu]
\end{equation}
\end{minipage}\\
\vspace{0.5cm}

\begin{minipage}{0.22\textwidth}
\includegraphics[width=3.3cm]{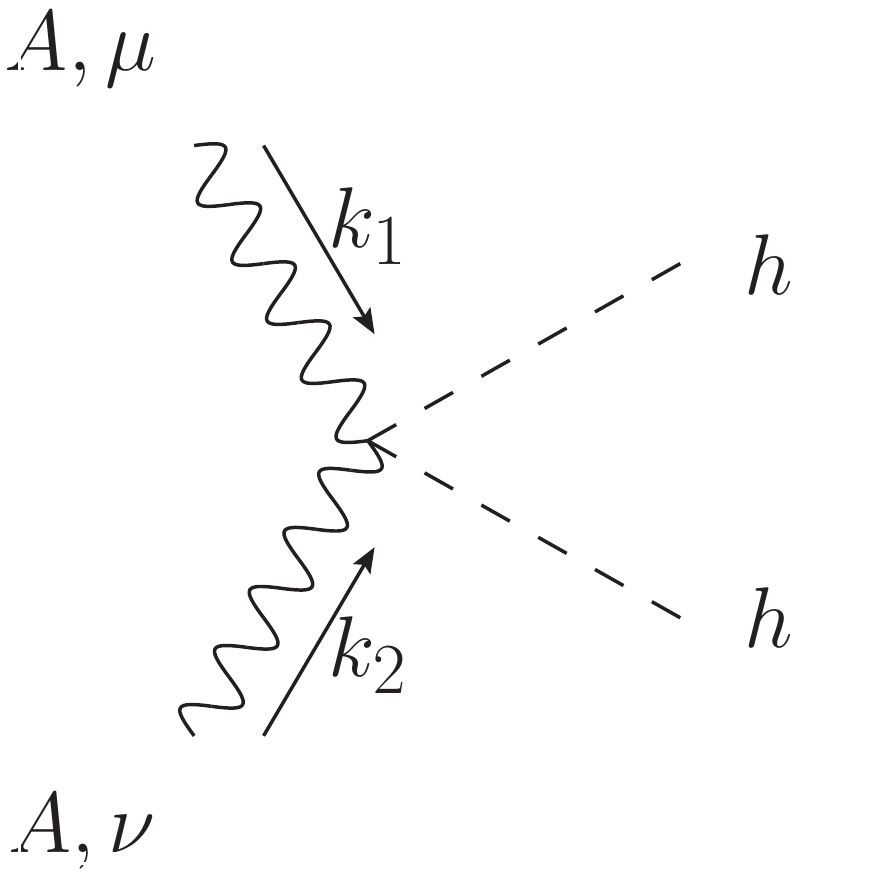}
\end{minipage}
\begin{minipage}{0.75\textwidth}
\begin{equation}
=i \dfrac{ d_7}{\Lambda^2}[-(k_1\cdot k_2) g^{\mu \nu} +k_1^\nu k_2^\mu]
\end{equation}
\end{minipage}\\
\vspace{0.5cm}

\begin{minipage}{0.22\textwidth}
\includegraphics[width=3.3cm]{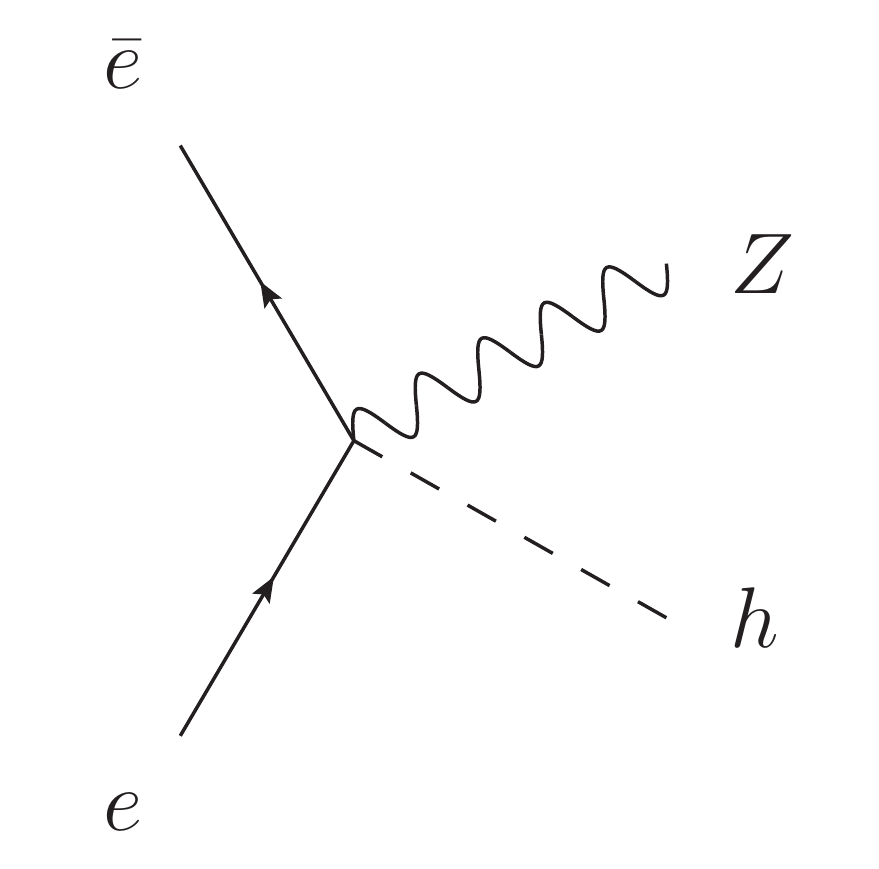}
\end{minipage}
\begin{minipage}{0.75\textwidth}
\begin{equation}
=i \dfrac{v}{\Lambda^2} \gamma^\mu (d_5 P_L + d_6 P_R)
\end{equation}
\end{minipage}\\
\vspace{0.5cm}

\begin{minipage}{0.22\textwidth}
\includegraphics[width=3.3cm]{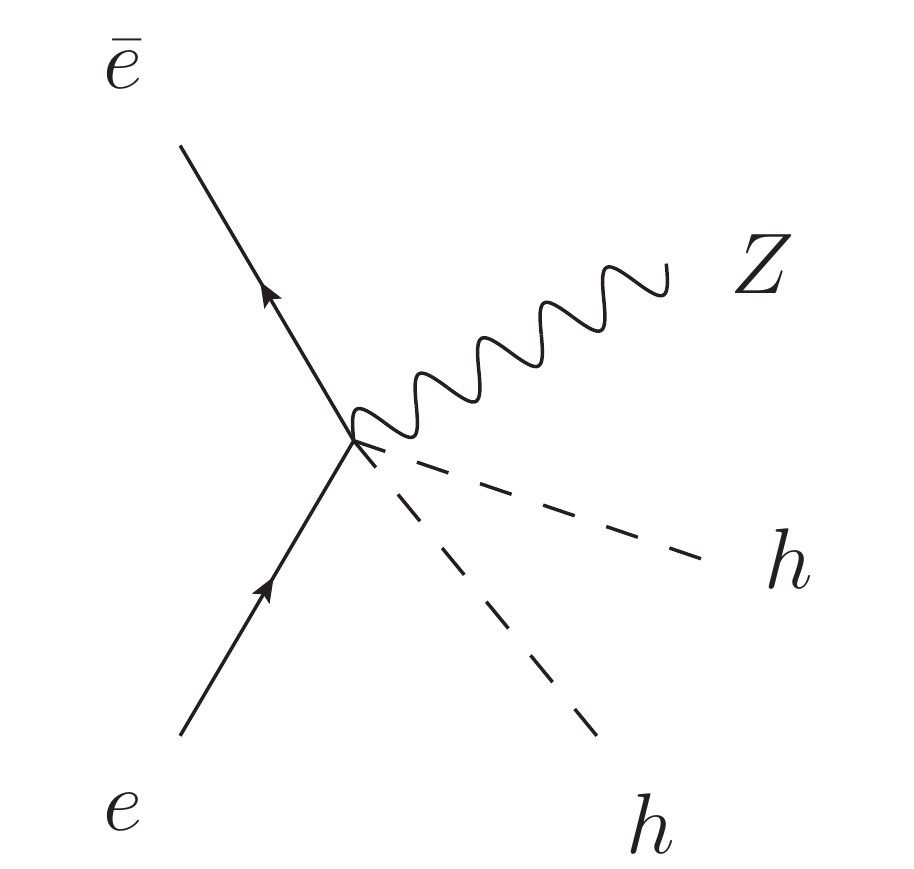}
\end{minipage}
\begin{minipage}{0.75\textwidth}
\begin{equation}
=i \dfrac{1}{\Lambda^2} \gamma^\mu (d_5 P_L + d_6 P_R)
\end{equation}
\end{minipage}\\
\vspace{0.5cm}

\begin{minipage}{0.22\textwidth}
\includegraphics[width=3.3cm]{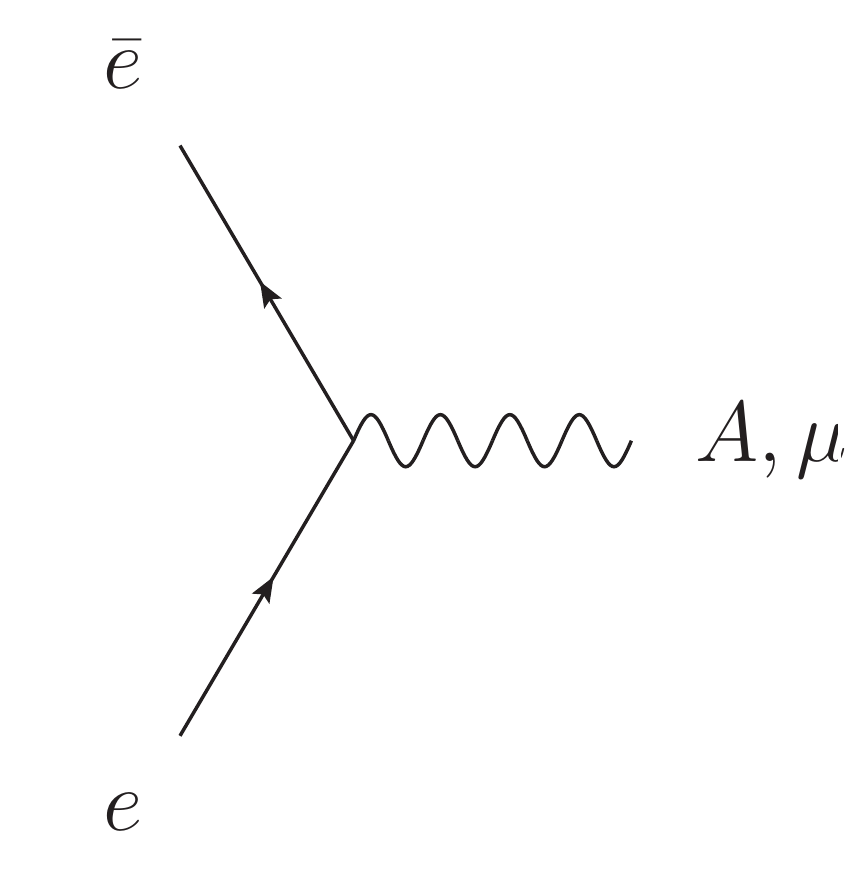}
\end{minipage}
\begin{minipage}{0.75\textwidth}
\begin{equation}
=-i \gamma^\mu e
\end{equation}
\end{minipage}\\
\vspace{0.5cm}

\begin{minipage}{0.22\textwidth}
\includegraphics[width=3.3cm]{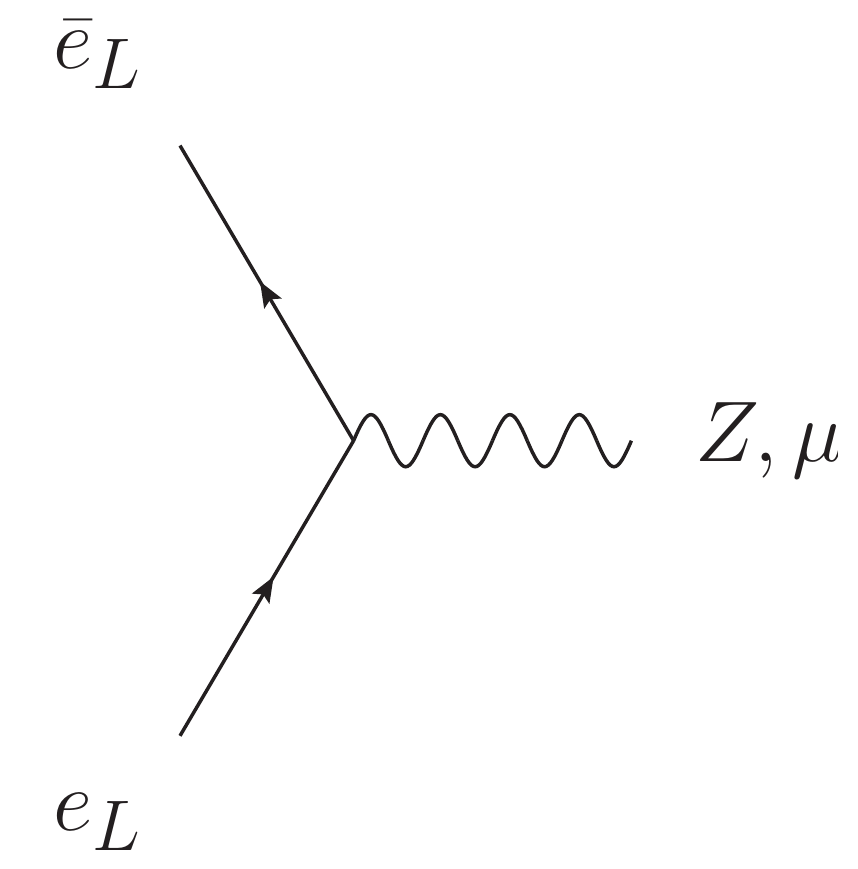}
\end{minipage}
\begin{minipage}{0.75\textwidth}
\begin{equation}
=i \gamma^\mu\Big(g_L \left(1+ \delta Z_Z +\dfrac{\delta g_Z}{g_Z} - \dfrac{2 s_{2w} \delta \theta_w}{c_{2w}}\right)-e \delta Z_X \Big)
\end{equation}
\end{minipage}\\
\vspace{0.5cm}

\begin{minipage}{0.22\textwidth}
\includegraphics[width=3.3cm]{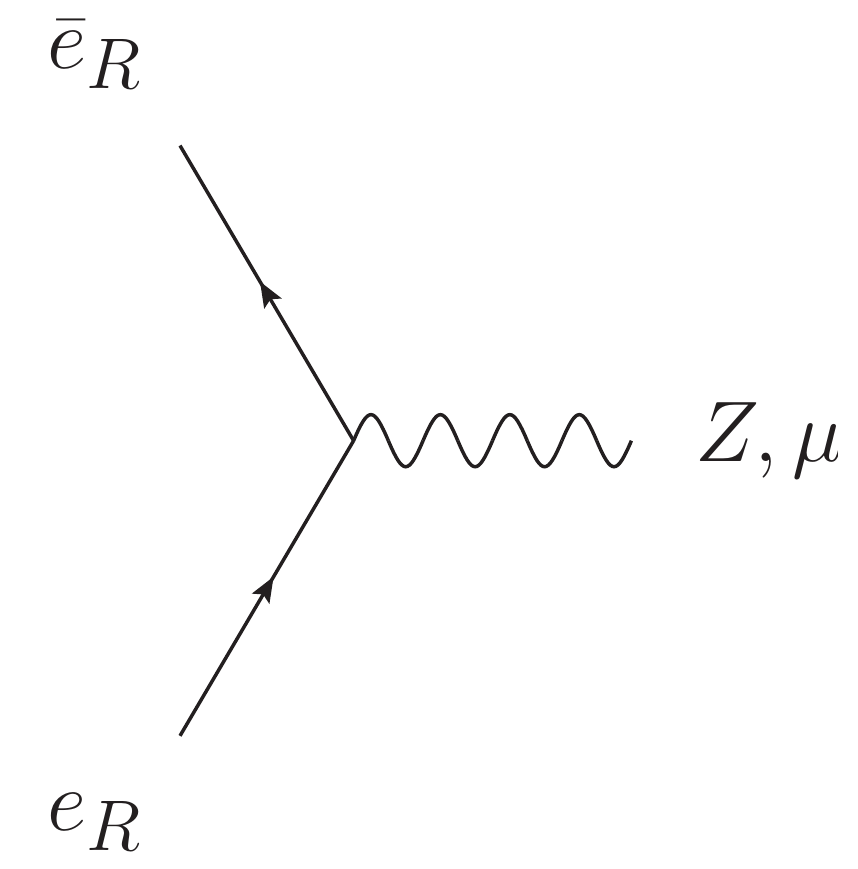}
\end{minipage}
\begin{minipage}{0.75\textwidth}
\begin{equation}
=i \gamma^\mu\Big(g_z \sin^2 \theta \left(1+\delta Z_Z + \dfrac{\delta g_Z}{g_Z} + \dfrac{2 c_w \delta \theta_w}{s_w}\right)-e \delta Z_X \Big)
\end{equation}
\end{minipage}
\vspace{0.5cm}

Here the Higgs and gauge fields have been rescaled to their canonical forms. The relevant coefficients are defined as
\begin{equation}
\begin{split}
d_1&=\dfrac{g_Z^2}{2} \left(\dfrac{1}{2}c_H +2 c_T\right)+g_Z^4(c^4_w c_{WW} + c_w^2 s_w^2 c_{WB} +s_w^4 c_{BB})\\
d_2&=4 g_Z^2 (c^4_w c_{WW} + c_w^2 s_w^2 c_{WB} +s_w^4 c_{BB})\\
d_3&=-3 g_Z^2 c_T -\dfrac{g_Z^2}{2} c_H + g_Z^4(c^4_w c_{WW} + c_w^2 s_w^2 c_{WB} +s_w^4 c_{BB})\\
d_4&=2 g_Z^2 c_w s_w (-2 s_w^2 c_{BB} -(c_w^2 - s_w^2)c_{WB} +2 c_w^2 c_{WW})\\
d_5&=-g_Z (c_L^{(3)l} + c_L^l)\\
d_6&=-g_Z c^e_R\\
d_7&=4 g_Z^2 c_w^2 s_w^2 (c_{WW}+c_{BB} -c_{WB})
\end{split}
\end{equation}

\section{Observables for Analysis: Numerical Formulae}
\label{App:formulae}

The formulae for calculating the contributions of the 6D operators to the observables at future $e^-e^+$ colliders are listed in the following. The formulae are obtained by using MadGraph and CalcHEP, with the model files generated by FeynRule, or by using Mathematica directly. The effect of renormalization group running from the cutoff to the $Z$ pole or the beam energy scales has been neglected for the Wilson coefficients. In the following, we use the simplified notation $\dfrac{c_i}{\Lambda^2} \equiv \dfrac{c_i}{(\Lambda / \text{TeV})^2}$

\begin{enumerate}

\item EWPOs.

\begin{itemize}

\item  $N_\nu$
\begin{equation}
\begin{split}
\dfrac{\Delta N_\nu}{N_\nu} =& 0.00585\dfrac{c_{WB}}{\Lambda^2} - 0.0107 \dfrac{c_{T}}{\Lambda^2}  +0.0139 \dfrac{c_{L}^{(3)l}}{\Lambda^2}-0.0213 \dfrac{c^{(3)l}_{LL}}{\Lambda^2} -\\
&0.250 \dfrac{c^{l}_{L}}{\Lambda^2} + 0.113 \dfrac{c^e_R}{\Lambda^2}
\end{split}  \label{Nnu}
\end{equation}

\item  $A_b$
\begin{equation}
\dfrac{\Delta A_b}{\mathcal{A}_b} =  - 0.00781 \dfrac{c_{WB}}{\Lambda^2}  + 0.0142 \dfrac{c_{T}}{\Lambda^2}  -0.0285 \dfrac{c^{(3)l}_L}{\Lambda^2} + 0.0285 \dfrac{c^{(3)l}_{LL}}{\Lambda^2}
\end{equation}

\item  $A_{FB}^{\mu}$
\begin{equation}
\begin{split}
\dfrac{\Delta A_{FB}^{\mu}}{\mathcal{A}_{FB}^{\mu}} =&  - 0.101 \dfrac{c_{WB}}{\Lambda^2}  + 0.184 \dfrac{c_{T}}{\Lambda^2}  -0.241 \dfrac{c^{(3)l}_L}{\Lambda^2} + 0.369 \dfrac{c^{(3)l}_{LL}}{\Lambda^2} +\\ & 0.128 \dfrac{c^l_{L}}{\Lambda^2} +0.146 \dfrac{c^e_{R}}{\Lambda^2}
\end{split}
\end{equation}

\item  $A_{FB}^b$
\begin{equation}
\begin{split}
\dfrac{\Delta A_{FB}^b}{A_{FB}^b} =&  - 0.625 \dfrac{c_{WB}}{\Lambda^2}  + 1.14 \dfrac{c_{T}}{\Lambda^2}  -1.50 \dfrac{c^{(3)l}_L}{\Lambda^2} + 2.28 \dfrac{c^{(3)l}_{LL}}{\Lambda^2} +\\ & 0.784 \dfrac{c^l_{L}}{\Lambda^2} +0.894 \dfrac{c^e_{R}}{\Lambda^2}
\end{split}
\end{equation}

\item  $R_b$
\begin{equation}
\begin{split}
\dfrac{\Delta R_b}{R_b} =& 0.00189 \dfrac{c_{WB}}{\Lambda^2} -0.00345 \dfrac{c_{T}}{\Lambda^2}  + 0.00691 \dfrac{c^{(3)l}_L}{\Lambda^2} - 0.00691 \dfrac{c^{(3)l}_{LL}}{\Lambda^2}
\end{split}
\end{equation}

\item  $R_\mu$
\begin{equation}
\begin{split}
\dfrac{\Delta R_\mu}{R_\mu} =& - 0.00969 \dfrac{c_{WB}}{\Lambda^2} +0.0177 \dfrac{c_{T}}{\Lambda^2}  -
 0.159 \dfrac{c^{(3)l}_L}{\Lambda^2} +\\&0.0353 \dfrac{c^{(3)l}_{LL}}{\Lambda^2} -0.124 \dfrac{c^l_{L}}{\Lambda^2} +0.109 \dfrac{c^e_{R}}{\Lambda^2}
\end{split}
\end{equation}

\item  $R_\tau$
\begin{equation}
\begin{split}
\dfrac{\Delta R_\tau}{R_\tau} =&-0.00970 \dfrac{c_{WB}}{\Lambda^2} + 0.0177 \dfrac{c_{T}}{\Lambda^2}  - 0.160 \dfrac{c^{(3)l}_L}{\Lambda^2} + \\& 0.0354 \dfrac{c^{(3)l}_{LL}}{\Lambda^2} -0.124 \dfrac{c^l_{L}}{\Lambda^2} +0.109 \dfrac{c^e_{R}}{\Lambda^2}
\end{split}
\end{equation}

\item  $\sin^2 \theta^{\text{lep}}_{\text{eff}}$
\begin{equation}
\begin{split}
\dfrac{\Delta s_w^2}{s_w^2} =& 0.0483 \dfrac{c_{WB}}{\Lambda^2} -0.0881 \dfrac{c_{T}}{\Lambda^2}  + 0.115 \dfrac{c^{(3)l}_L}{\Lambda^2} -0.176 \dfrac{c^{(3)l}_{LL}}{\Lambda^2} -\\&0.0612 \dfrac{c^l_{L}}{\Lambda^2} -0.0698 \dfrac{c^e_{R}}{\Lambda^2}
\end{split}  \label{leptheta}
\end{equation}

\item  $\Gamma_Z$
\begin{equation}
\begin{split}
\dfrac{\Delta \Gamma_Z}{\Gamma_Z} =&-0.0112 \dfrac{c_{WB}}{\Lambda^2} + 0.079 \dfrac{c_{T}}{\Lambda^2}  - 0.121 \dfrac{c^{(3)l}_L}{\Lambda^2} + \\& 0.158 \dfrac{c^{(3)l}_{LL}}{\Lambda^2} -0.0113	 \dfrac{c^l_{L}}{\Lambda^2} -0.0113 \dfrac{c^e_{R}}{\Lambda^2}
\end{split}
\end{equation}

\item  $m_W$
\begin{equation}
\begin{split}
\dfrac{\Delta m_W}{m_W} =&  - 0.0111 \dfrac{c_{WB}}{\Lambda^2}  +0.0433 \dfrac{c_{T}}{\Lambda^2}  -
 0.0264 \dfrac{c^{(3)l}_L}{\Lambda^2} +0.0264 \dfrac{c^{(3)l}_{LL}}{\Lambda^2}
\end{split}
\end{equation}

\end{itemize}

\item Total signal rates.

\begin{itemize}

\item  $e^+ e^- \rightarrow Z h $\\
(1) unpolarized 240 GeV
\begin{eqnarray}
\begin{split}
\dfrac{\Delta \sigma}{\sigma_0} =& 0.225 \dfrac{c_{WW}}{\Lambda^2} + 0.0554 \dfrac{c_{WB}}{\Lambda^2} + 0.0164 \dfrac{c_{BB}}{\Lambda^2} - 0.0500 \dfrac{c_{T}}{\Lambda^2} - 0.0606 \dfrac{c_{H}}{\Lambda^2} \\
& + 0.627 \dfrac{c^{(3)l}_L}{\Lambda^2} + 0.264 \dfrac{c^{(3)l}_{LL}}{\Lambda^2} + 0.891 \dfrac{c^l_{L}}{\Lambda^2} - 0.781 \dfrac{c^e_{R}}{\Lambda^2} - 0.00106 \dfrac{c_6}{\Lambda^2}
\end{split}
\end{eqnarray}

(2) polarized (-0.8, 0.3) 250 GeV
\begin{eqnarray}
\begin{split}
\dfrac{\Delta \sigma}{\sigma_0} =& 0.379 \dfrac{c_{WW}}{\Lambda^2} -0.0613 \dfrac{c_{WB}}{\Lambda^2} -0.0263 \dfrac{c_{BB}}{\Lambda^2} +0.0779 \dfrac{c_{T}}{\Lambda^2} - 0.0606 \dfrac{c_{H}}{\Lambda^2} \\
& + 1.12 \dfrac{c^{(3)l}_L}{\Lambda^2} + 0.520 \dfrac{c^{(3)l}_{LL}}{\Lambda^2} + 1.64 \dfrac{c^l_{L}}{\Lambda^2} - 0.0858 \dfrac{c^e_{R}}{\Lambda^2} - 0.000944 \dfrac{c_6}{\Lambda^2}
\end{split}
\end{eqnarray}

(3) polarized (-0.8, 0.3) 500 GeV
\begin{eqnarray}
\begin{split}
\dfrac{\Delta \sigma}{\sigma_0} =& 0.666 \dfrac{c_{WW}}{\Lambda^2} -0.0617 \dfrac{c_{WB}}{\Lambda^2} -0.0532 \dfrac{c_{BB}}{\Lambda^2} +0.0779 \dfrac{c_{T}}{\Lambda^2} - 0.0606 \dfrac{c_{H}}{\Lambda^2} \\
& + 6.02 \dfrac{c^{(3)l}_L}{\Lambda^2} + 0.520 \dfrac{c^{(3)l}_{LL}}{\Lambda^2} + 6.54 \dfrac{c^l_{L}}{\Lambda^2} - 0.343 \dfrac{c^e_{R}}{\Lambda^2} - 0.0000260 \dfrac{c_6}{\Lambda^2}
\end{split}
\end{eqnarray}

\item
 $e^+ e^- \rightarrow \nu_e \bar{\nu}_e h$ (240 GeV)
\begin{equation}
\begin{split}
\dfrac{\Delta \sigma}{\sigma_0} =& -0.0132 \dfrac{c_{WW}}{\Lambda^2}- 0.0283 \dfrac{c_{WB}}{\Lambda^2}  +0.349 \dfrac{c_{T}}{\Lambda^2} -0.0606 \dfrac{c_{H}}{\Lambda^2} \\
& - 0.730 \dfrac{c^{(3)l}_L}{\Lambda^2} + 0.577 \dfrac{c^{(3)l}_{LL}}{\Lambda^2} + 0.0136 \dfrac{c^l_{L}}{\Lambda^2} -0.000399 \dfrac{c_6}{\Lambda^2}
\end{split}
\end{equation}

\item
 $e^+ e^- \rightarrow \nu_e \bar{\nu}_e h$ (250 GeV)
\begin{equation}
\begin{split}
\dfrac{\Delta \sigma}{\sigma_0} =& -0.0139 \dfrac{c_{WW}}{\Lambda^2}- 0.0291 \dfrac{c_{WB}}{\Lambda^2}  +0.349 \dfrac{c_{T}}{\Lambda^2} -0.0606 \dfrac{c_{H}}{\Lambda^2} \\
& - 0.738 \dfrac{c^{(3)l}_L}{\Lambda^2} + 0.577 \dfrac{c^{(3)l}_{LL}}{\Lambda^2} + 0.0130 \dfrac{c^l_{L}}{\Lambda^2} -0.000398 \dfrac{c_6}{\Lambda^2}
\end{split}
\end{equation}

\item
 $e^+ e^- \rightarrow \nu_e \bar{\nu}_e h$ (350 GeV)
\begin{equation}
\begin{split}
\dfrac{\Delta \sigma}{\sigma_0} =& -0.0192 \dfrac{c_{WW}}{\Lambda^2}- 0.0340 \dfrac{c_{WB}}{\Lambda^2} +0.349 \dfrac{c_{T}}{\Lambda^2} -0.0606 \dfrac{c_{H}}{\Lambda^2} \\
& - 0.806 \dfrac{c^{(3)l}_L}{\Lambda^2} + 0.577 \dfrac{c^{(3)l}_{LL}}{\Lambda^2} + 0.00878 \dfrac{c^l_{L}}{\Lambda^2} -0.000387 \dfrac{c_6}{\Lambda^2}
\end{split}
\end{equation}

\item
 $e^+ e^- \rightarrow \nu_e \bar{\nu}_e h$ (500 GeV)
\begin{equation}
\begin{split}
\dfrac{\Delta \sigma}{\sigma_0} =&-0.0224 \dfrac{c_{WW}}{\Lambda^2} - 0.0372 \dfrac{c_{WB}}{\Lambda^2} +0.349 \dfrac{c_{T}}{\Lambda^2} -0.0606 \dfrac{c_{H}}{\Lambda^2} \\
& - 0.879 \dfrac{c^{(3)l}_L}{\Lambda^2} + 0.577 \dfrac{c^{(3)l}_{LL}}{\Lambda^2} + 0.00573 \dfrac{c^l_{L}}{\Lambda^2} -0.000380 \dfrac{c_6}{\Lambda^2}
\end{split}
\end{equation}

\item
 $e^+ e^- \rightarrow W^+ W^-$\\
 (1) Unpolarized 240 GeV
\begin{equation}
\begin{split}
\dfrac{\Delta \sigma}{\sigma_0} =& - 0.0287 \dfrac{c_{WB}}{\Lambda^2} +0.170 \dfrac{c_{T}}{\Lambda^2} -
 0.0741 \dfrac{c^{(3)l}_L}{\Lambda^2} + 0.338 \dfrac{c^{(3)l}_{LL}}{\Lambda^2}\\& -0.0282 \dfrac{c^l_{L}}{\Lambda^2} -0.0194 \dfrac{c^e_R}{\Lambda^2} + 0.000696 \dfrac{c_{3W}}{\Lambda^2}
\end{split} \label{WW1}
\end{equation}

 (2) Polarized (-0.8, 0.3) 250 GeV
\begin{equation}
\begin{split}
\dfrac{\Delta \sigma}{\sigma_0} =& - 0.0420 \dfrac{c_{WB}}{\Lambda^2} +0.172 \dfrac{c_{T}}{\Lambda^2} - 0.0740
\dfrac{c^{(3)l}_L}{\Lambda^2} + 0.343 \dfrac{c^{(3)l}_{LL}}{\Lambda^2}\\
& -0.0306 \dfrac{c^l_{L}}{\Lambda^2} -0.00115 \dfrac{c^e_R}{\Lambda^2}+ 0.000816 \dfrac{c_{3W}}{\Lambda^2}
\end{split} \label{WW2}
\end{equation}

 (2) Polarized (-0.8, 0.3) 500 GeV
\begin{equation}
\begin{split}
\dfrac{\Delta \sigma}{\sigma_0} =& - 0.0354 \dfrac{c_{WB}}{\Lambda^2} +0.173 \dfrac{c_{T}}{\Lambda^2} - 0.0364
\dfrac{c^{(3)l}_L}{\Lambda^2} + 0.346 \dfrac{c^{(3)l}_{LL}}{\Lambda^2}\\
& -0.0690 \dfrac{c^l_{L}}{\Lambda^2} -0.000884 \dfrac{c^e_R}{\Lambda^2}+ 0.00119 \dfrac{c_{3W}}{\Lambda^2}
\end{split} \label{WW3}
\end{equation}

\item
 $e^+ e^- \rightarrow Z h h$  (polarized beam (-0.8, 0.3) at 500 GeV)
\begin{equation}
\begin{split}
\dfrac{\Delta \sigma}{\sigma_0} =& 0.912 \dfrac{c_{WW}}{\Lambda^2} + 0.173 \dfrac{c_{WB}}{\Lambda^2} + 0.0339 \dfrac{c_{BB}}{\Lambda^2} - 0.312 \dfrac{c_{T}}{\Lambda^2} - 0.213 \dfrac{c_{H}}{\Lambda^2} +\\
& 1.69 \dfrac{c^{(3)l}_L}{\Lambda^2} + 0.417 \dfrac{c^{(3)l}_{LL}}{\Lambda^2} + 2.13 \dfrac{c^l_{L}}{\Lambda^2} - 1.36 \dfrac{c^e_{R}}{\Lambda^2} - 0.0345 \dfrac{c_6}{\Lambda^2}
\end{split}
\end{equation}

\end{itemize}

\item Angular observables. Here we set the SM value of $\sin \theta^{\text{lep}}_{\text{eff}} = 0.23124$ \cite{Craig:2015wwr}.

\begin{itemize}
\item  Unpolarized beam at 240 GeV
\begin{equation}
\begin{split}
\mathcal{A}_{\theta_1}&=-0.448 + 0.00671 \dfrac{c_{WW}}{\Lambda^2} +0.00180 \dfrac{c_{WB}}{\Lambda^2} +0.000474 \dfrac{c_{BB}}{\Lambda^2}\\
& -0.0000429 \dfrac{c^{(3)l}_L}{\Lambda^2} -0.0000429 \dfrac{c^l_{L}}{\Lambda^2} -0.0000369 \dfrac{c^e_{R}}{\Lambda^2}
\end{split}
\end{equation}

\begin{equation}
\begin{split}
\mathcal{A}_{c\theta_1,c\theta_2}&=0.00755 + 0.00953 \dfrac{c_{WW}}{\Lambda^2} -0.0121 \dfrac{c_{WB}}{\Lambda^2} -0.00285 \dfrac{c_{BB}}{\Lambda^2}\\
& +0.0161 \dfrac{c_{T}}{\Lambda^2}-0.0211 \dfrac{c^{(3)l}_L}{\Lambda^2}+0.0322 \dfrac{c^{(3)l}_{LL}}{\Lambda^2} +0.0112 \dfrac{c^l_{L}}{\Lambda^2} +0.0130 \dfrac{c^e_{R}}{\Lambda^2}
\end{split}
\end{equation}

\begin{equation}
\begin{split}
\mathcal{A}^{(3)}_{\phi}&=0.0136 + 0.0151 \dfrac{c_{WW}}{\Lambda^2} -0.0212 \dfrac{c_{WB}}{\Lambda^2} -0.00462 \dfrac{c_{BB}}{\Lambda^2}\\
& +0.0291 \dfrac{c_{T}}{\Lambda^2}-0.0367 \dfrac{c^{(3)l}_L}{\Lambda^2}+0.0582 \dfrac{c^{(3)l}_{LL}}{\Lambda^2} +0.0215 \dfrac{c^l_{L}}{\Lambda^2} +0.0222 \dfrac{c^e_{R}}{\Lambda^2}
\end{split}
\end{equation}

\begin{equation}
\begin{split}
\mathcal{A}^{(4)}_{\phi}&=0.0959 + 0.00201 \dfrac{c_{WW}}{\Lambda^2} +0.000540 \dfrac{c_{WB}}{\Lambda^2} +0.000142 \dfrac{c_{BB}}{\Lambda^2}\\
& -0.0000218 \dfrac{c^{(3)l}_L}{\Lambda^2} -0.0000218 \dfrac{c^l_{L}}{\Lambda^2} -0.0000188 \dfrac{c^e_{R}}{\Lambda^2}
\end{split}
\end{equation}

\item Polarized beam (-0.8, 0.3) at 250 GeV

\begin{equation}
\begin{split}
\mathcal{A}_{\theta_1}&=-0.462 + 0.0148 \dfrac{c_{WW}}{\Lambda^2} +0.000600 \dfrac{c_{WB}}{\Lambda^2} +0.00101 \dfrac{c_{BB}}{\Lambda^2}\\
& -0.0000708 \dfrac{c^{(3)l}_L}{\Lambda^2} -0.0000708 \dfrac{c^l_{L}}{\Lambda^2} -0.00000364 \dfrac{c^e_{R}}{\Lambda^2}
\end{split}
\end{equation}

\begin{equation}
\begin{split}
\mathcal{A}_{c\theta_1,c\theta_2}&=0.0443 + 0.00384 \dfrac{c_{WW}}{\Lambda^2} -0.0272 \dfrac{c_{WB}}{\Lambda^2} -0.000662 \dfrac{c_{BB}}{\Lambda^2}\\
& +0.0486 \dfrac{c_{T}}{\Lambda^2}-0.0637 \dfrac{c^{(3)l}_L}{\Lambda^2}+0.0973 \dfrac{c^{(3)l}_{LL}}{\Lambda^2} +0.0336 \dfrac{c^l_{L}}{\Lambda^2} +0.0391 \dfrac{c^e_{R}}{\Lambda^2}
\end{split}
\end{equation}

\begin{equation}
\begin{split}
\mathcal{A}^{(3)}_{\phi}&=0.0843 + 0.00206 \dfrac{c_{WW}}{\Lambda^2} -0.0518 \dfrac{c_{WB}}{\Lambda^2} -0.000803 \dfrac{c_{BB}}{\Lambda^2}\\
& +0.0925 \dfrac{c_{T}}{\Lambda^2}-0.119 \dfrac{c^{(3)l}_L}{\Lambda^2}+0.185 \dfrac{c^{(3)l}_{LL}}{\Lambda^2} +0.0663 \dfrac{c^l_{L}}{\Lambda^2} +0.0742 \dfrac{c^e_{R}}{\Lambda^2}
\end{split}
\end{equation}

\begin{equation}
\begin{split}
\mathcal{A}^{(4)}_{\phi}&=0.0919 + 0.00444 \dfrac{c_{WW}}{\Lambda^2} +0.000179 \dfrac{c_{WB}}{\Lambda^2} - 0.000304 \dfrac{c_{BB}}{\Lambda^2}\\
& -0.0000372 \dfrac{c^{(3)l}_L}{\Lambda^2} -0.0000372 \dfrac{c^l_{L}}{\Lambda^2} -0.00000192 \dfrac{c^e_{R}}{\Lambda^2}
\end{split}
\end{equation}

\end{itemize}

\end{enumerate}

\section{Normalized Correlation Matrices}
\label{sec:NCM}

The normalized correlation matrix for the 6D operators is defined as
\begin{eqnarray}
M_{ij} = \frac{\partial^2 \chi^2 }{ \partial c_i  \partial c_j }  \Big /  \sqrt {\frac{\partial^2 \chi^2 }{ (\partial c_i)^2 } \frac{\partial^2 \chi^2 }{ (\partial c_j)^2 } } \ .
\end{eqnarray}
Here $c_i$ and $c_j$ run over all Wilson coefficients. Obviously the correlation matrix is symmetric.

\begin{table}[H]
  \centering
  \resizebox{\textwidth}{!}{
    \begin{tabular}{c|ccccccccccc}
          & $c_{WW}$ & $c_{WB}$ & $c_{BB}$ & $c_{T}$ & $c_{H}$ & $c_{LL}^{(3)l}$ & $c_{L}^{(3)l}$ & $c_{L}^{l}$ & $c_{R}^{e}$ & $c_{6}$ & $c_{3W}$ \\
    \hline
   $c_{WW}$ & 1     & 0.0172 & 0.955 & -0.00485 & -0.981 & 0.0139 & 0.0562 & 0.22  & -0.186 & -0.995 & 0 \\
    $c_{WB}$ &       & 1     & 0.0179 & -0.883 & -0.0169 & -0.863 & 0.801 & -0.457 & -0.605 & -0.0173 & -0.459 \\
    $c_{BB}$ &       &       & 1     & -0.00538 & -0.956 & 0.0128 & 0.0554 & 0.213 & -0.182 & -0.968 & 0 \\
    $c_{T}$ &       &       &       & 1     & 0.00383 & 0.97  & -0.877 & 0.0977 & 0.216 & 0.00453 & 0.617 \\
    $c_{H}$ &       &       &       &       & 1     & -0.0145 & -0.0536 & -0.217 & 0.183 & 0.994 & 0 \\
    $c_{LL}^{(3)l}$ &       &       &       &       &       & 1     & -0.879 & 0.108 & 0.225 & -0.0141 & 0.654 \\
    $c_{L}^{(3)l}$ &       &       &       &       &       &       & 1     & -0.133 & -0.389 & -0.0555 & -0.25 \\
    $c_{L}^{l}$ &       &       &       &       &       &       &       & 1     & 0.559 & -0.22 & -0.252 \\
    $c_{R}^{e}$ &       &       &       &       &       &       &       &       & 1     & 0.186 & -0.16 \\
    $c_{6}$ &       &       &       &       &       &       &       &       &       & 1     & 0 \\
    $c_{3W}$ &       &       &       &       &       &       &       &       &       &       & 1 \\
    \end{tabular}%
}   \caption{Normalized correlation matrix in the $\chi^2$ fit at CEPC. } \label{tab:CEPC_cor}%
\end{table}%

\begin{table}[H]
  \centering
   \resizebox{\textwidth}{!}{
   \begin{tabular}{c|ccccccccccc}
          & $c_{WW}$ & $c_{WB}$ & $c_{BB}$ & $c_{T}$ & $c_{H}$ & $c_{LL}^{(3)l}$ & $c_{L}^{(3)l}$ & $c_{L}^{l}$ & $c_{R}^{e}$ & $c_{6}$ & $c_{3W}$ \\
           \hline
    $c_{WW}$ & 1     & 0.0133 & 0.954 & -0.00391 & -0.866 & 0.0114 & 0.0278 & 0.0598 & -0.0587 & -0.971 & 0 \\
    $c_{WB}$ &       & 1     & 0.0136 & -0.911 & -0.00988 & -0.825 & 0.687 & 0.161 & -0.402 & -0.0123 & -0.357 \\
    $c_{BB}$ &       &       & 1     & -0.00378 & -0.863 & 0.011 & 0.0267 & 0.0579 & -0.0576 & -0.954 & 0 \\
    $c_{T}$ &       &       &       & 1     & -0.00264 & 0.894 & -0.645 & -0.126 & 0.172 & 0.00108 & 0.443 \\
    $c_{H}$ &       &       &       &       & 1     & -0.0167 & -0.0159 & -0.0534 & 0.0524 & 0.959 & 0 \\
    $c_{LL}^{(3)l}$ &       &       &       &       &       & 1     & -0.683 & -0.162 & 0.222 & -0.0142 & 0.574 \\
    $c_{L}^{(3)l}$ &       &       &       &       &       &       & 1     & 0.752 & -0.8  & -0.0233 & -0.123 \\
    $c_{L}^{l}$ &       &       &       &       &       &       &       & 1     & -0.852 & -0.0589 & -0.0702 \\
    $c_{R}^{e}$ &       &       &       &       &       &       &       &       & 1     & 0.0579 & -0.0519 \\
    $c_{6}$ &       &       &       &       &       &       &       &       &       & 1     & 0 \\
    $c_{3W}$ &       &       &       &       &       &       &       &       &       &       & 1 \\
    \end{tabular}%
}  \caption{Normalized correlation matrix in the $\chi^2$ fit at FCC-ee. }  \label{tab:FCC_cori}%
\end{table}%

\begin{table}[H]
  \centering
   \resizebox{\textwidth}{!}{
\begin{tabular}{c|ccccccccccc}
          & $c_{WW}$ & $c_{WB}$ & $c_{BB}$ & $c_{T}$ & $c_{H}$ & $c_{LL}^{(3)l}$ & $c_{L}^{(3)l}$ & $c_{L}^{l}$ & $c_{R}^{e}$ & $c_{6}$ & $c_{3W}$ \\
           \hline    
    $c_{WW}$ & 1     & -0.00989 & -0.981 & 0.00315 & -0.876 & 0.0142 & 0.0775 & 0.314 & -0.0244 & -0.968 & 0 \\
    $c_{WB}$ &       & 1     & 0.0112 & -0.992 & 0.00961 & -0.868 & 0.949 & 0.334 & -0.126 & 0.00988 & -0.659 \\
    $c_{BB}$ &       &       & 1     & -0.00384 & 0.882 & -0.0151 & -0.0755 & -0.315 & 0.0209 & 0.969 & 0 \\
    $c_{T}$ &       &       &       & 1     & -0.00604 & 0.88  & -0.937 & -0.383 & 0.0416 & -0.00456 & 0.688 \\
    $c_{H}$ &       &       &       &       & 1     & -0.0164 & -0.0572 & -0.281 & 0.0228 & 0.966 & 0 \\
    $c_{LL}^{(3)l}$ &       &       &       &       &       & 1     & -0.763 & -0.519 & 0.0568 & -0.0156 & 0.94 \\
    $c_{L}^{(3)l}$ &       &       &       &       &       &       & 1     & 0.404 & -0.301 & -0.0704 & -0.515 \\
    $c_{L}^{l}$ &       &       &       &       &       &       &       & 1     & -0.307 & -0.308 & -0.589 \\
    $c_{R}^{e}$ &       &       &       &       &       &       &       &       & 1     & 0.0251 & -0.0345 \\
    $c_{6}$ &       &       &       &       &       &       &       &       &       & 1     & 0 \\
    $c_{3W}$ &       &       &       &       &       &       &       &       &       &       & 1 \\
    \end{tabular}%
} \caption{Normalized correlation matrix in the $\chi^2$ fit at ILC250. }   \label{tab:ILC250_cor}%
\end{table}%

\begin{table}[H]
  \centering
   \resizebox{\textwidth}{!}{
\begin{tabular}{c|ccccccccccc}
          & $c_{WW}$ & $c_{WB}$ & $c_{BB}$ & $c_{T}$ & $c_{H}$ & $c_{LL}^{(3)l}$ & $c_{L}^{(3)l}$ & $c_{L}^{l}$ & $c_{R}^{e}$ & $c_{6}$ & $c_{3W}$ \\
           \hline
    $c_{WW}$ & 1     & -0.0084 & -0.986 & 0.00193 & -0.565 & 0.00988 & 0.179 & 0.34  & -0.0494 & -0.452 & 0 \\
    $c_{WB}$ &       & 1     & 0.00944 & -0.99 & 0.0111 & -0.886 & 0.902 & 0.542 & -0.0943 & 0.00307 & -0.732 \\
    $c_{BB}$ &       &       & 1     & -0.0032 & 0.601 & -0.0111 & -0.18 & -0.35 & 0.0485 & 0.443 & 0 \\
    $c_{T}$ &       &       &       & 1     & -0.0147 & 0.919 & -0.866 & -0.619 & 0.0174 & -0.0033 & 0.796 \\
    $c_{H}$ &       &       &       &       & 1     & -0.0197 & -0.0273 & -0.186 & 0.0274 & 0.441 & 0 \\
    $c_{LL}^{(3)l}$ &       &       &       &       &       & 1     & -0.69 & -0.74 & 0.0206 & -0.0056 & 0.955 \\
    $c_{L}^{(3)l}$ &       &       &       &       &       &       & 1     & 0.466 & -0.287 & -0.0281 & -0.477 \\
    $c_{L}^{l}$ &       &       &       &       &       &       &       & 1     & -0.164 & -0.0812 & -0.818 \\
    $c_{R}^{e}$ &       &       &       &       &       &       &       &       & 1     & 0.012 & -0.042 \\
    $c_{6}$ &       &       &       &       &       &       &       &       &       & 1     & 0 \\
    $c_{3W}$ &       &       &       &       &       &       &       &       &       &       & 1 \\
    \end{tabular}%
} \caption{Normalized correlation matrix in the $\chi^2$ fit at ILC500. }   \label{tab:ILC500_cori}%
\end{table}%

\begin{table}[H]
  \centering
   \resizebox{\textwidth}{!}{
\begin{tabular}{c|ccccccccccc}
          & $c_{WW}$ & $c_{WB}$ & $c_{BB}$ & $c_{T}$ & $c_{H}$ & $c_{LL}^{(3)l}$ & $c_{L}^{(3)l}$ & $c_{L}^{l}$ & $c_{R}^{e}$ & $c_{6}$ & $c_{3W}$ \\
           \hline
    $c_{WW}$ & 1     & -0.00591 & -0.986 & 0.00176 & -0.565 & 0.00868 & 0.114 & 0.241 & -0.0147 & -0.452 & 0 \\
    $c_{WB}$ &       & 1     & 0.00665 & -0.923 & 0.00781 & -0.882 & 0.938 & -0.228 & -0.693 & 0.00216 & -0.516 \\
    $c_{BB}$ &       &       & 1     & -0.0029 & 0.601 & -0.0097 & -0.114 & -0.248 & 0.0144 & 0.443 & 0 \\
    $c_{T}$ &       &       &       & 1     & -0.0134 & 0.933 & -0.819 & -0.121 & 0.381 & -0.003 & 0.725 \\
    $c_{H}$ &       &       &       &       & 1     & -0.0173 & -0.0173 & -0.132 & 0.00815 & 0.441 & 0 \\
    $c_{LL}^{(3)l}$ &       &       &       &       &       & 1     & -0.753 & -0.138 & 0.441 & -0.0049 & 0.839 \\
    $c_{L}^{(3)l}$ &       &       &       &       &       &       & 1     & -0.305 & -0.751 & -0.0178 & -0.303 \\
    $c_{L}^{l}$ &       &       &       &       &       &       &       & 1     & 0.64  & -0.0574 & -0.579 \\
    $c_{R}^{e}$ &       &       &       &       &       &       &       &       & 1     & 0.00357 & -0.0125 \\
    $c_{6}$ &       &       &       &       &       &       &       &       &       & 1     & 0 \\
    $c_{3W}$ &       &       &       &       &       &       &       &       &       &       & 1 \\
    \end{tabular}%
} \caption{Normalized correlation matrix in the $\chi^2$ fit at ILC. }   \label{tab:ILC_cor}%
\end{table}%

\section{Parameter Marginalization in $\chi^2$}
\label{app:PM}

The introduction on parameter marginalization in $\chi^2$ can be found in various lecture notes (see, e.g.,~\cite{PM}). Below we will simply introduce this method.

Let's consider $n$ independent observables $O_1,O_2,\cdots,O_n$, with their measured values satisfying Gaussian distribution. These observables are all  linearly dependent on $m$ parameters or Wilson coefficients $C^T= \{c_1,c_2,\cdots,c_m\}$, namely, $O_i = O_i(c_1,c_2,\cdots,c_m)$ with $i =1, \cdots, n$ and $m \leq n$. Then their  probability distribution function (PDF) is given by
\begin{equation}
f(\delta_1,\delta_2,\cdots,\delta_n) = \dfrac{1}{(2 \pi)^{\frac{n}{2}}} \exp \Big(-\dfrac{1}{2}\sum_i^n \delta_i^2 \Big)
\end{equation}
Here $\delta_i$ represents a normalized deviation from the measured value. $f(\delta_1,\delta_2,\cdots,\delta_n)$ can be converted to the PDF of  the $m$ parameters
\begin{equation}
 g(c_1,c_2,\cdots,c_m) \propto \exp \Big(-\dfrac{1}{2}\sum_i^n \delta_i(c_j)^2 \Big) = \exp (\chi^2)
\end{equation}
with
\begin{equation}
\int dc_1 dc_2 \cdots dc_m \; g(c_1,c_2,\cdots,c_m) = 1
\end{equation}
Here
\begin{eqnarray}
\chi^2 = C^T M C
\end{eqnarray}
is a quadratic function of $C^T=\{c_1,c_2,\cdots,c_m\}$. $M$ is the correlation matrix of the $m$ parameters which is symmetric.

The marginalized distribution for a single parameter, say, $c_{m}$, is defined as
\begin{equation}
g_M(c_m) = \int d c_1 d c_2 \cdots d c_{m-1} g(c_1,c_2,\cdots,c_m)
\end{equation}
Separating $c_m$ from the other $m-1$ paramters we have
\begin{equation}
\chi^2 = y c_m^2 + c_{m} C_X^T Z + c_{m} Z^T C_X + C_X^T X C_X
\end{equation}
with $C_X=(c_1\; c_2\cdots c_{m-1})^T$. $X$, $Z$ and $y$ are the entries of the correlation matrix
\begin{equation}
M= \begin{bmatrix}
X & Z\\
Z^T& y
\end{bmatrix}  \ .
\end{equation}
The $(m-1) \times (m-1)$ matrix $X$ can be diagonalized by taking a unitary transformation $C_X \to C'_X$. In the new parameter basis, we have
\begin{eqnarray}
\chi^2 &=& y c_m^2 + c_m C_X^{'T} Z'+ c_m Z^{'T} C'_X + {C'}_X^T  X' C'_X  \nonumber \\
 &= &y c_m^2 + 2 c_m \sum_{i=1}^{m-1} c'_i z'_i + \sum_{i=1}^{m-1} x'_i c_i^{'2} \nonumber \\
&=& \Big(y -\sum_{i=1}^{m-1} \dfrac{z_i^{'2}}{x'_i} \Big) c_m^2 + \sum_{i=1}^{m-1}   x'_i \Big( c'_i + \dfrac{c_m z'_i}{x'_i} \Big)^2   \label{chi2ex} 
\end{eqnarray}
Here we define $C'_X=(c'_1\; c'_2\cdots c'_{m-1})^T$, $X' = \text{diag}(x'_1 \; x'_2\cdots x'_{m-1})$ and $Z'=(z'_1\; z'_2 \cdots z'_{m-1})^T$.
Integrating out $C'_X$, we obtain
\begin{equation}
\Delta \chi^2 (c_m)  = \Big(y - \sum_{i=1}^{m-1} \dfrac{z_i^{'2}}{x'_i}\Big) c_m^2 \ . \label{dchi}
\end{equation}
It defines the marginalized PDF of $c_m$ as
\begin{eqnarray}
g_M(c_m) \propto \exp(\Delta \chi^2(c_m))  \ .
\end{eqnarray}

Taking a further step, let's define the correlation matrix in the new parameter basis $({C'}_X^T, c_m)$ as
\begin{equation}
\begin{split}
M' &=
\begin{bmatrix}
X' & Z'\\
{Z'}^T& y
\end{bmatrix}
= \begin{bmatrix}
x'_1&  &  & z'_1\\
   &\ddots& & \vdots\\
   &  & x'_{m-1}& z'_{m-1}\\
z'_1& \cdots & z'_{m-1}&y
\end{bmatrix}
\end{split}   \ .
\end{equation}
The determinant of the correlation matrix $M$ can be calculated as
\begin{equation}
\det M= \det M'= (y-\sum_{i=1}^{m-1} \dfrac{z_i^{'2}}{x'_i} ) \prod_{i=1}^{m-1} x'_i   \ .
\end{equation}
With this relation, we immediately obtain
\begin{equation}
\Delta \chi^2 = c_m^2 \dfrac{\det M}{\det X}  \ .
\end{equation}
This relation indicates that, given a confidence level for the $\chi^2$ analysis, the constraints for $c_m$ is completely determined by the correlation matrix $M$. There is a geometric interpretation regrading this. Eq.(\ref{chi2ex}) defines a $(m-1)$-dimensional ellipsoid in a $m$-dimensional space which is expanded by $C^T=(C_X^T, c_m)$. Integrating out $C'_X$ is equivalent to imposing the conditions $c'_i + \frac{c_m z'_i}{ x'_i} = 0$ or equivalently, the conditions $\frac{\partial \chi^2}{\partial c'_i} = 0$, for $i=1, \cdots, m-1$. Therefore, the marginalization of $C_X$ is simply a projection of the ellipsoid to the $c_m$ axis in the $m$-dimensional space.

The discussions above can be generalized to the case with multiple variables. In this case, $\chi^2$ is defined as
\begin{equation}
\chi^2= (C_X^T \; C_Y^T) \begin{bmatrix}
X & Z\\
Z^T& Y
\end{bmatrix}\begin{bmatrix}
C_X\\
C_Y
\end{bmatrix}
\end{equation}
with $C_X = {c_1, \cdots, c_k}$ representing the $k$ parameters to marginalize. Here  $X$ and $Y$ are $k \times k$ and $(m-k) \times (m-k)$ matrices, respectively. With this setup, we have
\begin{equation}
\begin{split}
\chi^2 &= C_X^T X C_X + C_X^T Z C_Y+ C_Y^T Z^T C_X + C_Y^T Y C_Y\\
&= (C_X + X^{-1} Z C_Y)^T X (C_X + X^{-1} Z C_Y) - C_Y^T Z^T X^{-1} Z C_Y + C_Y^T Y C_Y\\
&=(C_X + X^{-1} Z C_Y)^T X (C_X + X^{-1} Z C_Y) + C_Y^T (Y - Z^T X^{-1}Z) C_Y
\end{split}  \label{chi2}
\end{equation}
$C_X$ is marginalized by integrating out the first term, yielding
\begin{eqnarray}
\Delta \chi^2 =  C_Y^T (Y - Z^T X^{-1}Z) C_Y  \label{dchi2}
\end{eqnarray}
Eq.(\ref{dchi2}) describes the correlation among the parameters in $C_Y$. At a given C.L., the value of $\Delta \chi^2$ depends on the number of parameters in $C_Y$. If $C_Y$ contains one parameter only, say, $c_m$, Eq.(\ref{dchi2}) is reduced to Eq.(\ref{dchi}), with $\Delta \chi^2 = 1$ at $1\sigma$ C.L..  Again marginalizing $C_X$ is equivalent to imposing the conditions $\partial \chi^2 / \partial c_{i}$ to Eq(\ref{chi2}), with $i$ running from $1$ to $k$. It can be geometrically interpreted as a projection of a $(m-1)$-dimensional ellipsoid in a $m$-dimensional space to its $(m-k)$-dimensional subspace, which are expanded by $(C_X, C_Y)$ and $C_Y$, respectively.

\begin{figure}[H]
\centering
\includegraphics[width=6cm]{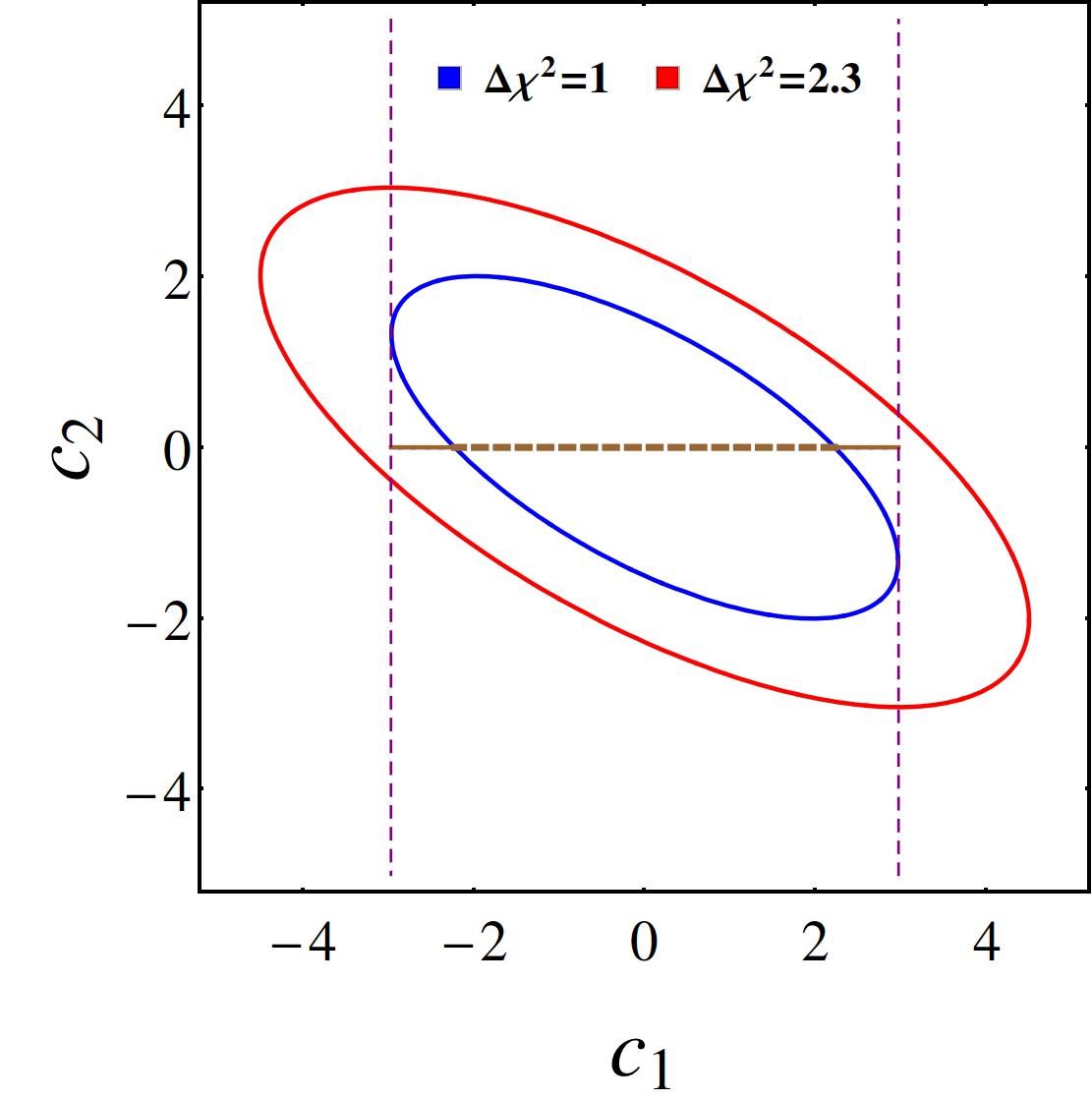}
\caption{Parameter marginalization in a two-parameter model.} \label{toy}
\end{figure}

The geometric interpretation of parameter marginalization in a simple model is presented in Figure~\ref{toy}. In this example there are two free parameters, say, $c_1$ and $c_2$, with $c_2$ being marginalized. Eq.(\ref{chi2ex}) defines an one-dimensional ellipsoid or ellipse at the $c_1 - c_2$ plane.  Integrating out $c'_2$ is equivalent to imposing a condition $\frac{\partial \chi^2}{\partial c'_2} = \frac{\partial \chi^2}{\partial c_2}  = 0$.  The $c_2$ marginalization is simply a projection of the ellipse to the $c_1$ axis. 
Here the size of the ellipse is determined by the $\Delta \chi^2$ value, which is equal to one at $1\sigma$ C.L.. In Figure~\ref{toy}, the allowed range for $c_1$ with a marginalized $c_2$ is indicated by the brown line ending at the purple lines. As a comparison, if $c_2$ is turned off, the constraint for $c_1$ becomes stronger, which is denoted by the brown dashed line ending at the blue ellipse. 

\section{2D $\chi^2$ Analysis}
\label{app:2D}

\begin{figure}[H]
\begin{center}
\includegraphics[width=15cm]{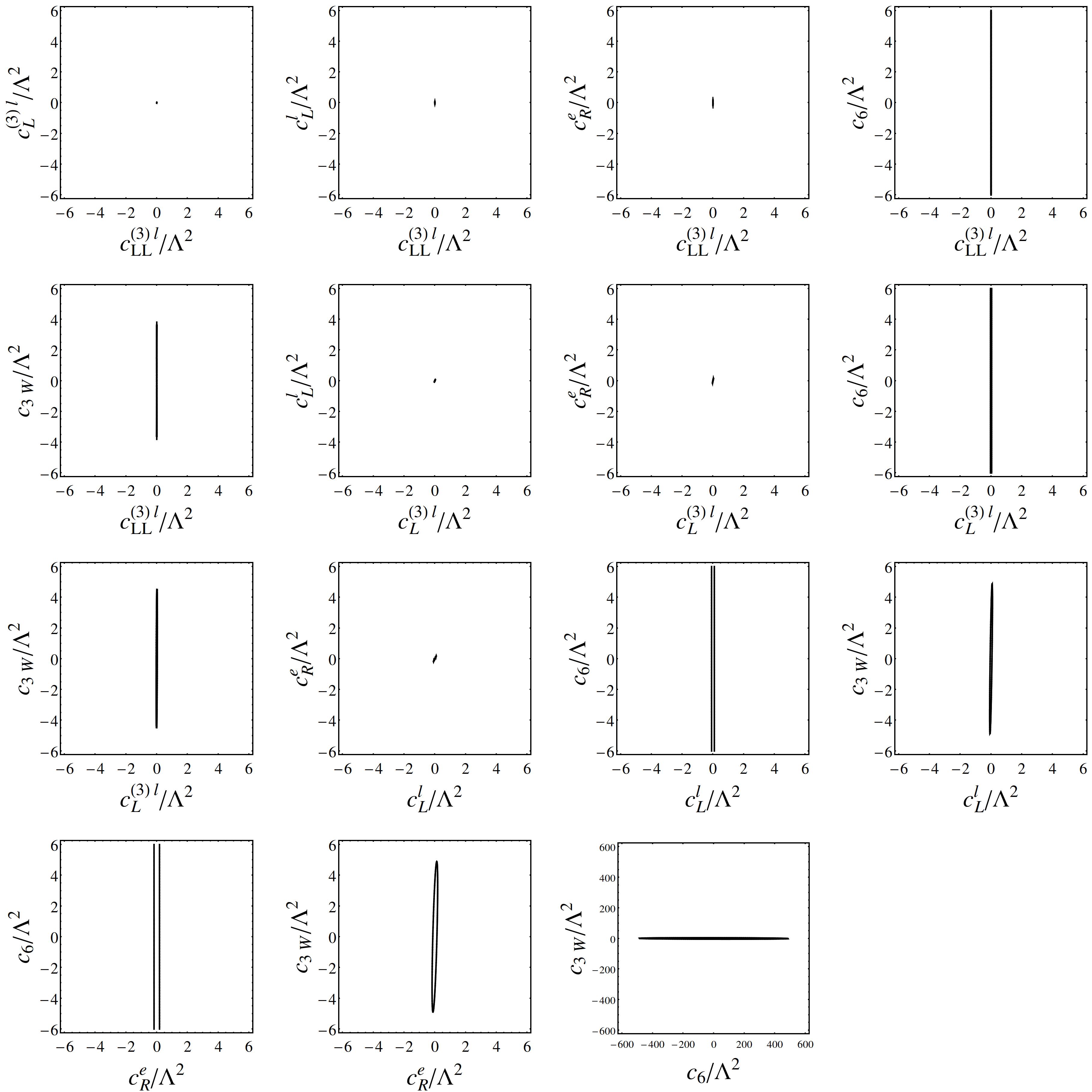}
\caption{2D sensitivity projection in the marginalized $\chi^2$ analysis. The coordinate axes are in the unit of (TeV)$^{-2}$. }  \label{fig_array1}
\end{center}
\end{figure}

\begin{figure} [H]
\begin{center}
\includegraphics[width=15cm]{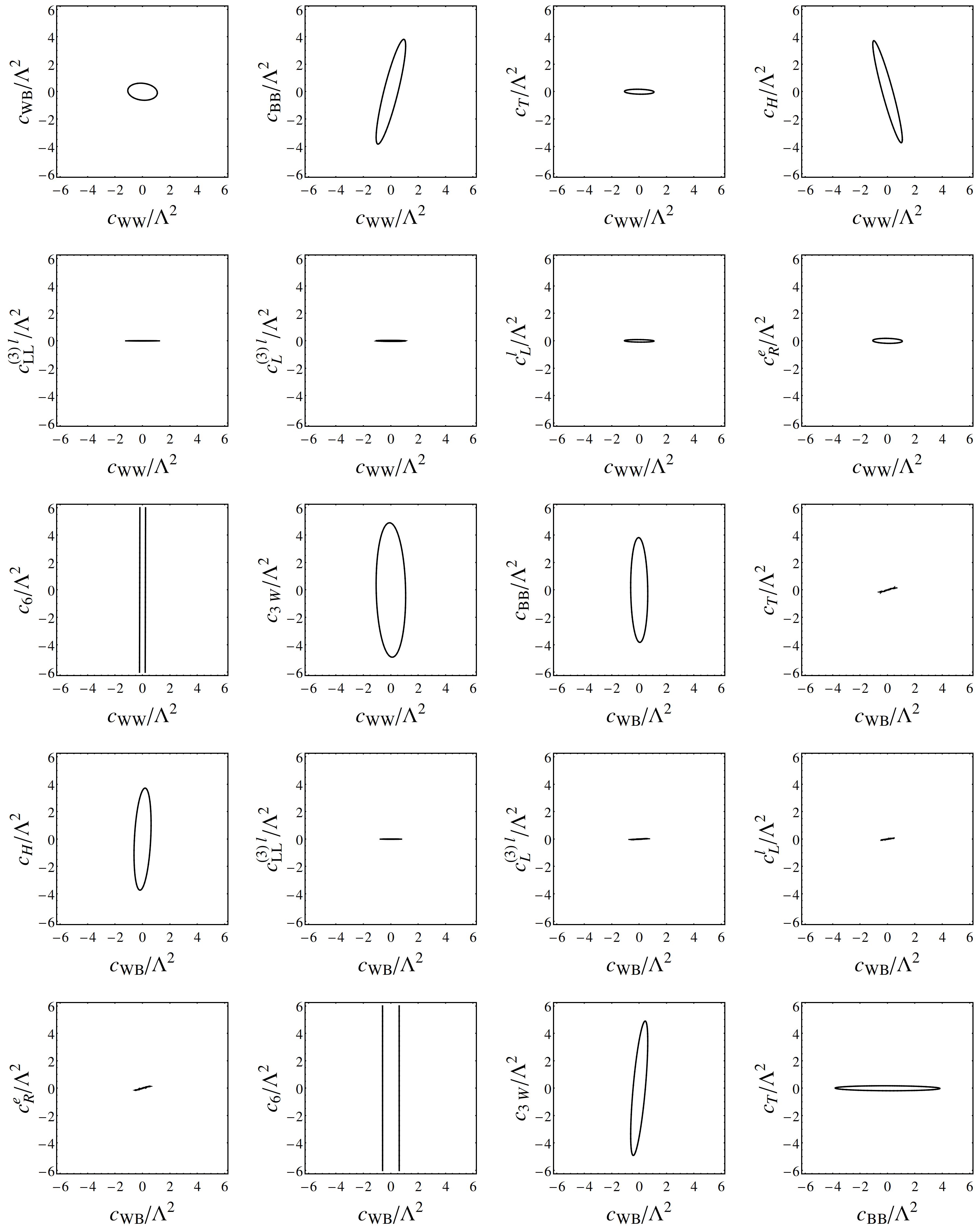}
\caption{2D sensitivity projection in the marginalized $\chi^2$ analysis.  The coordinate axes are in the unit of (TeV)$^{-2}$. }  \label{fig_array2}
\end{center}
\end{figure}

\begin{figure}[H]
\begin{center}
\includegraphics[width=15cm]{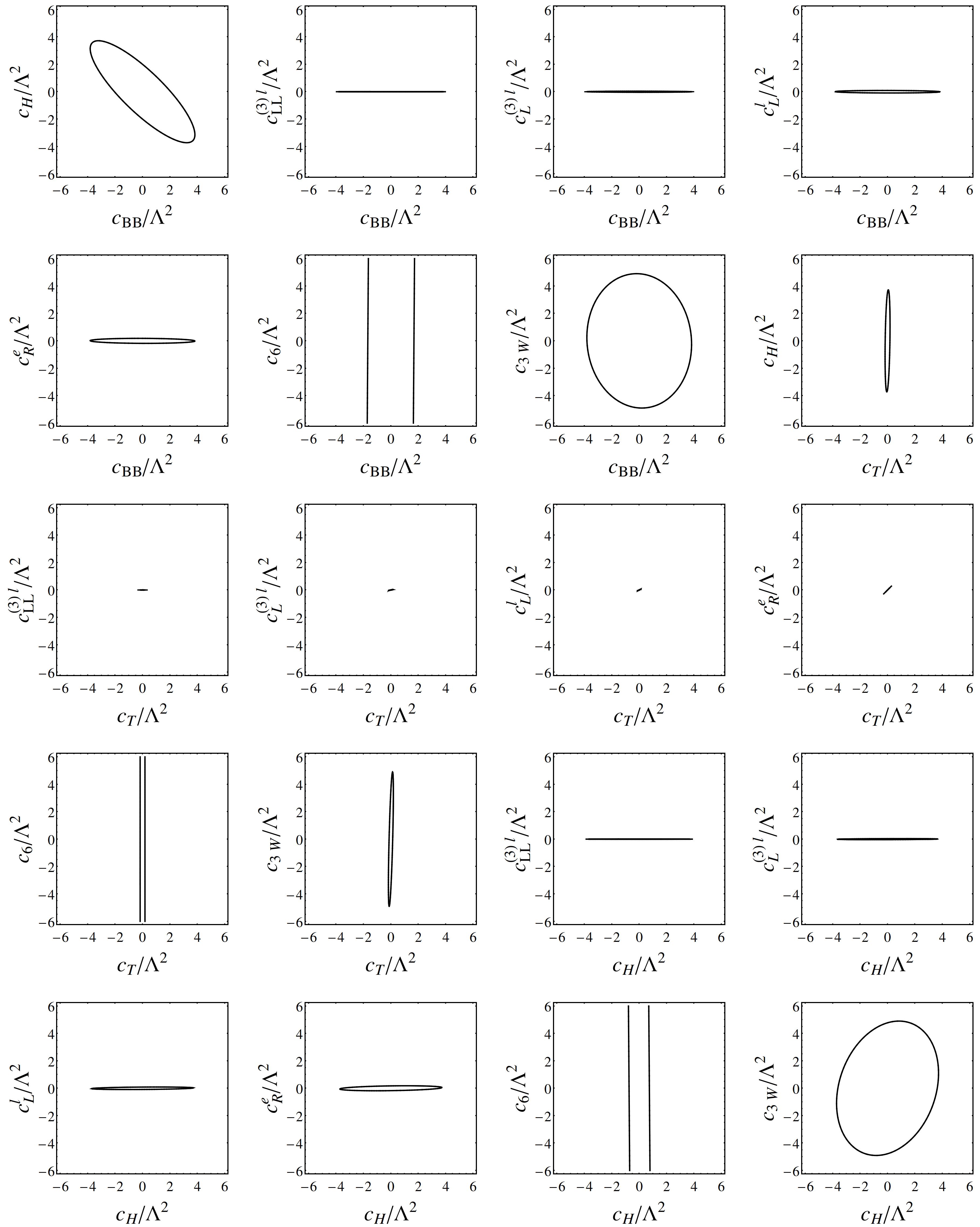}
\caption{2D sensitivity projection in the marginalized $\chi^2$ analysis.  The coordinate axes are in the unit of (TeV)$^{-2}$. }  \label{fig_array3}
\end{center}
\end{figure}

\begin{figure}[H]
  \centering
  \includegraphics[width=5.cm]{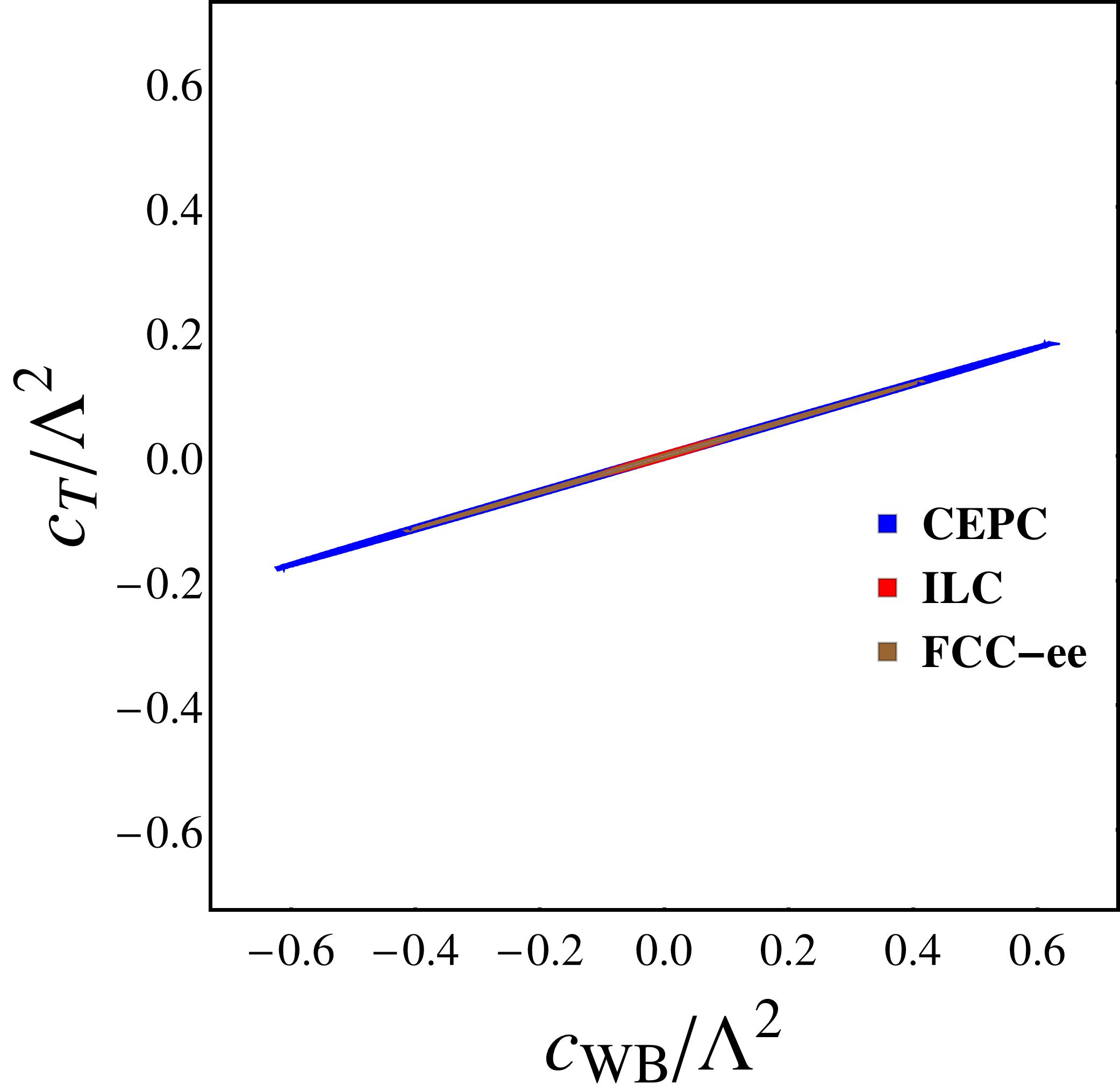}
     \includegraphics[width=5.5cm]{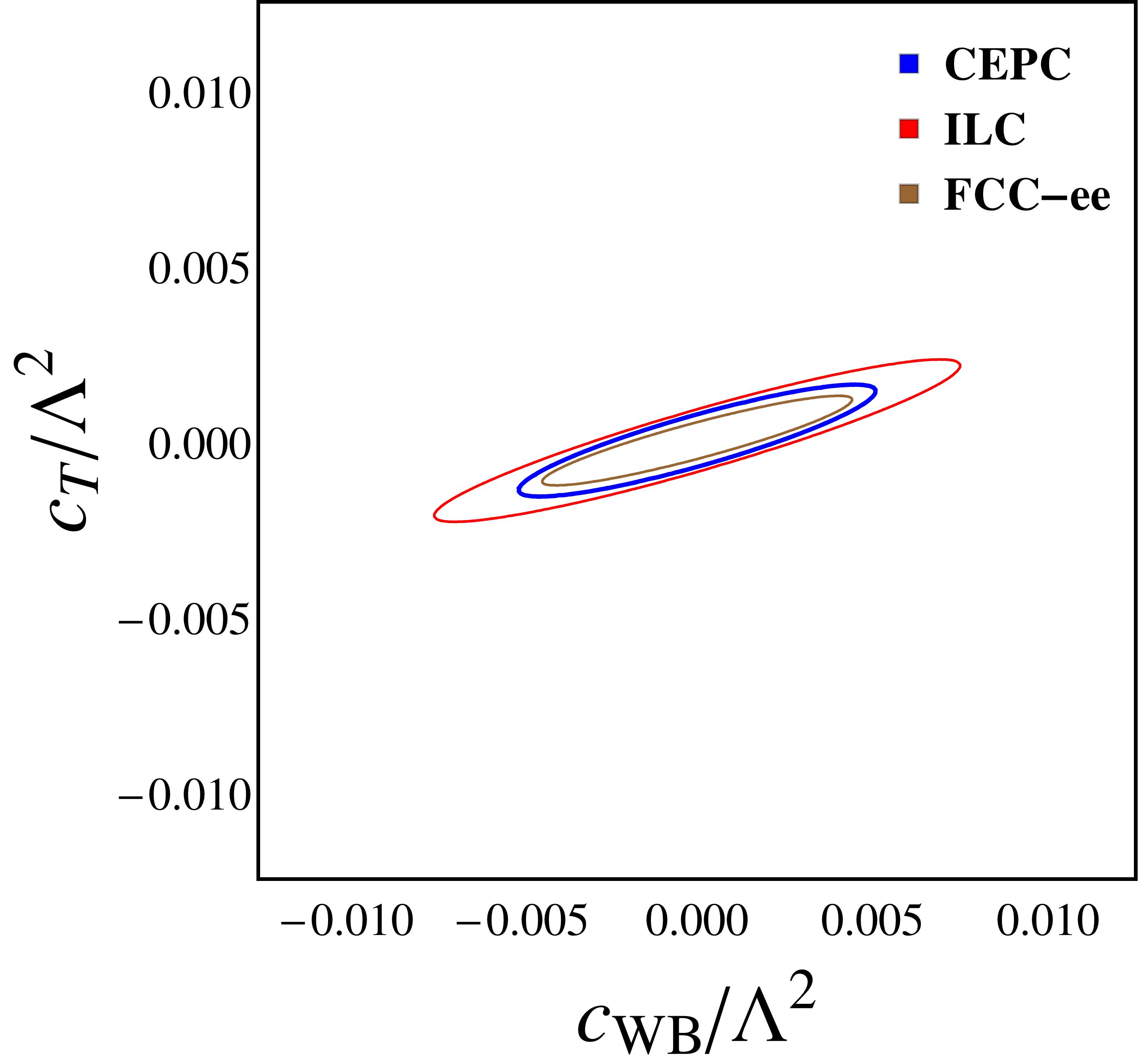}
  \includegraphics[width=5.cm]{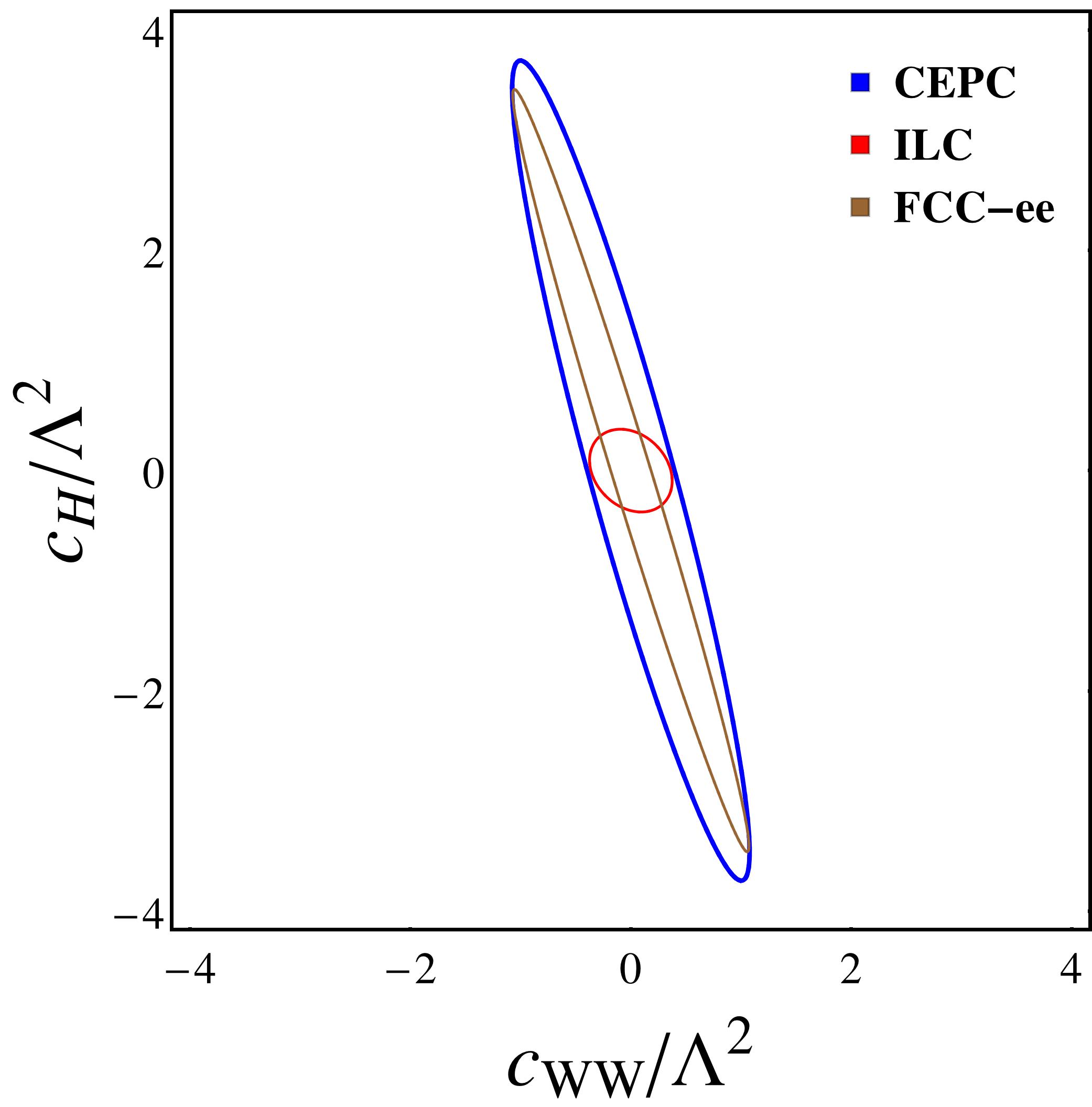}
    \includegraphics[width=5.cm]{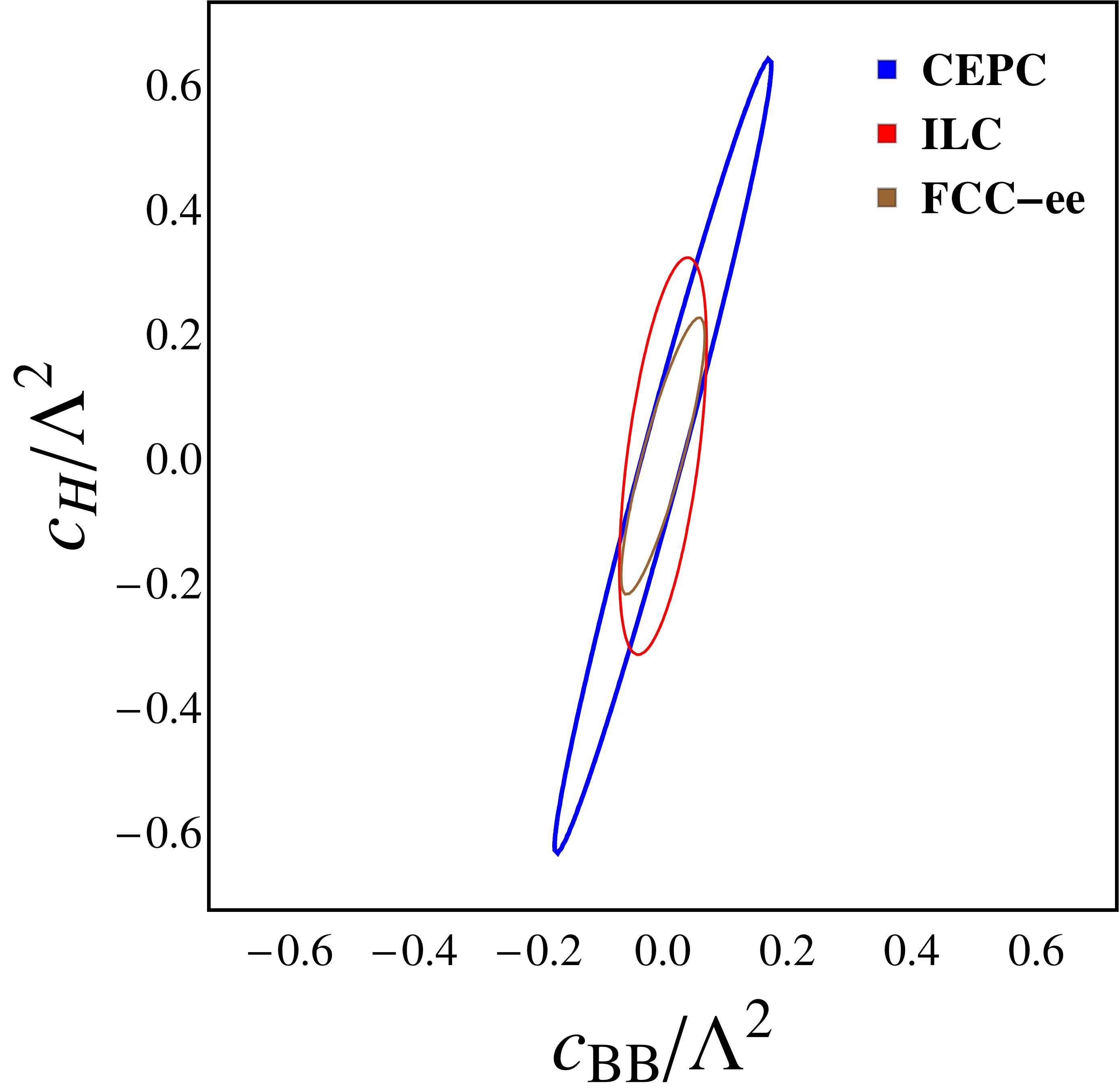}
  \includegraphics[width=5.cm]{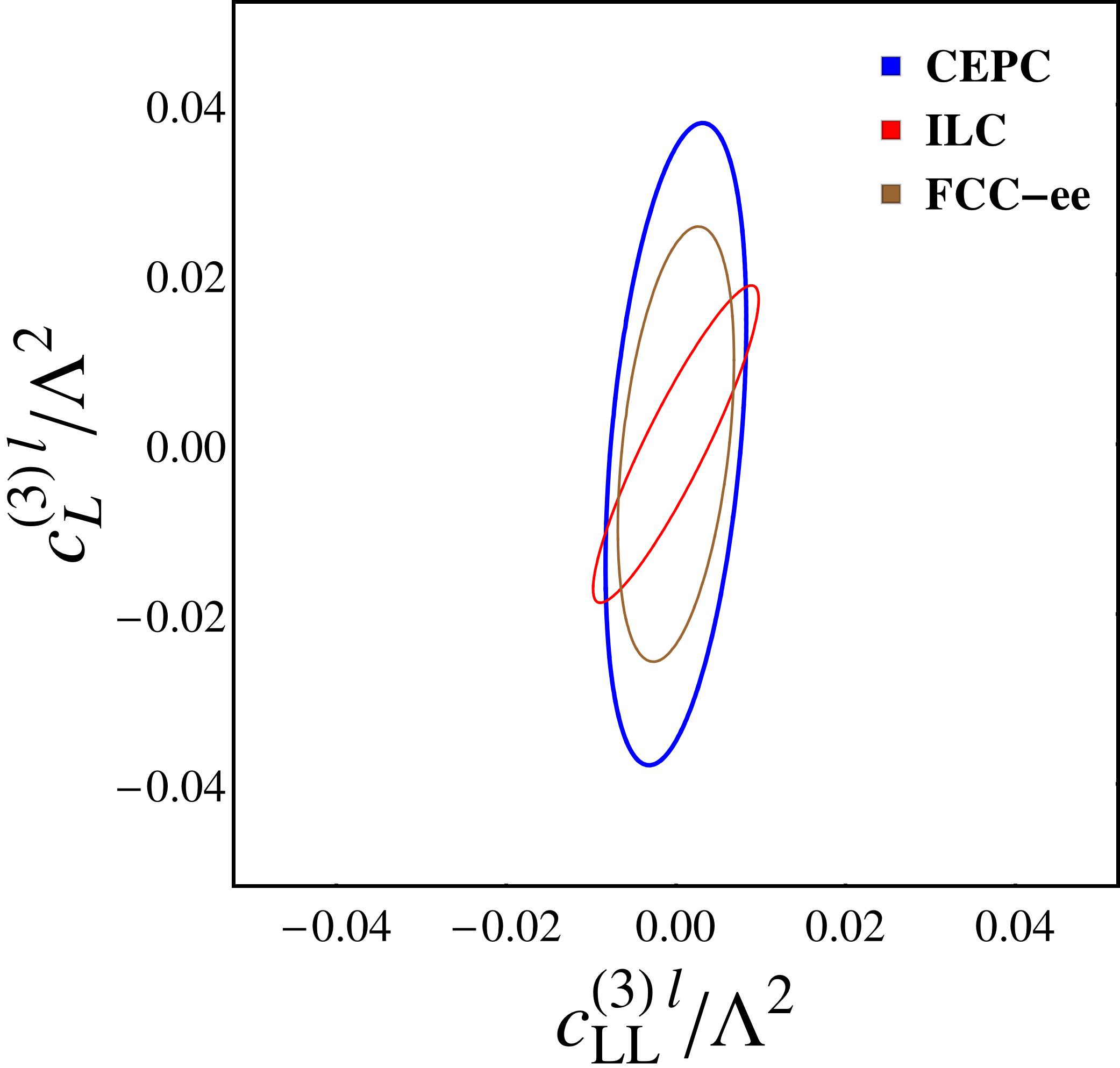}
   \includegraphics[width=5.2cm]{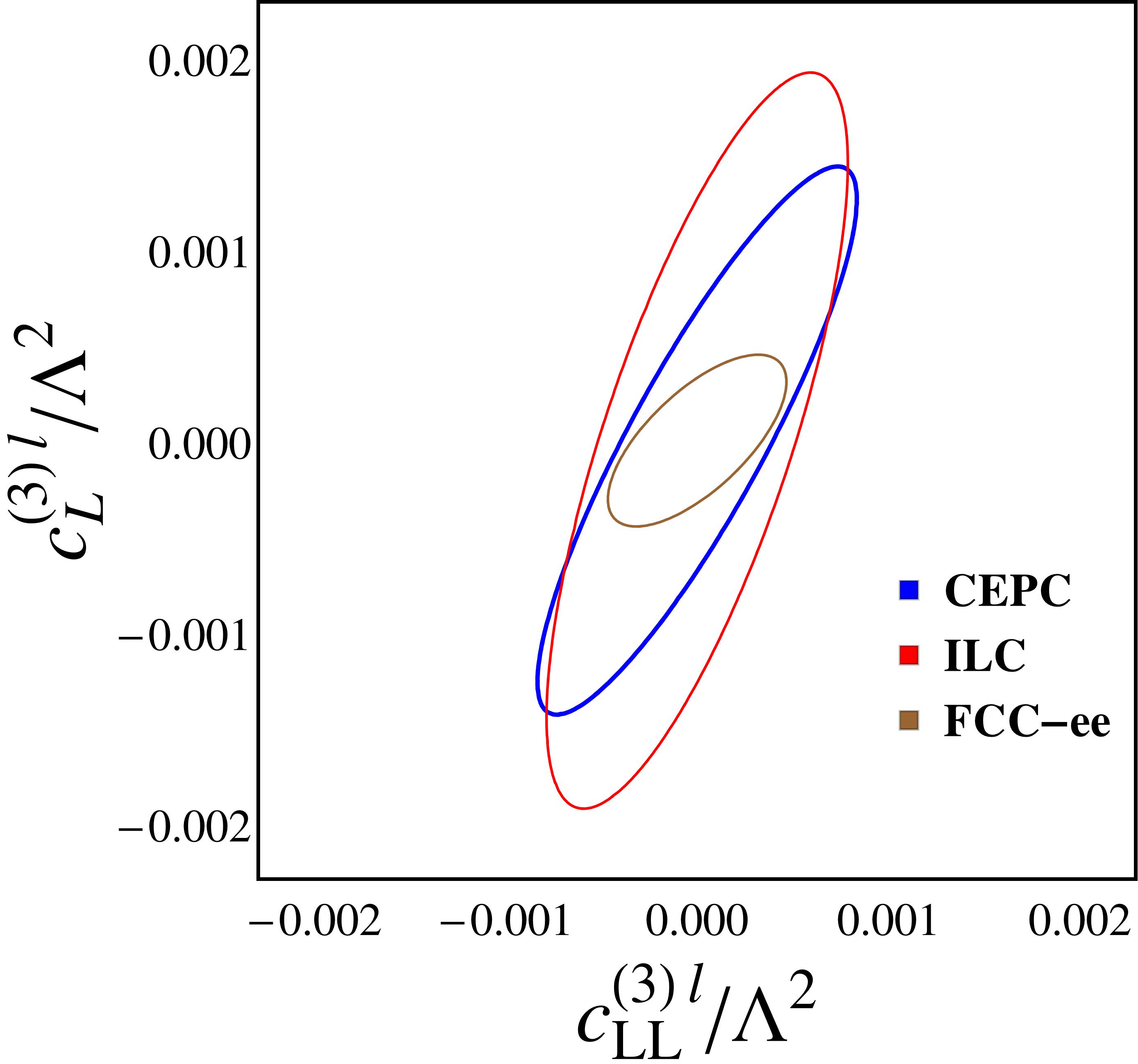}
  \includegraphics[width=5.cm]{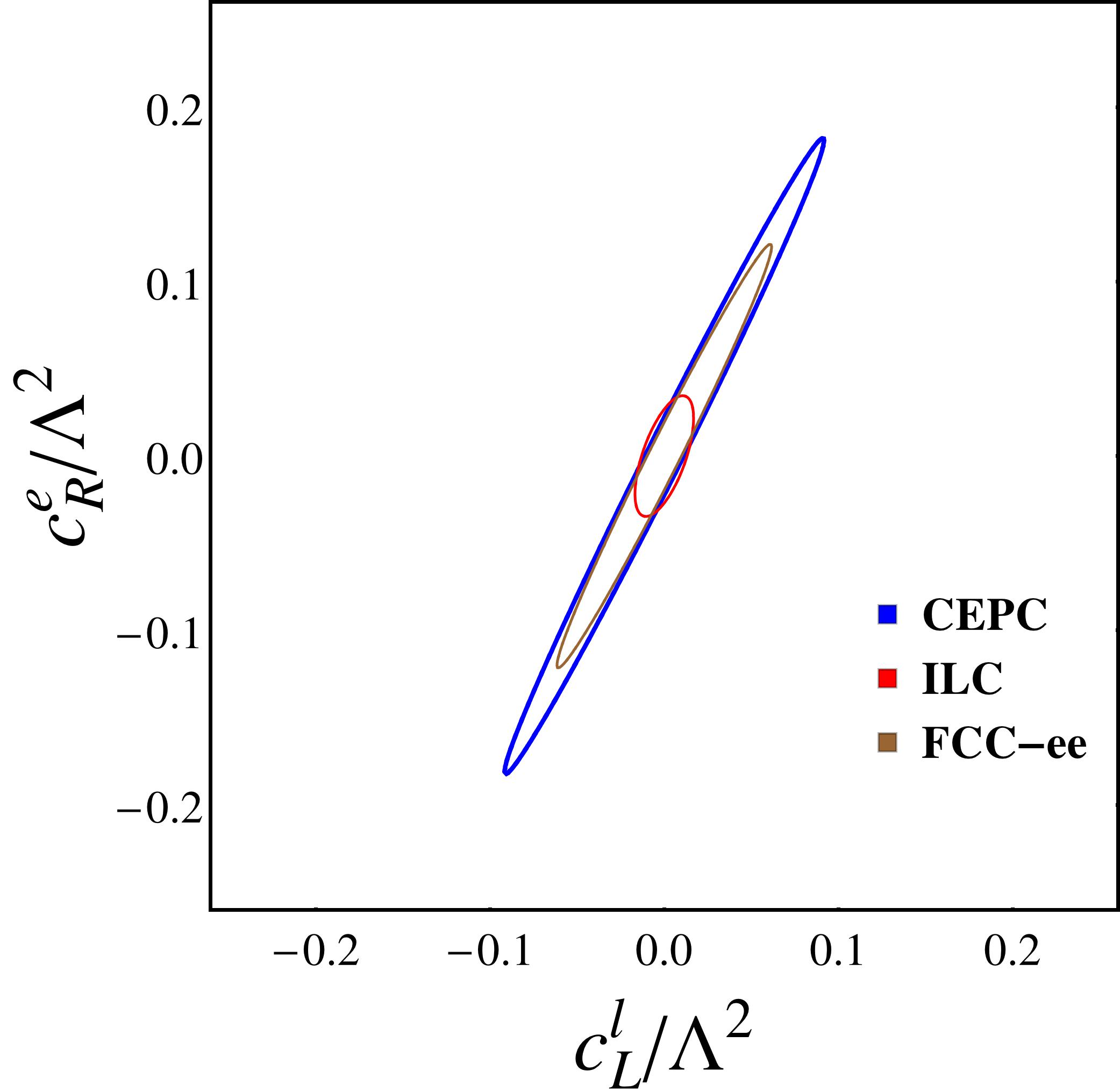}
  \includegraphics[width=5.3cm]{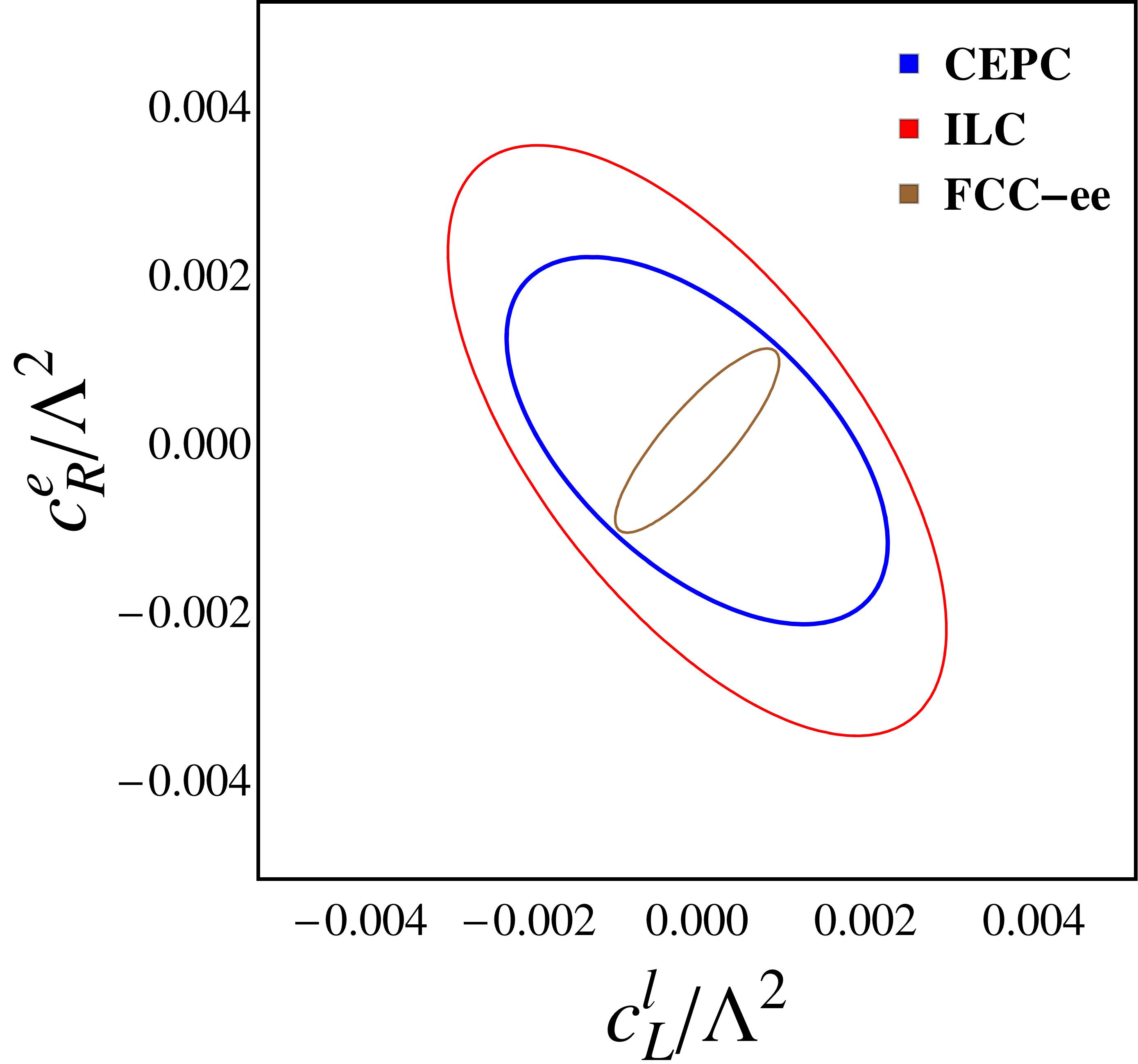}
\caption{``Optimistic'' (left panels) and ``conservative'' (right panels) 2D sensitivities at CEPC, FCC-ee and ILC. The coordinate axes are in the unit of (TeV)$^{-2}$.  In the ``optimistic'' analysis, only two 6D operators are turned on.  In the ``conservative''  analysis, all 6D operators listed in Table~\ref{tab:ops} are turned on, whereas the irrelevant Wilson coefficients are marginalized. } \label{comparison2D}
\end{figure}

\bibliographystyle{JHEP}
\bibliography{main}

\providecommand{\href}[2]{#2}\begingroup\raggedright\begin{thebibliography}{10}

\bibitem{Aad:2012tfa}
{\bf ATLAS} Collaboration, G.~Aad et~al., {\it {Observation of a new particle
  in the search for the Standard Model Higgs boson with the ATLAS detector at
  the LHC}},  {\em Phys. Lett.} {\bf B716} (2012) 1--29,
  [\href{http://arxiv.org/abs/1207.7214}{{\tt arXiv:1207.7214}}].

\bibitem{Chatrchyan:2012xdj}
{\bf CMS} Collaboration, S.~Chatrchyan et~al., {\it {Observation of a new boson
  at a mass of 125 GeV with the CMS experiment at the LHC}},  {\em Phys. Lett.}
  {\bf B716} (2012) 30--61, [\href{http://arxiv.org/abs/1207.7235}{{\tt
  arXiv:1207.7235}}].

\bibitem{Weiglein:2004hn}
{\bf LHC/LC Study Group} Collaboration, G.~Weiglein et~al., {\it {Physics
  interplay of the LHC and the ILC}},  {\em Phys. Rept.} {\bf 426} (2006)
  47--358, [\href{http://arxiv.org/abs/hep-ph/0410364}{{\tt hep-ph/0410364}}].

\bibitem{Baer:2013cma}
H.~Baer, T.~Barklow, K.~Fujii, Y.~Gao, A.~Hoang, S.~Kanemura, J.~List, H.~E.
  Logan, A.~Nomerotski, M.~Perelstein, et~al., {\it {The International Linear
  Collider Technical Design Report - Volume 2: Physics}},
  \href{http://arxiv.org/abs/1306.6352}{{\tt arXiv:1306.6352}}.

\bibitem{Asner:2013psa}
D.~M. Asner et~al., {\it {ILC Higgs White Paper}},  in {\em {Proceedings, 2013
  Community Summer Study on the Future of U.S. Particle Physics: Snowmass on
  the Mississippi (CSS2013): Minneapolis, MN, USA, July 29-August 6, 2013}},
  2013.
\newblock \href{http://arxiv.org/abs/1310.0763}{{\tt arXiv:1310.0763}}.

\bibitem{CEPC-SPPCStudyGroup:2015csa}
{\bf CEPC-SPPC} Collaboration, M.~Ahmad et~al., {\it {CEPC-SPPC Preliminary
  Conceptual Design Report. 1. Physics and Detector}},  2015.

\bibitem{Gomez-Ceballos:2013zzn}
{\bf TLEP Design Study Working Group} Collaboration, M.~Bicer et~al., {\it
  {First Look at the Physics Case of TLEP}},  {\em JHEP} {\bf 01} (2014) 164,
  [\href{http://arxiv.org/abs/1308.6176}{{\tt arXiv:1308.6176}}].

\bibitem{Hagiwara:1993ck}
K.~Hagiwara, S.~Ishihara, R.~Szalapski, and D.~Zeppenfeld, {\it {Low-energy
  effects of new interactions in the electroweak boson sector}},  {\em Phys.
  Rev.} {\bf D48} (1993) 2182--2203.

\bibitem{Grzadkowski:2010es}
B.~Grzadkowski, M.~Iskrzynski, M.~Misiak, and J.~Rosiek, {\it {Dimension-Six
  Terms in the Standard Model Lagrangian}},  {\em JHEP} {\bf 10} (2010) 085,
  [\href{http://arxiv.org/abs/1008.4884}{{\tt arXiv:1008.4884}}].

\bibitem{Giudice:2007fh}
G.~F. Giudice, C.~Grojean, A.~Pomarol, and R.~Rattazzi, {\it {The
  Strongly-Interacting Light Higgs}},  {\em JHEP} {\bf 06} (2007) 045,
  [\href{http://arxiv.org/abs/hep-ph/0703164}{{\tt hep-ph/0703164}}].

\bibitem{Elias-Miro:2013eta}
J.~Elias-MirÃ³, C.~Grojean, R.~S. Gupta, and D.~Marzocca, {\it {Scaling and
  tuning of EW and Higgs observables}},  {\em JHEP} {\bf 05} (2014) 019,
  [\href{http://arxiv.org/abs/1312.2928}{{\tt arXiv:1312.2928}}].

\bibitem{Elias-Miro:2013mua}
J.~Elias-Miro, J.~R. Espinosa, E.~Masso, and A.~Pomarol, {\it {Higgs windows to
  new physics through d=6 operators: constraints and one-loop anomalous
  dimensions}},  {\em JHEP} {\bf 11} (2013) 066,
  [\href{http://arxiv.org/abs/1308.1879}{{\tt arXiv:1308.1879}}].

\bibitem{Henning:2014wua}
B.~Henning, X.~Lu, and H.~Murayama, {\it {How to use the Standard Model
  effective field theory}},  {\em JHEP} {\bf 01} (2016) 023,
  [\href{http://arxiv.org/abs/1412.1837}{{\tt arXiv:1412.1837}}].

\bibitem{Pomarol:2013zra}
A.~Pomarol and F.~Riva, {\it {Towards the Ultimate SM Fit to Close in on Higgs
  Physics}},  {\em JHEP} {\bf 01} (2014) 151,
  [\href{http://arxiv.org/abs/1308.2803}{{\tt arXiv:1308.2803}}].

\bibitem{Craig:2014una}
N.~Craig, M.~Farina, M.~McCullough, and M.~Perelstein, {\it {Precision
  Higgsstrahlung as a Probe of New Physics}},  {\em JHEP} {\bf 03} (2015) 146,
  [\href{http://arxiv.org/abs/1411.0676}{{\tt arXiv:1411.0676}}].

\bibitem{Falkowski:2014tna}
A.~Falkowski and F.~Riva, {\it {Model-independent precision constraints on
  dimension-6 operators}},  {\em JHEP} {\bf 02} (2015) 039,
  [\href{http://arxiv.org/abs/1411.0669}{{\tt arXiv:1411.0669}}].

\bibitem{Corbett:2015ksa}
T.~Corbett, O.~J.~P. Eboli, D.~Goncalves, J.~Gonzalez-Fraile, T.~Plehn, and
  M.~Rauch, {\it {The Higgs Legacy of the LHC Run I}},  {\em JHEP} {\bf 08}
  (2015) 156, [\href{http://arxiv.org/abs/1505.05516}{{\tt arXiv:1505.05516}}].

\bibitem{Ellis:2015sca}
J.~Ellis and T.~You, {\it {Sensitivities of Prospective Future $e^+ e^-$
  Colliders to Decoupled New Physics}},  {\em JHEP} {\bf 03} (2016) 089,
  [\href{http://arxiv.org/abs/1510.04561}{{\tt arXiv:1510.04561}}].

\bibitem{Ge:2016zro}
S.-F. Ge, H.-J. He, and R.-Q. Xiao, {\it {Probing new physics scales from Higgs
  and electroweak observables at e$^{+}$ e$^{â}$ Higgs factory}},  {\em JHEP}
  {\bf 10} (2016) 007, [\href{http://arxiv.org/abs/1603.03385}{{\tt
  arXiv:1603.03385}}].

\bibitem{Ellis:2016yrj}
J.~Ellis, {\it {Prospects for Future Collider Physics}},  {\em Int. J. Mod.
  Phys.} {\bf A31} (2016), no.~33 1644002,
  [\href{http://arxiv.org/abs/1604.00333}{{\tt arXiv:1604.00333}}].

\bibitem{Durieux:2017rsg}
G.~Durieux, C.~Grojean, J.~Gu, and K.~Wang, {\it {The leptonic future of the
  Higgs}},  \href{http://arxiv.org/abs/1704.02333}{{\tt arXiv:1704.02333}}.

\bibitem{Jiang:2016czg}
Y.~Jiang and M.~Trott, {\it {On the non-minimal character of the SMEFT}},  {\em
  Phys. Lett.} {\bf B770} (2017) 108--116,
  [\href{http://arxiv.org/abs/1612.02040}{{\tt arXiv:1612.02040}}].

\bibitem{Amar:2014fpa}
G.~Amar, S.~Banerjee, S.~von Buddenbrock, A.~S. Cornell, T.~Mandal, B.~Mellado,
  and B.~Mukhopadhyaya, {\it {Exploration of the tensor structure of the Higgs
  boson coupling to weak bosons in e$^{+}$ e$^{â}$ collisions}},  {\em JHEP}
  {\bf 02} (2015) 128, [\href{http://arxiv.org/abs/1405.3957}{{\tt
  arXiv:1405.3957}}].

\bibitem{Jana:2017hqg}
S.~Jana and S.~Nandi, {\it {New Physics Scale from Higgs Observables with
  Effective Dimension-6 Operators}},
  \href{http://arxiv.org/abs/1710.00619}{{\tt arXiv:1710.00619}}.

\bibitem{Barklow:2017suo}
T.~Barklow, K.~Fujii, S.~Jung, R.~Karl, J.~List, T.~Ogawa, M.~E. Peskin, and
  J.~Tian, {\it {Improved Formalism for Precision Higgs Coupling Fits}},
  \href{http://arxiv.org/abs/1708.08912}{{\tt arXiv:1708.08912}}.

\bibitem{Barklow:2017awn}
T.~Barklow, K.~Fujii, S.~Jung, M.~E. Peskin, and J.~Tian, {\it
  {Model-Independent Determination of the Triple Higgs Coupling at e+e-
  Colliders}},  \href{http://arxiv.org/abs/1708.09079}{{\tt arXiv:1708.09079}}.

\bibitem{Gu:2017ckc}
J.~Gu, H.~Li, Z.~Liu, S.~Su, and W.~Su, {\it {Learning from Higgs Physics at
  Future Higgs Factories}},  \href{http://arxiv.org/abs/1709.06103}{{\tt
  arXiv:1709.06103}}.

\bibitem{FCC-ee}
\url{http://indico.ihep.ac.cn/event/6618/session/2/contribution/31/material/slides/0.pdf}.

\bibitem{Agashe:2004rs}
K.~Agashe, R.~Contino, and A.~Pomarol, {\it {The Minimal composite Higgs
  model}},  {\em Nucl. Phys.} {\bf B719} (2005) 165--187,
  [\href{http://arxiv.org/abs/hep-ph/0412089}{{\tt hep-ph/0412089}}].

\bibitem{Contino:2006qr}
R.~Contino, L.~Da~Rold, and A.~Pomarol, {\it {Light custodians in natural
  composite Higgs models}},  {\em Phys. Rev.} {\bf D75} (2007) 055014,
  [\href{http://arxiv.org/abs/hep-ph/0612048}{{\tt hep-ph/0612048}}].

\bibitem{ArkaniHamed:2002qy}
N.~Arkani-Hamed, A.~G. Cohen, E.~Katz, and A.~E. Nelson, {\it {The Littlest
  Higgs}},  {\em JHEP} {\bf 07} (2002) 034,
  [\href{http://arxiv.org/abs/hep-ph/0206021}{{\tt hep-ph/0206021}}].

\bibitem{Beneke:2014sba}
M.~Beneke, D.~Boito, and Y.-M. Wang, {\it {Anomalous Higgs couplings in angular
  asymmetries of $H \to Z\ell^{+} \ell^{-}$ and e$^{+}$ e$^{-} \to HZ$}},  {\em
  JHEP} {\bf 11} (2014) 028, [\href{http://arxiv.org/abs/1406.1361}{{\tt
  arXiv:1406.1361}}].

\bibitem{Craig:2015wwr}
N.~Craig, J.~Gu, Z.~Liu, and K.~Wang, {\it {Beyond Higgs Couplings: Probing the
  Higgs with Angular Observables at Future e$^{+}$ e$^{â}$ Colliders}},  {\em
  JHEP} {\bf 03} (2016) 050, [\href{http://arxiv.org/abs/1512.06877}{{\tt
  arXiv:1512.06877}}].

\bibitem{ALEPH:2005ab}
{\bf SLD Electroweak Group, DELPHI, ALEPH, SLD, SLD Heavy Flavour Group, OPAL,
  LEP Electroweak Working Group, L3} Collaboration, S.~Schael et~al., {\it
  {Precision electroweak measurements on the $Z$ resonance}},  {\em Phys.
  Rept.} {\bf 427} (2006) 257--454,
  [\href{http://arxiv.org/abs/hep-ex/0509008}{{\tt hep-ex/0509008}}].

\bibitem{dEnterria:2016sca}
D.~d'Enterria, {\it {Physics at the FCC-ee}},  in {\em {17th Lomonosov
  Conference on Elementary Particle Physics Moscow, Russia, August 20-26,
  2015}}, 2016.
\newblock \href{http://arxiv.org/abs/1602.05043}{{\tt arXiv:1602.05043}}.

\bibitem{Baak:2014ora}
{\bf Gfitter Group} Collaboration, M.~Baak, J.~CÃºth, J.~Haller, A.~Hoecker,
  R.~Kogler, K.~MÃ¶nig, M.~Schott, and J.~Stelzer, {\it {The global electroweak
  fit at NNLO and prospects for the LHC and ILC}},  {\em Eur. Phys. J.} {\bf
  C74} (2014) 3046, [\href{http://arxiv.org/abs/1407.3792}{{\tt
  arXiv:1407.3792}}].

\bibitem{Patrignani:2016xqp}
{\bf Particle Data Group} Collaboration, C.~Patrignani et~al., {\it {Review of
  Particle Physics}},  {\em Chin. Phys.} {\bf C40} (2016), no.~10 100001.

\bibitem{ATLAS:2014wva}
{\bf ATLAS, CDF, CMS, D0} Collaboration, {\it {First combination of Tevatron
  and LHC measurements of the top-quark mass}},
  \href{http://arxiv.org/abs/1403.4427}{{\tt arXiv:1403.4427}}.

\bibitem{Fan:2014vta}
J.~Fan, M.~Reece, and L.-T. Wang, {\it {Possible Futures of Electroweak
  Precision: ILC, FCC-ee, and CEPC}},  {\em JHEP} {\bf 09} (2015) 196,
  [\href{http://arxiv.org/abs/1411.1054}{{\tt arXiv:1411.1054}}].

\bibitem{Schael:2013ita}
{\bf DELPHI, OPAL, LEP Electroweak, ALEPH, L3} Collaboration, S.~Schael et~al.,
  {\it {Electroweak Measurements in Electron-Positron Collisions at
  W-Boson-Pair Energies at LEP}},  {\em Phys. Rept.} {\bf 532} (2013) 119--244,
  [\href{http://arxiv.org/abs/1302.3415}{{\tt arXiv:1302.3415}}].

\bibitem{Dam:2016ebi}
M.~Dam, {\it {Precision Electroweak measurements at the FCC-ee}},
  \href{http://arxiv.org/abs/1601.03849}{{\tt arXiv:1601.03849}}.

\bibitem{Baak:2013fwa}
M.~Baak et~al., {\it {Working Group Report: Precision Study of Electroweak
  Interactions}},  in {\em {Proceedings, 2013 Community Summer Study on the
  Future of U.S. Particle Physics: Snowmass on the Mississippi (CSS2013):
  Minneapolis, MN, USA, July 29-August 6, 2013}}, 2013.
\newblock \href{http://arxiv.org/abs/1310.6708}{{\tt arXiv:1310.6708}}.

\bibitem{Alloul:2013bka}
A.~Alloul, N.~D. Christensen, C.~Degrande, C.~Duhr, and B.~Fuks, {\it
  {FeynRules 2.0 - A complete toolbox for tree-level phenomenology}},  {\em
  Comput. Phys. Commun.} {\bf 185} (2014) 2250--2300,
  [\href{http://arxiv.org/abs/1310.1921}{{\tt arXiv:1310.1921}}].

\bibitem{Belyaev:2012qa}
A.~Belyaev, N.~D. Christensen, and A.~Pukhov, {\it {CalcHEP 3.4 for collider
  physics within and beyond the Standard Model}},  {\em Comput. Phys. Commun.}
  {\bf 184} (2013) 1729--1769, [\href{http://arxiv.org/abs/1207.6082}{{\tt
  arXiv:1207.6082}}].

\bibitem{Alwall:2011uj}
J.~Alwall, M.~Herquet, F.~Maltoni, O.~Mattelaer, and T.~Stelzer, {\it {MadGraph
  5 : Going Beyond}},  {\em JHEP} {\bf 06} (2011) 128,
  [\href{http://arxiv.org/abs/1106.0522}{{\tt arXiv:1106.0522}}].

\bibitem{Azatov:2016sqh}
A.~Azatov, R.~Contino, C.~S. Machado, and F.~Riva, {\it {Helicity selection
  rules and noninterference for BSM amplitudes}},  {\em Phys. Rev.} {\bf D95}
  (2017), no.~6 065014, [\href{http://arxiv.org/abs/1607.05236}{{\tt
  arXiv:1607.05236}}].

\bibitem{Goertz:2014qta}
F.~Goertz, A.~Papaefstathiou, L.~L. Yang, and J.~Zurita, {\it {Higgs boson pair
  production in the D=6 extension of the SM}},  {\em JHEP} {\bf 04} (2015) 167,
  [\href{http://arxiv.org/abs/1410.3471}{{\tt arXiv:1410.3471}}].

\bibitem{Azatov:2015oxa}
A.~Azatov, R.~Contino, G.~Panico, and M.~Son, {\it {Effective field theory
  analysis of double Higgs boson production via gluon fusion}},  {\em Phys.
  Rev.} {\bf D92} (2015), no.~3 035001,
  [\href{http://arxiv.org/abs/1502.00539}{{\tt arXiv:1502.00539}}].

\bibitem{Barr:2014sga}
A.~J. Barr, M.~J. Dolan, C.~Englert, D.~E. Ferreira~de Lima, and M.~Spannowsky,
  {\it {Higgs Self-Coupling Measurements at a 100 TeV Hadron Collider}},  {\em
  JHEP} {\bf 02} (2015) 016, [\href{http://arxiv.org/abs/1412.7154}{{\tt
  arXiv:1412.7154}}].

\bibitem{He:2015spf}
H.-J. He, J.~Ren, and W.~Yao, {\it {Probing new physics of cubic Higgs boson
  interaction via Higgs pair production at hadron colliders}},  {\em Phys.
  Rev.} {\bf D93} (2016), no.~1 015003,
  [\href{http://arxiv.org/abs/1506.03302}{{\tt arXiv:1506.03302}}].

\bibitem{Degrassi:2016wml}
G.~Degrassi, P.~P. Giardino, F.~Maltoni, and D.~Pagani, {\it {Probing the Higgs
  self coupling via single Higgs production at the LHC}},  {\em JHEP} {\bf 12}
  (2016) 080, [\href{http://arxiv.org/abs/1607.04251}{{\tt arXiv:1607.04251}}].

\bibitem{DiLuzio:2017tfn}
L.~Di~Luzio, R.~Grber, and M.~Spannowsky, {\it {Maxi-sizing the trilinear
  Higgs self-coupling: how large could it be?}},  {\em Eur. Phys. J.} {\bf C77}
  (2017), no.~11 788, [\href{http://arxiv.org/abs/1704.02311}{{\tt
  arXiv:1704.02311}}].

\bibitem{Sanz:2017tco}
V.~Sanz and J.~Setford, {\it {Composite Higgs Models after Run 2}},  {\em Adv.
  High Energy Phys.} {\bf 2018} (2018) 7168480,
  [\href{http://arxiv.org/abs/1703.10190}{{\tt arXiv:1703.10190}}].

\bibitem{Banerjee:2017wmg}
A.~Banerjee, G.~Bhattacharyya, N.~Kumar, and T.~S. Ray, {\it {Constraining
  Composite Higgs Models using LHC data}},  {\em JHEP} {\bf 03} (2018) 062,
  [\href{http://arxiv.org/abs/1712.07494}{{\tt arXiv:1712.07494}}].

\bibitem{ATLAS:2016btu}
{\bf ATLAS} Collaboration, T.~A. collaboration, {\it {Search for new phenomena
  in $t\bar{t}$ final states with additional heavy-flavour jets in $pp$
  collisions at $\sqrt{s}=13$ TeV with the ATLAS detector}}, .

\bibitem{ATLAS:2017lvm}
{\bf ATLAS} Collaboration, T.~A. collaboration, {\it {Search for pair
  production of vector-like top quarks in events with one lepton and an
  invisibly decaying Z boson in $\sqrt{s} = 13$ TeV pp collisions at the ATLAS
  detector}}, .

\bibitem{Sirunyan:2017pks}
{\bf CMS} Collaboration, A.~M. Sirunyan et~al., {\it {Search for pair
  production of vector-like quarks in the bW$\overline{\mathrm{b}}$W channel
  from proton-proton collisions at $\sqrt{s} =$ 13 TeV}},  {\em Phys. Lett.}
  {\bf B779} (2018) 82--106, [\href{http://arxiv.org/abs/1710.01539}{{\tt
  arXiv:1710.01539}}].

\bibitem{Sirunyan:2017usq}
{\bf CMS} Collaboration, A.~M. Sirunyan et~al., {\it {Search for pair
  production of vector-like T and B quarks in single-lepton final states using
  boosted jet substructure in proton-proton collisions at $\sqrt{s}=13$ TeV}},
  {\em JHEP} {\bf 11} (2017) 085, [\href{http://arxiv.org/abs/1706.03408}{{\tt
  arXiv:1706.03408}}].

\bibitem{PM}
\url{http://astronomy.swin.edu.au/~cblake/StatsLecture3.pdf}.

\end{thebibliography}\endgroup

\end{document}